\documentclass[aps,nofootinbib,notitlepage,superscriptaddress]{revtex4-1}
\usepackage{amsmath}
\usepackage{amssymb}
\usepackage{amsfonts}
\usepackage{graphicx}
\usepackage{hyperref}
\usepackage{color}
\usepackage{multirow}
\usepackage{mathtools}
\usepackage{booktabs}
\usepackage{amsthm}
\usepackage{mathtools}
\usepackage{color}
\usepackage{tabularx}
\usepackage{url}

\renewcommand{\thetable}{\arabic{table}}%

\hypersetup{pdftitle=Network geometry and market instability}

\begin{document}
\title{Network geometry and market instability}

\author{Areejit Samal}
\thanks{These authors contributed equally to this work}
\affiliation{The Institute of Mathematical Sciences (IMSc), 
Homi Bhabha National Institute (HBNI), Chennai 600113 India}
\author{Hirdesh Kumar Pharasi}
\thanks{These authors contributed equally to this work}
\affiliation{Instituto de Ciencias F\'{i}sicas, Universidad 
Nacional Aut\'{o}noma de M\'{e}xico, Cuernavaca 62210, M\'{e}xico}
\author{Sarath Jyotsna Ramaia}
\affiliation{Department of Applied Mathematics and Computational
Sciences, PSG College of Technology, Coimbatore 641004, India}
\author{Harish Kannan}
\affiliation{Department of Mathematics,  University of California San Diego, 
La Jolla, California  92093, USA}
\author{Emil Saucan}
\affiliation{Department of Applied Mathematics, ORT Braude College, 
Karmiel 2161002, Israel}
\author{J\"{u}rgen Jost}
\affiliation{Max Planck Institute for Mathematics in the Sciences, Leipzig 04103
Germany}
\affiliation{The Santa Fe Institute, Santa Fe, New Mexico 87501, USA}
\author{Anirban Chakraborti}
\affiliation{School of Computational and Integrative Sciences,
Jawaharlal Nehru University, New Delhi 110067, India}
\affiliation{Centre for Complexity Economics, Applied Spirituality and 
Public Policy (CEASP), Jindal School of Government and Public Policy, 
O.P. Jindal Global University, Sonipat 131001, India}
\affiliation{Centro Internacional de Ciencias, Cuernavaca 62210,
M\'{e}xico}

\begin{abstract}
The complexity of financial markets arise from the strategic interactions 
among agents trading stocks, which manifest in the form of vibrant correlation 
patterns among stock prices. Over the past few decades, complex financial 
markets have often been represented as networks whose interacting pairs 
of nodes are stocks, connected by edges that signify 
the correlation strengths. However, we often have interactions that occur in 
groups of three or more nodes, and these cannot be described simply by pairwise 
interactions but we also need to take the relations between these interactions 
into account. Only recently, researchers have started devoting attention to the 
higher-order architecture of complex financial systems, that can significantly 
enhance our ability to estimate systemic risk as well as measure the robustness 
of financial systems in terms of market efficiency. Geometry-inspired network 
measures, such as the Ollivier-Ricci curvature and Forman-Ricci curvature, can 
be used to capture the network fragility and continuously monitor financial 
dynamics. Here, we explore the utility of such discrete Ricci curvatures 
in characterizing the structure of financial systems, and further, evaluate 
them as generic indicators of the market instability. For this purpose, we 
examine the daily returns from a set of stocks comprising the USA S\&P-500 and 
the Japanese Nikkei-225 over a 32-year period, and monitor the changes in the 
edge-centric network curvatures. We find that the different geometric measures 
capture well the system-level features of the market and hence we can distinguish
between the normal or `business-as-usual' periods and all the major market 
crashes. This can be very useful in strategic designing of financial systems and 
regulating the markets in order to tackle financial instabilities.
\end{abstract}

\maketitle

\section{Introduction}

For centuries science had thrived on the method of reductionism-- considering 
the units of a system in isolation, and then trying to understand and infer 
about the whole system. However, the simple method of reductionism has severe 
limitations \cite{Anderson1972}, and fails to a large extent when it comes to 
the understanding and modeling the collective behavior of the components of a 
`complex system'. More and more systems are now being identified as complex 
systems, and hence scientists are now embracing the idea of complexity as one
 of the governing principles of the world we live in. Any deep understanding of 
a complex system has to be based on a system-level description, since a key 
ingredient of any complex system is the rich interplay of nonlinear interactions 
between the system components. The financial market is truly a spectacular 
example of such a complex system, where the agents interact strategically to 
determine the best prices of the assets. So new tools and interdisciplinary 
approaches are needed \cite{Vemuri1978,Gellmann1995}, and already there has 
been an influx of ideas from econophysics and complexity theory \cite{Mantegna2007,Bouchaud2003,Sinha2010,Chakraborti2011a,Chakraborti2015} 
to explain and understand economic and financial markets.

The traditional economic theories, based on axiomatic approaches and consequently 
less predictive power, could not foresee event like the sub-prime crisis of 2007-2008 
or the long-lasting effects of such a critical financial crash on the global economy.
Researchers advocated that new concepts and techniques \cite{battiston2016complexity} 
like tipping points, feedback, contagion, network analysis along with the use of 
complexity models \cite{lux2009economics} could help in better understanding of 
highly interconnected economic and financial systems, as well as monitoring them. 
There have been numerous papers in the past that have addressed similar concerns 
and tried to adopt new approaches for studying financial systems. Since the 
correlations among stocks change with time, the underlying market dynamics generate 
very interesting correlation-based networks evolving over time. The study of empirical 
cross-correlations among stock prices goes back to more than two decades 
\cite{mantegna1999information,Mantegna1999hierarchical,laloux_1999,plerou_1999,
Gopikrishnan_pre2001,kullmann2002time}. One of commonly adopted approaches for the 
modeling and analysis of complex financial systems has been correlation-based networks, 
and it has emerged as an important tool \cite{mantegna1999information, Mantegna1999hierarchical,Plerou2002,Onnela2003,Tumminello2005,Pharasi2018,Pharasi2019,
Chakraborti2020}. 

A network or graph consists of nodes connected by edges. In real-world 
networks, nodes represent the components or entities, while edges represent the interactions or relationships between nodes. In the context of financial markets, the nodes represent 
the stocks and the edges characterize the correlation strengths (or their transformations 
into distance measures). The network formed by connecting stocks of highly correlated prices, 
price returns, and trading volumes are all scale-free, with a relatively small number of 
stocks influencing the majority of the stocks in the market \cite{chi2010network}. 
Hierarchical clustering has been used to cluster stocks into sectors and sub-sectors, 
and their network analysis provides additional information on the interrelationships 
among the stocks \cite{tumminello2007correlation,tumminello2010correlation}. The 
cross-correlations among stock returns allow one to construct other correlation-based 
networks such as minimum spanning tree (MST)  \cite{mantegna1999information,Mantegna1999hierarchical,Onnela2003,micciche2003degree} 
or a threshold network  \cite{Kumar2012}. Another approach to monitor the correlation-based 
networks over time, referred to as structural entropy, quantifies the structural 
changes of the network as a single parameter. It takes into account the number of 
communities as well as the size of the communities \cite{almog2019} 
to determine the structural entropy, which is then used to continuously monitor the market. 
The thermodynamical entropy  \cite{wang2019thermodynamic} can also be used to describe 
the dynamics of stock market networks as it acts like an indicator for the financial system. 
Very recently, based on the distribution properties of the eigenvector centralities 
of correlation matrices, Chakraborti et al. \cite{chakraborti2020phase} have proposed 
a computationally cheap yet uniquely defined and non-arbitrary eigen-entropy measure, 
to show that the financial market undergoes `phase separation' and there exists a new 
type of scaling behavior (data collapse) in financial markets. Further, a recent review 
by Kukreti et al. \cite{kukreti2020} critically examines correlation-based networks and 
entropy approaches in evolving financial systems. To understand the topology of the 
correlation-based networks as well as to define the complexity, a volume-based dimension 
has also been proposed by Nie et al. \cite{nie2018relationship}. There have also been some novel 
studies where the financial market has been considered as a quasi-stationary system, 
and then the ensuing dynamics have been studied \cite{stepanov2015stability,rinn2015dynamics, chetalova2015zooming,Heckens_2020,wang2020quasi}. 

Introduced long ago by Gauss and Riemann, curvature is a central concept in 
geometry that quantifies the extent to which a space is curved \cite{Jost2017}. 
In geometry, the primary invariant is curvature in its many forms. While 
curvature has connections to several essential aspects of the underlying space, 
in a specific case, curvature has a connection to the Laplacian, and hence, to 
the `heat kernel' on a network. Curvature also has connections to the Brownian 
motion and entropy growth on a network. Moreover, curvature is also related to algebraic topological aspects, such as the homology groups and Betti numbers, 
which are relevant, for instance, for persistent homology and topological data 
analysis \cite{Carlsson2009}. 
Recently, there has been immense interest in geometrical characterization 
of complex networks \cite{Krioukov2010,Bianconi2015,Sandhu2015,Sreejith2016, Bianconi2017}. 
Network geometry can reveal higher-order correlations 
between nodes beyond pairwise relationships captured by edges connecting 
two nodes in a graph \cite{Kartun-Giles2019,Iacopini2019,Kannan2019}. From 
the point of view of structure and dynamics of complex networks, edges are 
more important than nodes, since the nodes by themselves cannot constitute a 
meaningful network. Hence, it may be more important to develop edge-centric 
measures rather than node-centric measures to characterize the structure of 
complex networks \cite{Sreejith2016,Samal2018}. 

Surprisingly, geometrical concepts, especially, discrete notions of Ricci 
curvature, have only very recently been used as edge-centric network measures 
\cite{Sandhu2015,Ni2015,Sreejith2016,Sandhu2016,Samal2018,Ni2019}. Furthermore, 
curvature has deep connections to related evolution equations that can be used 
to predict the long-time evolution of networks. Although the importance of 
geometric measures like curvature have been understood for quite some time, 
yet there has been limited number of applications in the context of complex 
financial networks. In particular, Sandhu et al. \cite{Sandhu2016} studied the 
evolution of Ollivier-Ricci curvature \cite{Ollivier2007,Ollivier2009} in 
threshold networks for the USA S\&P-500 market over a 15-year span (1998-2013) 
and showed that the Ollivier-Ricci curvature is correlated to the increase 
in market network fragility. Consequently, Sandhu et al. \cite{Sandhu2016} 
suggested that the Ollivier-Ricci curvature can be employed as an indicator of 
market fragility and study the designing of (banking) systems and framing 
regulation policies to combat financial instabilities such as the sub-prime 
crisis of 2007-2008. In this paper, we expand the study of geometry-inspired 
network measures for characterizing the structure of the financial systems to 
four notions of discrete Ricci curvature, and evaluate the curvature measures as 
generic indicators of the market instability. 

It is noteworthy that in the present paper, the term `curvature' refers to 
four notions of discrete Ricci curvature investigated here, which are as 
such intrinsic curvatures, and not extrinsic curvatures as has been considered elsewhere 
in the context of complex networks (see e.g., Aste et al. \cite{aste2005complex}). 
Recall that extrinsic geometry is given by embedding the networks in a suitable 
ambient space (which in practice is the hyperbolic plane or space), and thereafter, 
the geometric properties induced by the embedding space are studied (see, e.g. 
\cite{Saucan2020}). While this approach is intuitive and conducive to simple 
illustrations, such network embeddings are distorting, except for the special case 
of isometric embeddings. In contrast, the intrinsic approach to networks is independent 
of any specific embedding, and hence, of the necessary additional computations and 
any distortion. Moreover, such an intrinsic approach allows for the independent study of such 
powerful tools as the Ricci flow, without the vagaries associated with the 
embedding in an ambient space of certain dimension (see, e.g. \cite{saucan2012isometric}). 
Furthermore, the Ollivier-Ricci curvature has been employed to show that the 
`backbone' of certain real-world networks is indeed tree-like, hence 
intrinsically hyperbolic \cite{Ni2015}. Specific to financial networks, 
Sandhu et al. \cite{Sandhu2016} have shown that Ollivier-Ricci curvature, 
which is of course an intrinsic curvature, presents a powerful tool in the 
detection of financial market crashes. In this work, we have considered 
three additional notions \cite{Sreejith2016,Saucan2020} of discrete Ricci 
curvature for the study of financial networks.

In the present paper, we examine the daily returns from a set of stocks 
comprising the USA S\&P-500 and the Japanese Nikkei-225 over a 32-year 
period, and monitor the changes in the edge-centric geometric curvatures. 
A major goal of this research is to evaluate different notions of discrete 
Ricci curvature for their ability to unravel the structure of complex 
financial networks and serve as indicators of market instabilities. 
Our study confirms that during a normal period the market is very modular 
and heterogeneous, whereas during an instability (crisis) the market is more 
homogeneous, highly connected and less modular 
\cite{Onnela2003,Scheffer2012,Pharasi2019,Chakraborti2020}. Further, we 
find that the discrete Ricci curvature measures, especially Forman-Ricci curvature \cite{Sreejith2016,Samal2018}, capture well the system-level features of 
the market and hence we can distinguish between the normal or `business-as-usual' 
periods and all the major market crises (bubbles and crashes).  
Importantly, among four Ricci-type curvature measures, the
Forman-Ricci curvature of edges correlates highest with the traditional 
market indicators and acts as an excellent indicator for the system-level 
fear (volatility) and fragility (risk) for both the markets. We also find 
using these geometric measures that there are succinct and inherent differences 
in the two markets, USA S\&P-500 and Japan Nikkei-225. These new insights 
will help us to understand tipping points, systemic risk, and resilience in 
financial networks, and enable us to develop monitoring tools required for the 
highly interconnected financial systems and perhaps forecast future financial 
crises and market slowdowns.


\section{Ricci-type curvatures for edge-centric analysis of networks}

The classical notion of Ricci curvature applies to smooth manifolds, and its 
classical definition requires tensors and higher-order derivatives \cite{Jost2017}. 
Thus, the classical definition of Ricci curvature is not immediately applicable 
in the discrete context of graphs or networks. Therefore, in order to develop 
any meaningful notion of Ricci curvature for networks, one has to inspect the 
essential geometric properties captured by this curvature notion, and find their 
proper analogues for discrete networks. To this end, it is essential to recall 
that Ricci curvature quantifies two essential geometric properties of the manifold, 
namely, volume growth and dispersion of geodesics. 
See Electronic Supplementary Material (ESM) Figure S1 for a schematic 
illustration of the Ricci curvature. 
Further, since classical Ricci curvature is associated to a vector (direction) 
in smooth manifolds \cite{Jost2017}, 
in the discrete case of networks, it is naturally assigned to edges \cite{Samal2018}. 
Thus, notions of discrete Ricci curvatures are associated to edges rather than 
vertices or nodes in networks \cite{Samal2018}. Note that no discretization of 
Ricci curvature for networks can capture the full spectrum of properties of the 
classical Ricci curvature defined on smooth manifolds, and thus, each discretization 
can shed a different light on the analyzed networks \cite{Samal2018}. In this work, 
we apply four notions of discrete Ricci curvature for networks to study the 
correlation-based networks of stock markets.


\subsection*{Ollivier-Ricci curvature}

Ollivier's discretization \cite{Ollivier2007,Ollivier2009} of the classical 
Ricci curvature has been extensively used to analyze graphs or networks 
\cite{Lin2010,Lin2011,Bauer2012,Jost2014,Ni2015,Sandhu2015,Sandhu2016,Samal2018,
Ni2019,Sia2019}. Ollivier's definition is based on the following observation. 
In spaces of positive curvature, balls are closer to each other on the average 
than their centers, while in spaces of negative curvature, balls are farther 
away on the average than their centers 
(ESM Figure S2).
Ollivier's definition extends this observation from balls (volumes) to measures 
(probabilities). More precisely, the Ollivier-Ricci (OR) curvature of an edge 
$e$ between nodes $u$ and $v$ is defined as 
\begin{equation} 
\label{OllivierRicciEdge}
\mathbf{O}(e)  = 1 - \frac{W_1(m_u,m_v)}{d(u,v)}\,
\end{equation}
where $m_u$ and $m_v$ represent measures concentrated at nodes $u$ and $v$, 
respectively, $W_1$ denotes the Wasserstein distance \cite{Vaserstein1969} 
(also known as the earth mover's distance) between the discrete probability 
measures $m_u$ and $m_v$, and the cost $d(u,v)$ is the distance between nodes 
$u$ and $v$, respectively. Moreover, the Wasserstein distance $W_1(m_u, m_v)$ 
which gives the transportation distance between the two measures $m_u$ and 
$m_v$, is given by
\begin{equation}
\label{Wasserstein}
W_1(m_u, m_v)=\inf_{\mu_{u,v}\in \prod(m_u, m_v)}\sum_{(u',v')\in V\times V}
d(u', v')\mu_{u,v}(u', v'),
\end{equation}
with $\prod(m_u, m_v)$ being the set of probability measures $\mu_{u,v}$ that 
satisfy
\begin{equation}
\sum_{v'\in V}\mu_{u,v}(u', v')=m_u(u'), \,\,\sum_{u'\in V}\mu_{u,v}(u', v')
=m_v(v')
\end{equation}
where $V$ is the set of nodes in the graph. The above equation represents all 
the transportation possibilities of the mass $m_u$ to $m_v$. $W_1(m_u, m_v)$ 
is the minimal cost or distance to transport the mass of $m_u$ to that of $m_v$. 
Note that the distance $d(u',v')$ in Eq. \ref{Wasserstein} is taken to be the 
path distance in the unweighted or weighted graph. Furthermore, the probability 
distribution $m_u$ for $u \in V$ has to be specified, and this is chosen to be 
uniform over neighbouring nodes of $u$ \cite{Lin2011}. 

Simply stated, to determine the OR curvature of an edge $e$, in Eq. 
\ref{OllivierRicciEdge} one compares the average distance between the neighbours 
of the nodes $u$ and $v$ anchoring the edge $e$ in an optimal arrangement with 
the distance between $u$ and $v$ itself. Importantly, the average distance 
between neighbours of $u$ and $v$ is evaluated as an optimal transport problem 
wherein the neighbours of $u$ are coupled with those of $v$ in such a manner that 
the average distance is as small as possible. In the setting of discrete graphs 
or networks, OR curvature by definition captures the volume growth aspect of the 
classical notion for smooth manifolds, see e.g. \cite{Samal2018} for details. 
In this work, we have computed the average OR curvature of edges (ORE) in 
undirected and weighted networks using Eq. \ref{OllivierRicciEdge}. 


\subsection*{Forman-Ricci curvature}

Forman's approach to the discretization of Ricci curvature \cite{Forman2003} 
is more algebraic in nature and is based on the relation between the 
Riemannian Laplace operator and Ricci curvature. While devised originally for 
a much larger class of discrete geometric objects than graphs, an adaptation 
to network setting was recently introduced by some of us \cite{Sreejith2016}. 
The Forman-Ricci (FR) curvature $\mathbf{F}(e)$ of an edge $e$ in an 
undirected network with weights assigned to both edges and nodes is given by 
\cite{Sreejith2016}
\begin{equation}
\label{FormanRicciEdge}
\mathbf{F}(e) = w_e \left( \frac{w_{v_1}}{w_e} +  \frac{w_{v_2}}{w_e}  
- \sum_{e_{v_1}\ \sim\ e,\ e_{v_2}\ \sim\ e} 
\left[\frac{w_{v_1}}{\sqrt{w_e w_{e_{v_1} }}} 
+ \frac{w_{v_2}}{\sqrt{w_e w_{e_{v_2} }}} \right] \right)\,
\end{equation}
where $e$ denotes the edge under consideration between nodes $v_1$ and $v_2$, 
$w_e$ denotes the weight of the edge $e$, $w_{v_1}$ and $w_{v_2}$ denote the 
weights associated with the nodes $v_1$ and $v_2$, respectively, $e_{v_1} \sim 
e$ and $e_{v_2} \sim e$ denote the set of edges incident on nodes $v_1$ and 
$v_2$, respectively, after excluding the edge $e$ under consideration which 
connects the two nodes $v_1$ and $v_2$. Furthermore, some of us have also 
extended the notion of FR curvature to directed networks \cite{Saucan2019a}.
In case of discrete networks, FR curvature captures the geodesic dispersal 
property of the classical notion \cite{Samal2018}. 
In ESM Figure S3, we illustrate, using a simple example, the 
computation of FR curvature in an undirected graph.
In this work, we have computed the average FR curvature of edges (FRE) in 
undirected and weighted networks using Eq. \ref{FormanRicciEdge}.

From a geometric perspective, the FR curvature quantifies the information 
spread at the ends of edges in a network (Figure \ref{fig:schematic}; ESM Figure S3). 
The higher the information spread at the ends of an edge, the more negative will 
be the value of its FR curvature. Specifically, an edge with high negative FR 
curvature is likely to have several neighbouring edges connected to both 
anchoring nodes, and moreover, such an edge can be seen as a funnel at both 
ends, connecting many other nodes. Intuitively, such an edge with high 
negative FR curvature can be expected to have high edge betweenness centrality 
as many shortest paths between other nodes, including those quite far in the 
network, are also likely to pass through this edge. Previously, some of us 
have empirically shown a high statistical correlation between FR curvature and 
edge betweenness centrality in diverse networks \cite{Sreejith2017,Samal2018}.


\subsection*{Menger-Ricci curvature}

The remaining two curvatures studied here are adaptations of curvatures for 
metric spaces to discrete graphs. Indeed, both unweighted and weighted graphs 
can be viewed as a metric space where the distance between any two nodes can 
be specified by the path length between them. Among notions of metric, and 
indeed, discrete curvature, Menger \cite{Menger1930} has proposed the simplest 
and earliest definition whereby he defines the curvature of metric triangles 
$T$ formed by three points in the space as the reciprocal $1/R(T)$ of the 
radius $R(T)$ of the circumscribed circle of a triangle $T$. Recently, some of 
us \cite{Saucan2019b,Saucan2020} have adapted Menger's definition to networks. 
Let $(M,d)$ be a metric space and $T = T(a,b,c)$ be a triangle with sides 
$a,\ b,\ c$, then the Menger curvature of $T$ is given by
\begin{equation}
K_M(T) = \frac{\sqrt{p(p - a)(p - b)(p - c)}}{a \cdot b \cdot c}\,
\end{equation}
where $p = (a + b + c)/2$. In the particular case of a combinatorial triangle 
with each side of length 1, the above formula gives $K_M(T)=\sqrt{3}/2$. 
Furthermore, it is clear from the above formula that Menger curvature is 
always positive. Following the differential geometric approach, the Menger-Ricci 
(MR) curvature of an edge $e$ in a network can be defined as 
\cite{Saucan2019b,Saucan2020}
\begin{equation}
\label{MengerRicciEdge}
\kappa_M(e) = \sum_{T_e \sim e}\kappa_{M}(T_e)\,,
\end{equation}
where $T_e \sim e$ denote the triangles adjacent to the edge $e$. Intuitively, 
if an edge is part of several triangles in the network, such an edge will have 
high positive MR curvature (Figure \ref{fig:schematic}). 
In ESM Figure S4, we illustrate, using a simple example, the 
computation of MR curvature in an undirected graph.
In this work, we have computed the average MR curvature of edges (MRE) in 
undirected financial networks by ignoring the edge weights and using 
Eq. \ref{MengerRicciEdge}. 


\subsection*{Haantjes-Ricci curvature}

We have also applied another notion of metric curvature to networks which is 
based on the suggestion of Finsler and was developed by his student Haantjes 
\cite{Haantjes1947}. Haantjes defined the curvature of a metric curve as the 
ratio between the length of an arc of the curve and that of the chord it 
subtends. More precisely, given a curve $c$ in a metric space $(M,d)$, and 
given three points $p, q, r$ on $c$, $p$ between $q$ and $r$, the Haantjes 
curvature at the point $p$ is defined as
\begin{equation}                         
\kappa_{H}^2(p) = 24\lim_{q,r \rightarrow p}\frac{l(\widehat{qr})-d(q,r)}
{\big(d(q,r)\big)^3}\,,
\end{equation}
where $l(\widehat{qr})$ denotes the length, in the intrinsic metric induced 
by $d$, of the arc $\widehat{qr}$. In networks, $\widehat{qr}$ can be replaced 
by a path $\pi = v_0,v_1,\ldots,v_n$ between two nodes $v_0$ and $v_n$, and the 
subtending chord by the edge $e = (v_0,v_n)$ between the two nodes. Recently, 
some of us \cite{Saucan2019b,Saucan2020} have defined the Haantjes curvature of 
such a simple path $\pi$ as
\begin{equation}                         
\kappa_{H}^2(\pi) = \frac{l(\pi)-l(v_0,v_n)}{l(v_0,v_n)^3}\,,
\end{equation}
where, if the graph is a metric graph, $l(v_0,v_n) = d(v_0,v_n)$, that is the 
shortest path distance between nodes $v_0$ and $v_n$. In particular, for the 
combinatorial metric (or unweighted graphs), we obtain that $\kappa_H(\pi) = 
\sqrt{n-1}$, where $\pi = v_0,v_1,\ldots,v_n$ is as above. Note that considering 
simple paths in graphs concords with the classical definition of Haantjes curvature, 
since a metric arc is, by its very definition, a simple curve. Thereafter, the 
Haantjes-Ricci (HR) curvature of an edge $e$ \cite{Saucan2019b,Saucan2020} can 
be defined as
\begin{equation}
\label{HaantjesRicciEdge}
\kappa_H(e) = \sum_{\pi \sim e}\kappa_{H}(\pi)\,,
\end{equation}
where $\pi \sim e$ denote the paths that connect the nodes anchoring the edge 
$e$. Note that while MR curvature considers only triangles or simple paths of 
length 2 between two nodes anchoring an edge in unweighted graphs, the HR 
curvature considers even longer paths between the same two nodes anchoring an 
edge (Figure \ref{fig:schematic}). Moreover, for triangles endowed with the 
combinatorial metric, the two notions by Menger and Haantjes coincide, up to a 
universal constant. 
In ESM Figure S4, we illustrate, using a simple example, the 
computation of HR curvature in an undirected graph.
In this work, we have computed the average HR curvature of 
edges (HRE) in undirected financial networks by ignoring the edge weights and 
using Eq. \ref{HaantjesRicciEdge}. Moreover, due to computational constraints, 
we only consider simple paths $\pi$ of length $\le 4$ between the two vertices 
at the ends of any edge while computing its HR curvature using Eq.
\ref{HaantjesRicciEdge} in analyzed networks. Note that both Menger and Haantjes 
curvature are positive in undirected networks, and they capture the (absolute 
value of) geodesics dispersal rate of the classical Ricci curvature. 


\section{Data and Methods}

\subsection*{Data description}

The data was collected from the public domain of Yahoo finance database 
\cite{Yahoo_finance} for two markets: USA S\&P-500 index and Japanese 
Nikkei-225 index. The investigation in this work spans a 32-year period from 
2 January 1985 (02-01-1985) to 30 December 2016 (30-12-2016). We analyzed the 
daily closure price data of $N=194$ stocks for $T=8068$ days for USA S\&P-500 
and $N=165$ stocks for $T=7998$ days for Japanese Nikkei-225 markets. 
ESM Tables S1 and S2 give the lists of 194 and 165 stocks (along with their 
sectors) for the USA S\&P-500 and Japanese Nikkei-225 markets, respectively, 
and these stocks are present in the two markets for the entire 32-year period 
considered here. 


\subsection*{Cross-correlation and distance matrices}

We present a study of time evolution of the cross-correlation structures 
of return time series for $N$ stocks (Figure \ref{fig:schematic}). The daily 
return time series is constructed as $r_k(t)=\ln P_k(t)-\ln P_k(t -1)$, 
where $P_k(t)$ is the adjusted closing price of the $k^{\text{th}}$ stock at time $t$ 
(trading day). Then, the cross-correlation matrix is constructed using 
equal-time Pearson cross-correlation coefficients,
\begin{equation}
\nonumber 
C_{ij}(t) = (\langle r_i r_j \rangle - \langle r_i \rangle \langle r_j \rangle)/\sigma_i\sigma_j,
\end{equation}
where $i,j=1, \dots, N$, $t$ indicates the end date of the epoch of size $\tau$ 
days, and the means $\langle \dots \rangle $ as well as the standard deviations 
$\sigma_k$ are computed over that epoch.

Instead of working with the correlation coefficient $C_{ij}(t)$, we use the 
`ultrametric' distance measure: 
\begin{equation}
\nonumber
D_{ij} (t)=\sqrt{2(1-C_{ij}(t))},
\end{equation}
such that $0 \leq D_{ij}(t) \leq 2$, which can be used for the construction of 
networks \cite{mantegna1999information,Mantegna1999hierarchical,Onnela2003,
Kumar2012}.

Here, we computed daily return cross-correlation matrix $\boldsymbol{C}_{\tau}(t)$
over the short epoch of $\tau=22$ days and shift of the rolling window by 
$\Delta \tau=5$ days, for (a) $N=194$ stocks of USA S\&P-500 for a return series 
of $T=8068$ days, and (b) $N=165$ stocks of Japan Nikkei-225 for $T=7998$ days, 
during the 32-year period from 1985 to 2016. We use epochs of $\tau=22$ days 
(one trading month) to obtain a balance between choosing short epochs for detecting 
changes and long ones for reducing fluctuations. In the main text, we show 
results for networks constructed from correlation matrices with overlapping windows 
of $\Delta \tau=5$ days, while in ESM, we show results for networks constructed 
from correlation matrices with non-overlapping windows of $\Delta \tau=22$ days. 


\subsection*{Network construction}

For a given time window of $\tau$ days ending on trading day $t$, the distance 
matrix $\boldsymbol{D}_{\tau}(t)$ constructed from the correlation matrix 
between the 194 stocks in USA S\&P-500 index or the 165 stocks in Japan 
Nikkei-225 index, can be viewed as an undirected complete graph $G_{\tau}(t)$ 
where the weight of an edge between stocks $i$ and $j$ is given by the distance 
$D_{ij}(t)$. For the time window of $\tau$ days ending on trading day $t$, we start 
with this edge weighted complete graph $G_{\tau}(t)$ and create the minimum 
spanning tree (MST) $T_{\tau}(t)$ using Prim's algorithm \cite{Prim1957}. 
Thereafter, we add edges in $G_{\tau}(t)$ with $C_{ij}(t) \ge 0.75$ to $T_{\tau}(t)$ 
to obtain the graph $S_{\tau}(t)$ (Figure \ref{fig:schematic}). We will use the 
graph $S_{\tau}(t)$ to compute different discrete Ricci curvatures and other 
network measures. We remark that the procedure used here to construct the graph 
$S_{\tau}(t)$ follows previous works \cite{Onnela2003,Sandhu2016} on analysis of 
correlation-based networks of stock markets. 

Intuitively, the motivation behind the above method of graph construction can be 
understood as follows. Firstly, the MST method gives a connected (spanning) graph 
between all nodes (stocks) in the specific market. Secondly, the addition of edges 
between nodes (stocks) with correlation $C_{ij}(t) \ge 0.75$ ensures that the important 
edges are also captured in the graph $S_{\tau}(t)$. 


\subsection*{Common network measures}

Given an undirected graph $G(V,E)$ with the sets of vertices or nodes $V$ and 
edges $E$, the number of edges is given by the cardinality of set $E$, that is 
$m=|E|$, and the number of nodes is given by the cardinality of set $V$, that 
is $n=|V|$. The edge density of such a graph is given by the ratio of the 
number of edges $m$ divided by the number of possible edges, that is, 
$\frac{2m}{n(n-1)}$. The average degree $\langle k \rangle$ of the graph gives 
the average number of edges per node, that is, $\langle k \rangle = \frac{m}{n}$. 
In case of an edge-weighted graph where $a_{ij}$ denotes the weight of the edge 
between nodes $i$ and $j$, one can also compute its average weighted degree 
$\langle k_w \rangle$ which gives the average of the sum of the weights of the 
edges connected to nodes, that is, $\langle k_w \rangle = \frac{m_w}{n}$ where 
$m_w = \sum_{i,j \in V} a_{ij}$. For any pair of nodes $i$ and $j$ in the graph, 
one can compute the shortest path length $d_{ij}$ between them. Thereafer, the 
average shortest path length $\langle L \rangle$ is given by the average of the
shortest path lengths between all pairs of nodes in the graph, that is, 
\begin{equation}
\nonumber
\langle L \rangle = \frac{1}{n(n-1)} \sum_{i\neq j \in V} d_{ij}.
\end{equation}
The diameter is given by the maximum of the shortest paths between all pairs 
of nodes in the graph, i.e., $\text{max}\{ d_{ij}\ \forall i,j \in V \}$. The 
communication efficiency \cite{Latora2001} of a graph is an indicator of its 
global ability to exchange information across the network. The communication 
efficiency $CE$ of a graph is given by 
\begin{equation}
\nonumber
CE = \frac{1}{n(n-1)} \sum_{i\neq j \in V} \frac{1}{d_{ij}}.
\end{equation}
Modularity measures the extent of community structure in the network and community 
detection algorithms aim to partition the graph into communities such that the 
modularity $Q$ attains the maximum value \cite{Girvan2002}. The modularity $Q$ is 
given by the equation \cite{Girvan2002,Blondel2008}
\begin{equation}
\nonumber
Q=\frac{1}{2m_w}\sum_{i\neq j \in V} [\, a_{ij} - \frac{k_{i}k_{j}}{2m_w}] \delta(c_i,c_j)\,
\end{equation}
where $k_i$ and $k_j$ give the sum of weights of edges attached to nodes $i$ and $j$, 
respectively, $c_i$ and $c_j$ give the communities of $i$ and $j$, respectively, and 
$\delta(c_i,c_j)$ is equal to 1 if $c_i = c_j$ else 0. Here, we use Louvain method 
\cite{Blondel2008} to compute the modularity of the edge-weighted networks. Network 
entropy is an average measure of graph heterogeneity as it quantifies the diversity 
of edge distribution using the remaining degree distribution $q_k$ \cite{Sole2004}. 
$q_k$ denotes the probabibility of a node to have remaining (excess) degree $k$ and 
is given by $q_k= \frac{(k+1)p_{k+1}}{<k>}$ where $p_{k+1}$ denotes the probability 
of a node to have degree $k+1$. The network entropy $H(q)$ of a graph is then given 
by 
\begin{equation}
\nonumber
H(q) = - \sum_{k} q_k log(q_k).
\end{equation}
The above-mentioned network measures were computed in stock market networks 
using the \texttt{python} package \texttt{NetworkX} \cite{Hagberg2008}.


\subsection*{GARCH($p,q$) process}

The generalized ARCH process GARCH($p,q$) was introduced by Bollerslev 
\cite{Bollerslev1986}. The variable $x_{t}$, a strong white noise process, can 
be written in terms of a time-dependent standard deviation $\sigma_t$, such that 
$x_{t}\equiv\eta_t\sigma_t$, where $\eta_t$ is a random Gaussian process with 
zero mean and unit variance. 

The simplest GARCH process is the GARCH(1,1) process, with Gaussian conditional 
 probability distribution function 
\begin{equation}
\sigma_t^2=\alpha_0+\alpha_1 x_{t-1}^{2}+\beta_{1}\sigma_{t-1}^{2} \, ,
\label{garch11}
\end{equation}
where $\alpha _{0}>0$ and $\alpha _{1}\geq 0$; $\beta_{1}$ is an additional 
control parameter. One can rewrite Eq. \ref{garch11} as a random multiplicative 
process 
\begin{equation}
\sigma_t^2=\alpha_0+(\alpha_1 \eta_{t-1}^{2}+\beta_{1})\sigma_{t-1}^{2} \, .
\label{garch12}
\end{equation}
For calculating this we have used an in-built function from \texttt{MATLAB garch} 
(\url{https://in.mathworks.com/help/econ/garch.html}).


\subsection*{Minimum Risk Portfolio}

We calculated the minimum risk portfolio in the Markowitz framework, as a 
measure of risk-aversion of each investor with maximized expected returns 
and minimized variance. In this model, the variance of a portfolio shows the 
importance of effective diversification of investments to minimize the total 
risk of a portfolio. The Markowitz model minimizes $\mathrm{w}^{\prime} 
\Omega \mathrm{w}- \phi R^{\prime} \mathrm{w}$ with respect to the normalized 
weight vector $\mathrm{w}$, where $\Omega$ is the covariance matrix calculated 
from the stock log-returns, $\phi$ is the measure of risk appetite of investor 
and  $R^{\prime}$ is the expected return of the assets. We set short-selling constraint, $\phi=0$ and $\mathrm{w}_i \geq 0$ which entails a convex 
combination of stock return for finding the minimum risk portfolio.
For calculating this we have used an in-built function from \texttt{MATLAB 
Portfolio} (\url{https://in.mathworks.com/help/finance/portfolio.html}).


\section{Results and Discussion}

\begin{figure}[]
\begin{center}
\includegraphics[width=0.79\linewidth]{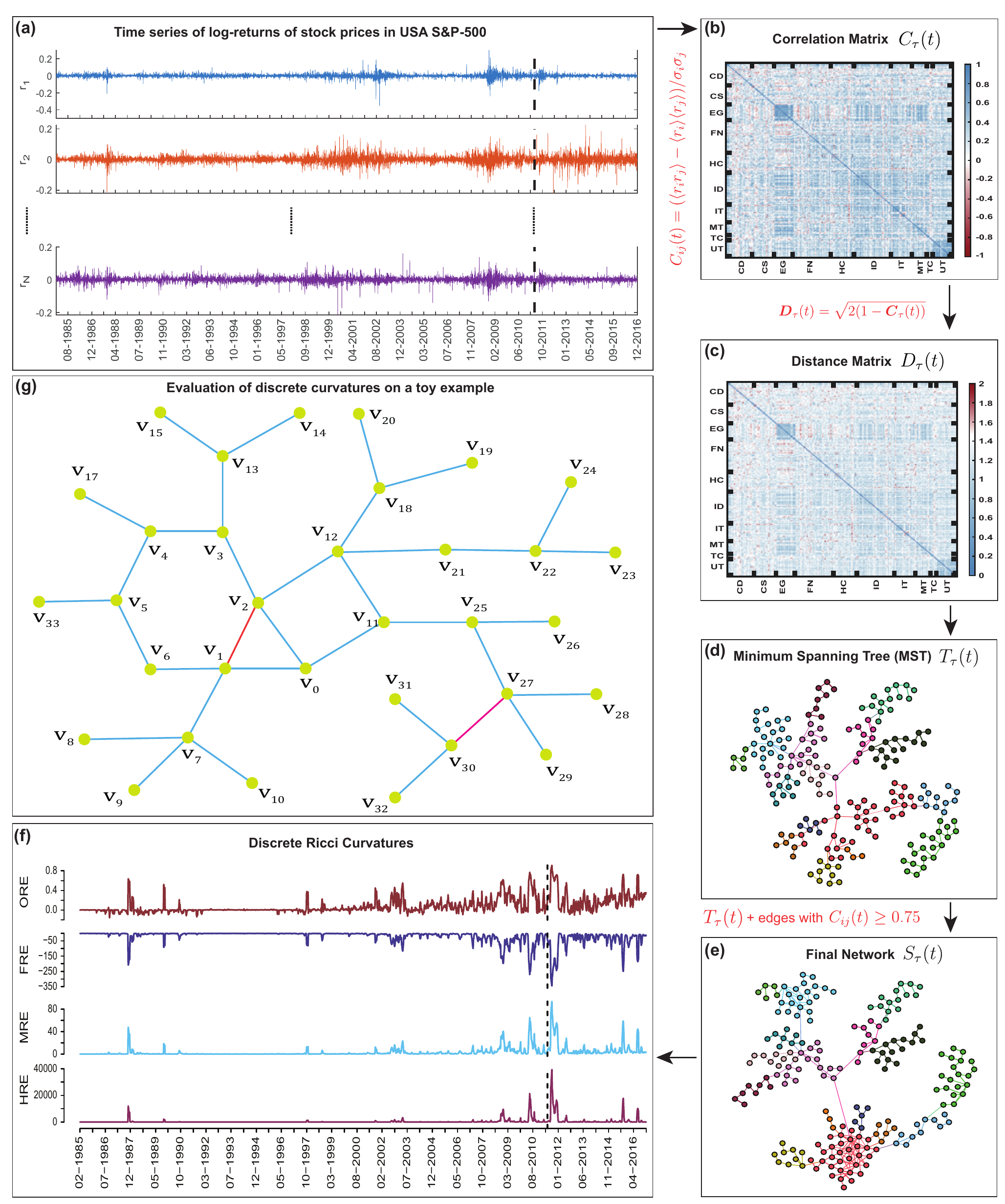}
\end{center}
\caption{Schematic diagram describing the evaluation of discrete 
Ricci curvatures in correlation-based networks constructed from 
log-returns of USA S\&P-500 market stocks. \textbf{(a)} Time series 
of log-returns over a 32-year period (1985-2016). \textbf{(b)} An 
arbitrarily chosen cross-correlation matrix $\boldsymbol C_\tau(t)$ 
for epoch ending on 04-05-2011. \textbf{(c)} Corresponding distance 
matrix $\boldsymbol D_\tau(t)=\sqrt{2(1-\boldsymbol C_\tau(t))}$ 
used for the construction of the threshold network. \textbf{(d)} 
Minimum spanning tree (MST) $T_\tau(t)$ constructed using the 
distance matrix $D_\tau(t)$. \textbf{(e)} Threshold network 
$S_\tau(t)$ constructed by adding edges with $C_{ij}(t)\geq 0.75$ to 
the MST $T_\tau(t)$. \textbf{(f)} Evolution of the average of four 
discrete Ricci curvatures for edges, namely, Ollivier-Ricci (ORE), 
Forman-Ricci (FRE), Menger-Ricci (MRE) and Haantjes-Ricci (HRE), 
computed using the threshold networks $S_\tau(t)$ constructed from 
correlation matrices over time epochs of $\tau=22$ days and 
overlapping shift of $\Delta \tau=5$ days. In this figure, $C_\tau(t)$, 
$D_\tau(t)$, $T_\tau(t)$ and $S_\tau(t)$ shown in (b)-(e) correspond 
to the correlation frame denoted by vertical dashed line in (a). 
\textbf{(g)} Evaluation of discrete Ricci curvatures on a toy 
example network which is undirected and unweighted. Here, the edge 
between $v_{27}$ and $v_{30}$ has a highly negative FR curvature as 
it depends on the degree of the two nodes or number of neighbouring 
edges. However, the edge between $v_{27}$ and $v_{30}$ has MR and HR 
curvature equal to zero as the edge under consideration is not part 
of any triangles or cycles, respectively. Moreover, the edge between 
$v_{1}$ and $v_{2}$ also has a highly negative FR curvature as the 
degree of both anchoring vertices is 4. In contrast, the edge between 
$v_{1}$ and $v_{2}$ has positive MR and HR curvature as the edge is 
part of a triangle which contributes to MR curvature and the edge is 
part of a triangle, a pentagon and a hexagon which contribute to HR 
curvature. For both the edges between $v_{27}$ and $v_{30}$ and 
between $v_{1}$ and $v_{2}$, one can compute OR curvature, however, 
only triangles, quadrangles and pentagons make positive contribution 
to the OR curvature in unweighted and undirected networks. Specifically, 
the edge between $v_{1}$ and $v_{2}$ is part of a triangle, a pentagon 
and a hexagon, however, only the triangle and pentagon make positive 
contribution to OR curvature.}
\label{fig:schematic}
\end{figure}

We analyze here the time series of the logarithmic returns of the stocks 
in the USA S\&P-500 and Japanese Nikkei-225 markets over a period of 32 
years (1985-2016) by constructing the corresponding Pearson cross-correlation 
matrices $\boldsymbol C_\tau(t)$. We then use cross-correlation matrices
$\boldsymbol C_\tau(t)$ computed over time epochs of size $\tau = 22$ days 
with either overlapping or non-overlapping windows (i.e. shifts of 
$\Delta \tau = 5$ or $22$ days, respectively) and ending on trading days $t$
to study the evolution of the correlation-based networks $S_\tau(t)$ and 
corresponding network properties, especially edge-centric geometric measures.
Figure~\ref{fig:schematic} gives an overview of our evaluation of discrete 
Ricci curvatures in correlation-based threshold networks constructed from 
log-returns of market stocks. Figure \ref{fig:schematic}(a) shows the daily 
log-returns over the 32-year period (1985-2016). An arbitrarily chosen 
cross-correlation matrix $\boldsymbol C_{\tau}(t)$ over time epoch of $\tau=22$ 
days and $\Delta \tau=5$ days ending on 04-05-2011 and corresponding distance 
matrix $\boldsymbol D_{\tau}(t)= \sqrt{2(1-\boldsymbol C_{\tau}(t))}$ are shown 
in Figure~\ref{fig:schematic}(b) and (c), respectively. The minimum spanning 
tree (MST) $T_{\tau}(t)$ constructed from the distance matrix $D_{\tau}(t)$ is 
shown in Figure~\ref{fig:schematic}(d). Thereafter, a threshold network 
$S_{\tau}(t)$ is constructed using MST $T_{\tau}(t)$ and edges with $C_{ij}(t) 
\geq 0.75$, as shown in Figure~\ref{fig:schematic}(e). The discrete Ricci curvatures 
are computed from the threshold networks. In Figure~\ref{fig:schematic}(f), we 
show the evolution of the discrete curvatures in threshold networks over the 
32-year period. In Figure~\ref{fig:schematic}(g), we motivate the four discrete 
Ricci curvatures considered here using a simple example network. 

\begin{figure}[]
\begin{center}
\includegraphics[width=0.79\linewidth]{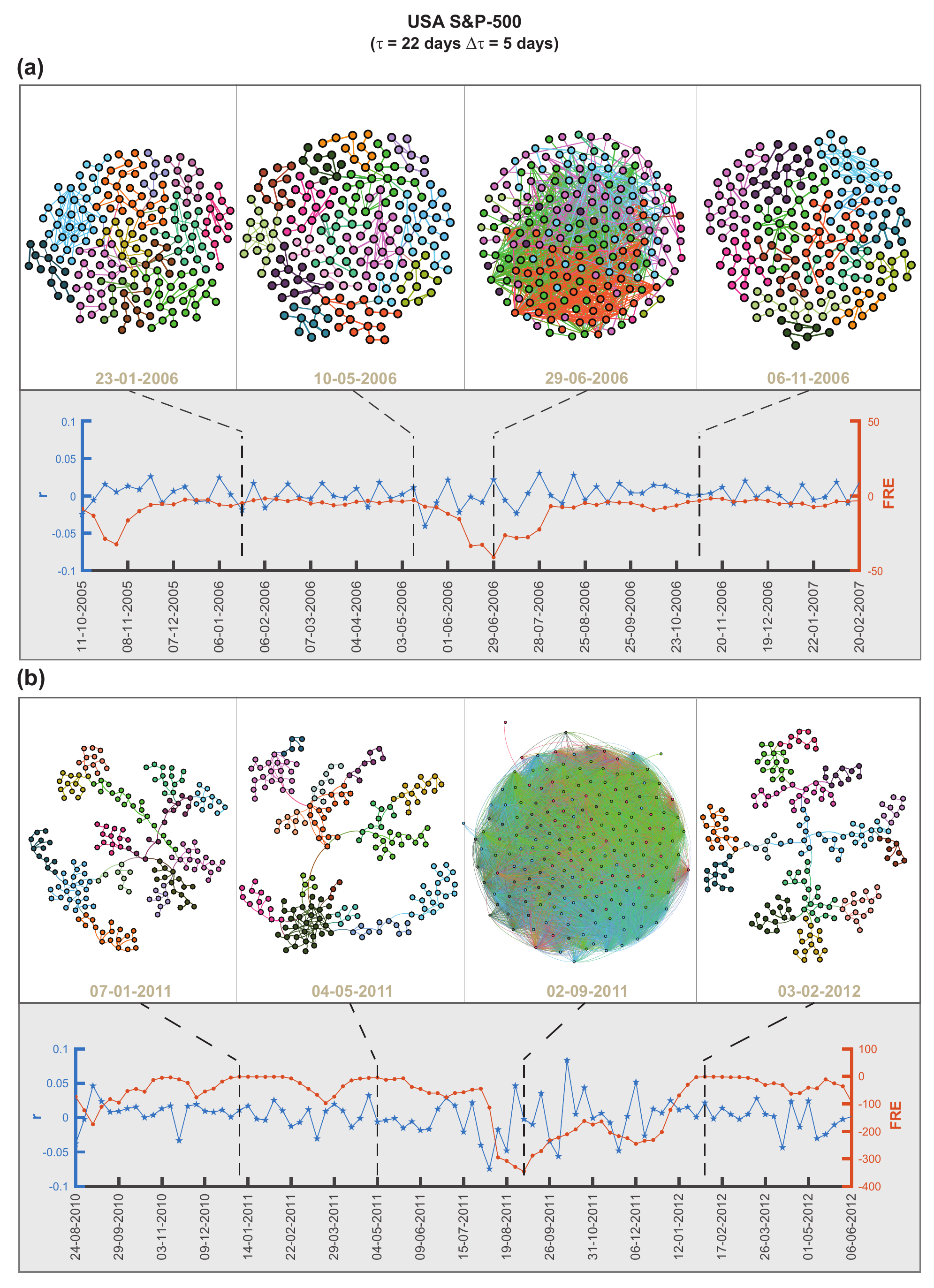}
\end{center}
\caption{\textbf{(a)} (Upper panel) Visualization of threshold networks 
for USA S\&P-500 market around the US Housing bubble period (2006-2007) 
at four distinct epochs of $\tau=22$ days ending on trading days 23-01-2006, 
10-05-2006, 29-06-2006, and 06-11-2006, with threshold $C_{ij}(t)\geq 0.75$. 
Here, the colour of the nodes correspond to the different communities 
determined by Louvain method for community detection. Threshold networks 
show higher number of edges and lower number of communities during a bubble. 
(Lower panel) Plot shows the evolution of log-returns $r$ of S\&P-500 
index (blue color line) and average Forman-Ricci curvature of edges (FRE) 
(sienna color line) for the period around the US Housing bubble. The FRE 
measure, constructed from threshold networks, is sensitive to both local 
(sectoral) and global fluctuations of the market, and shows a local minimum 
(more negative) during the bubble, whereas not much variation is seen in 
$r$ (low volatility).  \textbf{(b)} (Upper panel) Visualization of threshold 
networks for USA S\&P-500 market around the August 2011 stock markets fall 
at four distinct epochs of $\tau=22$ days ending on 07-01-2011, 04-05-2011, 
02-09-2011, and 03-02-2012 with threshold $C_{ij}(t)\geq 0.75$. Here, the 
threshold network shows significantly higher number of edges and lower 
number of communities during the crash. (Lower panel) Plot shows the evolution 
of log-returns $r$ of S\&P 500 index (blue color line) and FRE (sienna color 
line) for the period around the August 2011 stock markets fall. During 
the crash $r$ has high fluctuations (high volatility) and FRE decreases 
significantly (local minima).}
\label{fig:corr-networks_both}
\end{figure}

A major goal of this research is to evaluate different notions of discrete 
Ricci curvature for their ability to unravel the structure of complex 
financial networks and serve as indicators of market instabilities. 
Previously, Sandhu et al. \cite{Sandhu2016} have analyzed the USA S\&P-500
market over a period of 15 years (1998-2013) to show that the average 
Ollivier-Ricci (OR) curvature of edges (ORE) in threshold networks increases 
during periods of financial crisis. Here, we extend the analysis by Sandhu 
et al. \cite{Sandhu2016} to (a) two different stock markets, namely, USA 
S\&P-500 and Japanese Nikkei-225, (b) a span of 32 years (1985-2016), (c) 
four traditional market indicators (namely, index log-returns $r$, mean 
market correlation $\mu$, volatility of the market index $r$ estimated using 
GARCH(1,1) process, and risk $\sigma_P$ corresponding to the minimum risk 
Markowitz portfolio of all the stocks in the market), and (d) four notions 
of discrete Ricci curvature for networks. Since discretizations of Ricci 
curvature are unable to capture the entire properties of the classical Ricci 
curvature defined on continuous spaces, the four discrete Ricci curvatures 
evaluated here can shed light on different properties of analyzed networks
\cite{Samal2018}. In particular, some of us have introduced another 
discretization, Forman-Ricci (FR) curvature, to the domain of networks 
\cite{Sreejith2016}. Note that OR curvature captures the volume growth property 
of classical Ricci curvature while FR curvature captures the geodesic dispersal 
property \cite{Samal2018}. Nevertheless, our empirical analysis has shown that 
the two discrete notions, OR and FR curvature, are highly correlated in model 
and real-world networks \cite{Samal2018}. Importantly, in large networks, 
computation of the OR curvature is intensive while that of the FR curvature is 
simple as the later depends only on immediate neighbours of an edge 
\cite{Samal2018}. Therefore, we started by investigating the ability of FR 
curvature to capture the structure of complex financial networks.  

Figure~\ref{fig:corr-networks_both} shows the comparisons of threshold 
networks, as well as the behaviour of index log-returns $r$ and average
FR curvature of edges (FRE), for (a) bubble and (b) crash periods, of the 
USA S\&P-500 market. The upper panel of Figure~\ref{fig:corr-networks_both}(a) 
shows the threshold networks near the US Housing bubble period (2006-2007) 
at four distinct epochs of $\tau=22$ days ending on trading days $t$ equal 
to 23-01-2006, 10-05-2006, 29-06-2006 and 06-11-2006, with threshold 
$C_{ij}(t)\geq 0.75$. Number of edges and communities in these four threshold 
networks are $251, 220, 996, 220$ and $13, 16, 11, 14$, respectively. 
The colour of the nodes correspond to the different communities determined by 
Louvain method \cite{Blondel2008} for community detection. The plots of 
log-returns of S\&P-500 index $r$ (blue color line) and FRE (sienna color line) 
around the US Housing bubble period are shown in the lower panel of
Figure~\ref{fig:corr-networks_both}(a). Threshold networks show higher number 
($996$) of edges and lower number ($11$) of communities for high (negative) 
values of FRE, but there is not much variation of $r$. In ESM Figure S5, we 
show that the FRE captures the same features for three other thresholds 
$C_{ij}(t)\geq 0.55$, $C_{ij}(t)\geq 0.65$, and $C_{ij}(t)\geq 0.85$, and the numbers 
of edges and communities for each threshold is listed in ESM Table S3. The 
measure FRE is sensitive to both local (sectoral) and global (market) 
fluctuations, and shows a local minimum during bubble. Note that during a bubble, 
only a few sectors of the market perform well compared to the others (the stocks 
within the well-performing sectors are highly correlated, but the inter-sectoral 
correlations are low). It is hard to identify bubble by only monitoring the market 
index as the returns do not show much volatility. 
Figure~\ref{fig:corr-networks_both}(b) shows the same for the period around the 
August 2011 stock markets fall at four distinct epochs of $\tau=22$ days ending 
on trading days $t$ equal to 07-01-2011, 04-05-2011, 02-09-2011 and 03-02-2012, 
with threshold $C_{ij}(t)\geq 0.75$. Number of edges and communities in these four 
threshold networks are $197, 245, 16004, 198$ and $14, 16, 4, 15$, respectively. 
During the crash, the threshold network shows sufficiently higher number of edges 
and extremely low number of communities. In ESM Figure S6, we show 
that the FRE captures the same features for three other thresholds $C_{ij}(t)\geq 0.55$, 
$C_{ij}(t)\geq 0.65$, and $C_{ij}(t)\geq 0.85$, and the numbers of edges and communities 
for each threshold is listed in ESM Table S3. The plots of log-returns $r$ of 
S\&P-500 index (blue color line) and FRE (sienna color line) are shown around the
August 2011 stock markets fall period in the lower panel of 
Figure~\ref{fig:corr-networks_both}(b). Note that during a market crash $r$ displays 
high volatility and FRE shows a significant decrease (local minimum). Earlier Sandhu 
et al. \cite{Sandhu2016} had focussed on OR curvature as an indicator of crashes.   
Here, we additionally show that discrete Ricci curvatures, especially FR curvature, 
are sensitive and can detect both crash (market volatility high) and 
bubble (market volatility low). 


\begin{figure}[]
\centering
\includegraphics[width=0.79\linewidth]{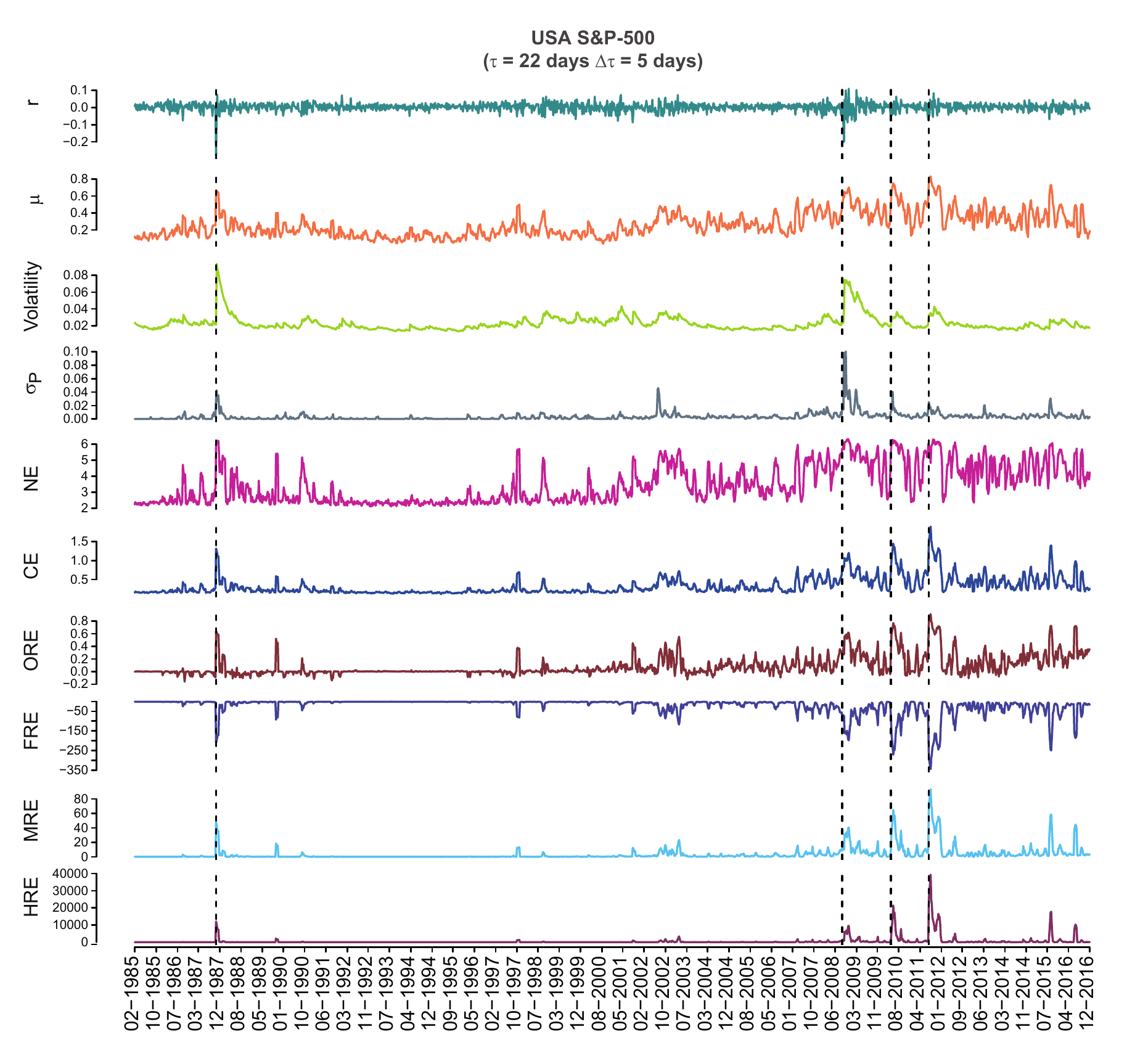}
\caption{Evolution of the market indicators and edge-centric geometric 
curvatures for the USA S\&P-500 market. From top to bottom, we plot the 
index log-returns $r$,  mean market correlation $\mu$, volatility of the 
market index $r$ estimated using GARCH(1,1) process, risk $\sigma_P$ 
corresponding to the minimum risk Markowitz portfolio of all the stocks 
in the market, network entropy (NE), communication efficiency (CE), 
average of Ollivier-Ricci (ORE), Forman-Ricci (FRE), Menger-Ricci (MRE), 
and Haantjes-Ricci (HRE) curvature of edges evaluated from the correlation 
matrices $\boldsymbol C_\tau(t)$ of window size $\tau=22$ days and an 
overlapping shift of $\Delta\tau=5$ days. Four vertical dashed lines 
indicate the epochs of four important crashes: Black Monday 1987, Lehman 
Brothers crash 2008, DJ Flash crash 2010, and August 2011 stock markets 
fall (see Table~\ref{tab:events}).}
\label{Fig:USA_TS}
\end{figure}

\begin{figure}[]
\centering
\includegraphics[width=0.79\linewidth]{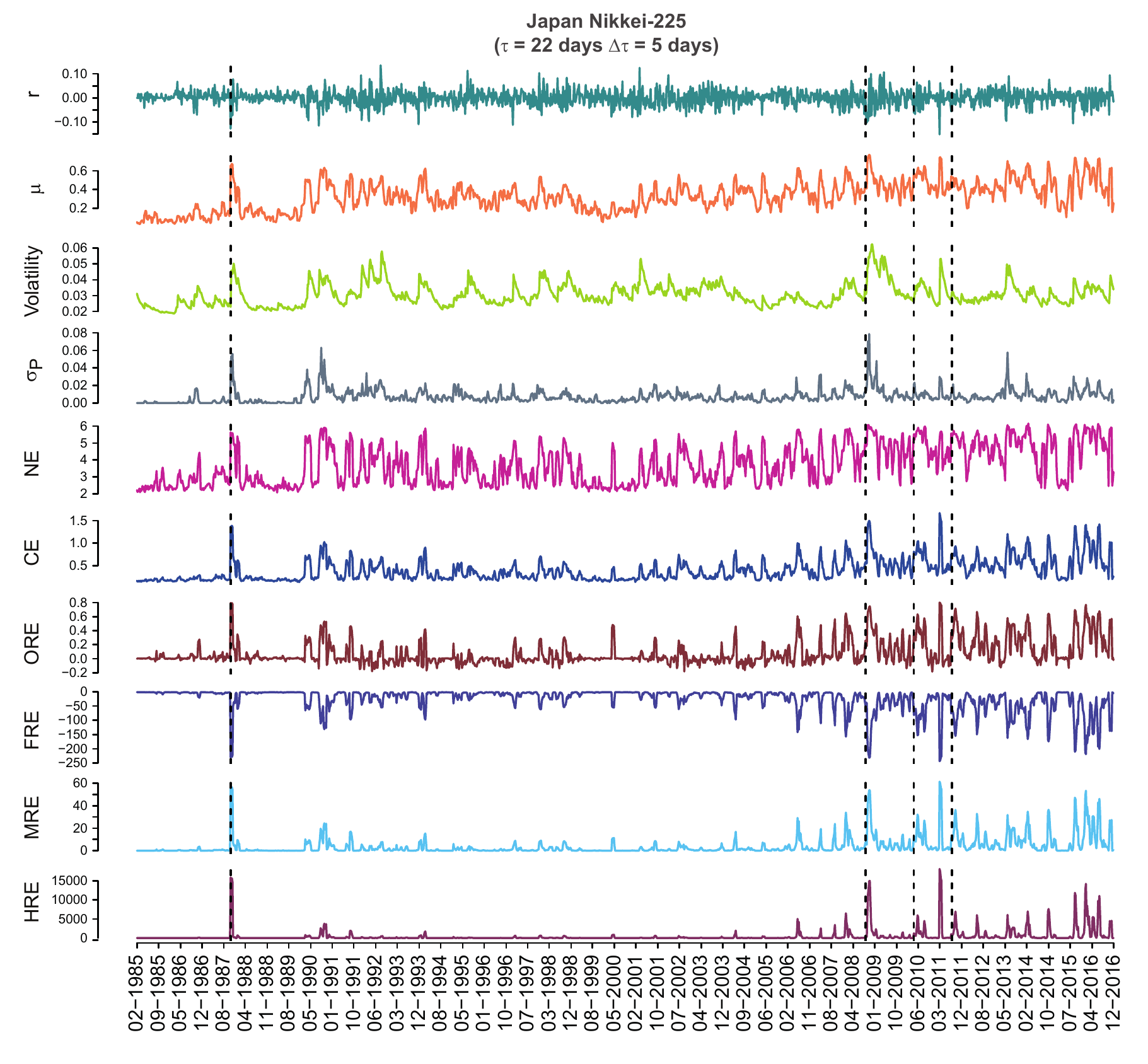}
\caption{Evolution of the market indicators and edge-centric geometric 
curvatures for the Japanese Nikkei-225 market. From top to bottom, we plot 
the index log-returns $r$,  mean market correlation $\mu$, volatility of 
the market index $r$ estimated using GARCH(1,1) process, risk $\sigma_P$ 
corresponding to the minimum risk Markowitz portfolio of all the stocks 
in the market, network entropy (NE), communication efficiency (CE), 
average of Ollivier-Ricci (ORE), Forman-Ricci (FRE), Menger-Ricci (MRE), 
and Haantjes-Ricci (HRE) curvature of edges evaluated from the correlation 
matrices $\boldsymbol C_\tau(t)$ of window size $\tau=22$ days and an 
overlapping shift of $\Delta\tau=5$ days. Four vertical dashed lines 
indicate the epochs of four important crashes: Black Monday 1987, Lehman 
Brothers crash 2008, DJ Flash crash 2010, and August 2011 stock markets fall 
(see Table~\ref{tab:events}).}
\label{Fig:JPN_TS}
\end{figure}

It is often difficult to gauge the state of the market by simply monitoring 
the market index or its log-returns. There exist no simple definitions of a 
market crash or a market bubble. The market becomes extremely correlated and 
volatile during a crash, but a bubble is even harder to detect as the volatility 
is relatively low and only certain sectors perform very well (stocks show high 
correlation) but the rest of the market behaves like normal or `business-as-usual'. 
Traditionally, the volatility of the market captures the `fear' and the evaluated 
risk captures the `fragility' of the market. Some of us showed in our earlier 
papers that the mean market correlation and the spectral properties of the 
cross-correlation matrices can be used to study the market states 
\cite{Pharasi2018} and identify the precursors of market instabilities 
\cite{Chakraborti2020}. A goal of this study is to show that the state of the 
market can be continuously monitored with certain network-based measures. Thus, 
we next performed a comparative investigation of several network measures, 
especially, the four discrete notions of Ricci curvature. 

Figures \ref{Fig:USA_TS} and \ref{Fig:JPN_TS} show for USA S\&P-500 market and 
Japanese Nikkei-225 market, respectively, the temporal evolution of the market
indicators and network measures, mainly edge-centric Ricci curvatures computed 
from the correlation matrices $\boldsymbol C_\tau(t)$ of epoch size $\tau=22$ 
days and overlapping shift of $\Delta \tau =5$ days, over a 32-year period 
(1985-2016). From top to bottom, the plots represent index log-returns $r$, 
mean market correlation $\mu$, volatility of the market index $r$ estimated 
using GARCH(1,1) process, risk $\sigma_P$ corresponding to the minimum risk 
Markowitz portfolio of all the stocks in the market, network entropy (NP), 
communication efficiency (CE), average of OR, FR, MR and HR curvature of edges. 
We find that the four Ricci-type curvatures, namely, ORE, FRE, MRE and HRE, 
along with the other important indicators of the markets, viz., the log-returns 
$r$, volatility, minimum risk $\sigma_P$ and mean market correlation $\mu$, are 
excellent indicators of market instabilities (bubbles and crashes). We highlight 
that the four discrete Ricci curvatures can capture important crashes and 
bubbles listed in Table \ref{tab:events} in the two markets during the 32-year 
period studied here.

In ESM Figure S7, we show the temporal evolution of the four discrete Ricci 
curvatures computed in threshold networks $S_\tau(t)$ obtained using three 
different thresholds, $C_{ij}(t) \geq 0.65$ (cyan color), $C_{ij}(t) \geq 0.75$ (dark 
blue color) and  $C_{ij}(t) \geq 0.85$ (sienna color), for the two markets. It is 
seen that the absolute value of ORE, FRE, MRE and HRE decreases with the 
increase in the threshold $C_{ij}(t)$ used to construct $S_\tau(t)$. Regardless 
of the three thresholds used to construct the threshold networks $S_\tau(t)$, 
we show that the four discrete Ricci curvatures are fine indicators of market
instabilities. 

In previous work, Sandhu et al. \cite{Sandhu2016} had contrasted the temporal
evolution of ORE in threshold networks for USA S\&P-500 market with NE, graph 
diameter and average shortest path length. Here, we have studied the temporal 
evolution of a larger set of network measures in threshold networks for USA 
S\&P-500 and Japanese Nikkei-225 markets computed from the correlation matrices 
$\boldsymbol C_\tau(t)$ of epoch size $\tau=22$ days and overlapping shift of 
$\Delta \tau =5$ days, over a 32-year period (1985-2016). From Figures 
\ref{Fig:USA_TS} and \ref{Fig:JPN_TS}, it is seen that NE and CE are also 
excellent indicators of market instabilities. In fact, we find that common 
network measures such as number of edges, edge density, average degree, average 
shortest path length, graph diameter, average clustering coefficient and 
modularity are also good indicators of market instabilities (ESM Figure S8). 
    
In ESM Figures S9 and S10, we show the temporal evolution of the market indicators 
and several network measures (including edge-centric Ricci curvatures) computed 
from the correlation matrices $\boldsymbol C_\tau(t)$ of epoch size $\tau=22$ 
days and non-overlapping shift of $\Delta \tau =22$ days, over a 32-year period
(1985-2016) in the two markets. It can be seen that our results are also not 
dependent on the choice of overlapping or non-overlapping shift used to 
construct the cross-correlation matrices and threshold networks.


\begin{figure}[]
\begin{center}
\includegraphics[width=0.61\linewidth]{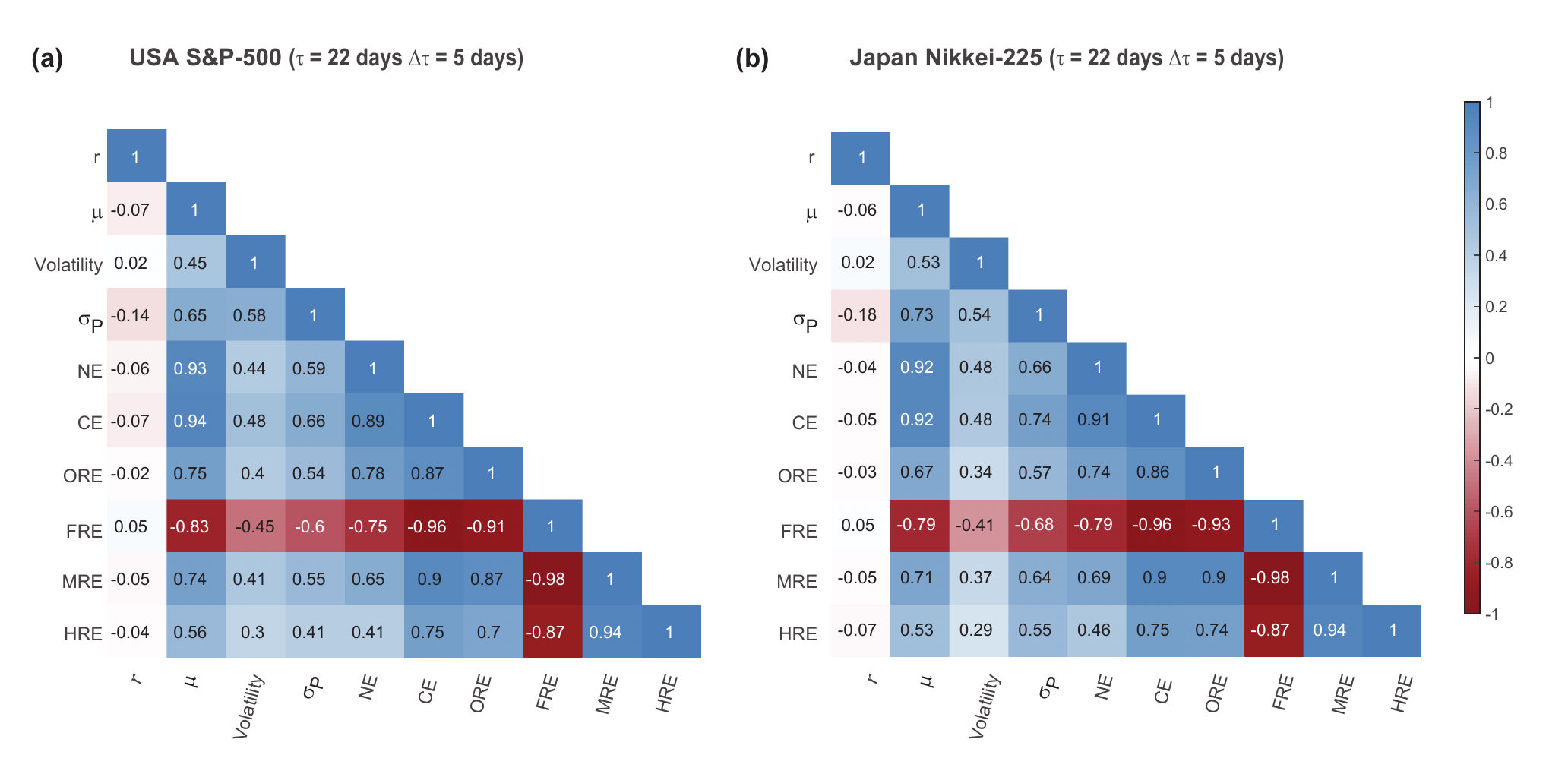}
\end{center}
\caption{Correlogram plots of \textbf{(a)} USA S\&P-500 and \textbf{(b)} 
Japan Nikkei-225 markets, for the traditional market indicators (index 
returns $r$, mean market correlation $\mu$, volatility, and minimum 
risk portfolio $\sigma_P$), global network properties (network entropy NE 
and communication efficiency CE) and discrete Ricci curvatures for edges 
(Ollivier-Ricci ORE, Forman-Ricci FRE, Menger-Ricci MRE, and Haantjes-Ricci 
HRE), computed for epochs of size $\tau=22$ days and overlapping shift 
$\Delta \tau=5$ days.}
\label{fig:correlogram}
\end{figure}

Figure~\ref{fig:correlogram} shows the correlogram plots of (a) USA S\&P-500 
and (b) Japanese Nikkei-225 markets, for the traditional market indicators 
(index returns $r$, mean market correlation $\mu$, volatility, and minimum 
portfolio risk $\sigma_P$), network properties (NE and CE) and discrete Ricci
curvatures (ORE, FRE, MRE and HRE), computed for epoch size $\tau=22$ days 
and overlapping shift of $\Delta \tau=5$ days. In ESM Figure S11, 
we show the correlogram plots for the traditional market indicators and network properties
including discrete Ricci curvatures computed for epoch size $\tau=22$ days and 
non-overlapping shift of $\Delta \tau=22$ days in the two markets. Notably, FRE 
shows the highest correlation among the four discrete Ricci curvatures with the 
four traditional market indicators in the two markets, and thus, FRE is an 
excellent indicator for market risk that captures local to global system-level
fragility of the markets. Furthermore, both NE and CE also have high correlation 
with the four traditional market indicators. Therefore, these measures can be 
used to monitor the health of the financial system and forecast market crashes 
or downturns. Overall, we show that FRE is a simple yet powerful tool for 
capturing the correlation structure of a dynamically changing network.

\begin{table}
\caption{List of major crashes and bubbles in stock markets of USA and 
Japan between 1985-2016 \cite{CrashList,Bullmarkets,UShousingbubble,
CrashHistory,Selloff,marketfall2011}.}
\begin{tabular}{|l|l|l|l|}
\hline
\textbf{Serial number} & \textbf{Major crashes and bubbles} & \textbf{Period} & \textbf{Affected region} \\ \hline
		1    & Black Monday                             & 19-10-1987      & USA, Japan                 \\ \hline
		2    & Friday the $13^{th}$ mini crash          & 13-10-1989      & USA                     \\ \hline
		3    & Early 90s recession                      & 1990            & USA                     \\ \hline
		4    & Mini crash due to Asian financial crisis & 27-10-1997      & USA                     \\ \hline
		5    & Lost decade                              & 2001-2010       & Japan                     \\ \hline
		6    & 9/11 financial crisis                    & 11-09-2001      & USA, Japan                 \\ \hline
		7    & Stock market downturn of 2002            & 09-10-2002      & USA, Japan                 \\ \hline
		8    & US Housing bubble                        & 2005-2007       & USA                     \\ \hline
		9    & Lehman Brothers crash                    & 16-09-2008      & USA, Japan                 \\ \hline
		10   & Dow Jones (DJ) Flash crash               & 06-05-2010      & USA, Japan                 \\ \hline
		11   & Tsunami and Fukushima disaster           & 11-03-2011      & Japan                     \\ \hline
		12   & August 2011 stock markets fall           & 08-08-2011      & USA, Japan  \\ \hline
		13   & Chinese Black Monday and 2015-2016 sell off  & 24-08-2015  & USA                     \\ \hline
\end{tabular}
\label{tab:events}
\end{table}

\section{Conclusion}

In this paper, we have employed geometry-inspired network curvature measures 
to characterize the correlation structures of the financial systems and used 
them as generic indicators for detecting market instabilities (bubbles and 
crashes). We reiterate here that it is often difficult to gauge the state of 
the market by simply monitoring the market index or its log-returns. There 
exist no simple definitions of a market crash or a market bubble. The market 
becomes extremely correlated and volatile during a crash, but a bubble is even 
harder to detect as the volatility is relatively low and only certain sectors 
perform very well (stocks show high correlation) but the rest of the market 
behaves like normal or `business-as-usual'. We have examined the daily returns 
from a set of stocks comprising the USA S\&P-500 and the Japanese Nikkei-225 
over a 32-year period, and monitored the changes in the edge-centric geometric
curvatures. Our results are very robust as we have studied two very different 
markets, and for a very long period of 32 years with several interesting market 
events (bubbles and crashes; see Table \ref{tab:events}). We showed that the 
results are not very sensitive to the choice of overlapping or non-overlapping 
windows used to construct the cross-correlation matrices and threshold networks 
(Figures \ref{Fig:USA_TS}-\ref{Fig:JPN_TS}; ESM Figures S8-S10). Further, the 
choice of the thresholds for constructing networks also has little influence on 
their behaviour as indicators (ESM Figures S5-S7). 
In addition, to test the robustness of our methodology in the current paper, 
we have added small amounts of Gaussian noise to the empirical correlation matrices 
for the USA S\&P-500 market, and reproduced the evolution of the topological 
properties as well as the geometric curvature measures over the 32-year period. 
Specifically, we have found that the results are not sensitive to small amounts 
of noise or random reshuffling of data (ESM Figure S12).
We found that the four different notions of discrete Ricci curvature captured 
well the system-level features of the market and hence we were able to distinguish 
between the normal or `business-as-usual' periods and all the major market crises 
(bubbles and crashes) using the network-centric indicators. Our studies confirmed 
that during a normal period the market is very modular and heterogeneous, whereas 
during an instability (crisis) the market is more homogeneous, highly connected 
and less modular. 

Interestingly, our methodology picks up many peaks other than the major crashes 
and bubbles; these are neither spurious nor false positives. Unlike the major 
crashes and bubbles which are well-documented in the financial literature (or 
listed in internet sources, see Table~\ref{tab:events}), many of these peaks 
correspond to interesting events 
that are not well understood or recorded in the literature. In fact the financial 
markets are often driven by endogenous and exogenous factors. Moreover, there are 
often multiple reasons leading to a market crash or a bubble burst. The study and 
characterization of such market events, including exogenous shocks, bubble bursts, 
and anomalies, corresponding to such peaks has already been done in our earlier 
papers \cite{Pharasi2018,Pharasi2019,Chakraborti2020,chakraborti2020phase}. 
The findings of the present paper are in concordance with the earlier ones.

It is important to note that partial correlations can detect direct as opposed to 
plausibly indirect connections among components of the stock market. In the 
Econophysics literature (see e.g. Refs. \cite{Kenett2010,San2012,Millington2020,Sharma2017,chakraborti2020phase}), 
researchers have used partial correlations for analyzing the dynamics and 
constructing networks of stock markets. Partial correlations are particularly 
relevant when people study eigenvalue spectra (market, group and random modes), 
or network centrality measures, by first filtering out the spurious correlations. 
However, it has been observed \cite{Millington2020,chakraborti2020phase} that 
partial correlations are less successful in picking the cluster or group dynamics, 
and the networks arising from partial correlations are also less stable. In this 
contribution, we are more interested in the market indicators and the use of 
discrete Ricci curvatures as generic indicators, for which we prefer to work with 
the more stable correlation matrices.

Also, we find from these geometric measures that there are succinct and inherent 
differences in the two markets, USA S\&P-500 and Japan Nikkei-225. Importantly, 
among four Ricci-type curvature measures, the Forman-Ricci curvature of edges (FRE)
correlates highest with the traditional market indicators and acts as an excellent
indicator for the system-level fear (volatility) and fragility (risk) for both the
markets. These new insights may help us in future to  better understand tipping 
points, systemic risk, and resilience in financial networks, and enable us to 
develop monitoring tools required for the highly interconnected financial systems 
and perhaps forecast future financial crises and market slowdowns. These can be 
further generalized to study other economic systems, and may thus enable us to understand 
the highly complex and interconnected economic-financial systems.


\subsection*{Author contributions}
A.S. and A.C. designed research; A.S., H.K.P., S.J.R., H.K., 
E.S., J.J. and A.C. performed research and analyzed data;  A.S., H.K.P. and 
S.J.R. prepared the figures; A.S. and A.C. supervised the research; A.S., E.S., 
J.J. and A.C. wrote the manuscript with input from the other authors. All 
authors have read and approved the manuscript.

\subsection*{Acknowledgement}
A.S. acknowledges financial support from Max Planck Society Germany 
through the award of a Max Planck Partner Group in Mathematical Biology. 
E.S. and J.J. acknowledge support from the German-Israeli Foundation (GIF) 
Grant I-1514-304.6/2019. 
H.K.P. is grateful for financial support provided by UNAM-DGAPA and 
CONACYT Proyecto Fronteras 952. A.C. and H.K.P. acknowledge support from the 
projects UNAM-DGAPA-PAPIIT AG100819 and IN113620, and CONACyT Project 
Fronteras 201.

\subsection*{Data Availability}
All data used are openly available for download on the websites of 
the relevant sources mentioned in the text and stated in the references 
section. All relevant data and codes for this study have been uploaded 
and made publicly available via the GitHub repository: 
\url{https://github.com/asamallab/StockMarkNetIndicator}.

\vspace{0.5cm}
\noindent \textbf{Correspondence to:} Areejit Samal (asamal@imsc.res.in ) 
or Anirban Chakraborti (anirban@jnu.ac.in)

%

\newpage
\section*{Electronic Supplementary Material (ESM)}
\setcounter{table}{0}
\renewcommand{\thetable}{S\arabic{table}}%
\setcounter{figure}{0}
\renewcommand{\thefigure}{S\arabic{figure}}%

\begin{figure}[!htbp]
\begin{center}
\includegraphics[width=0.43\linewidth]{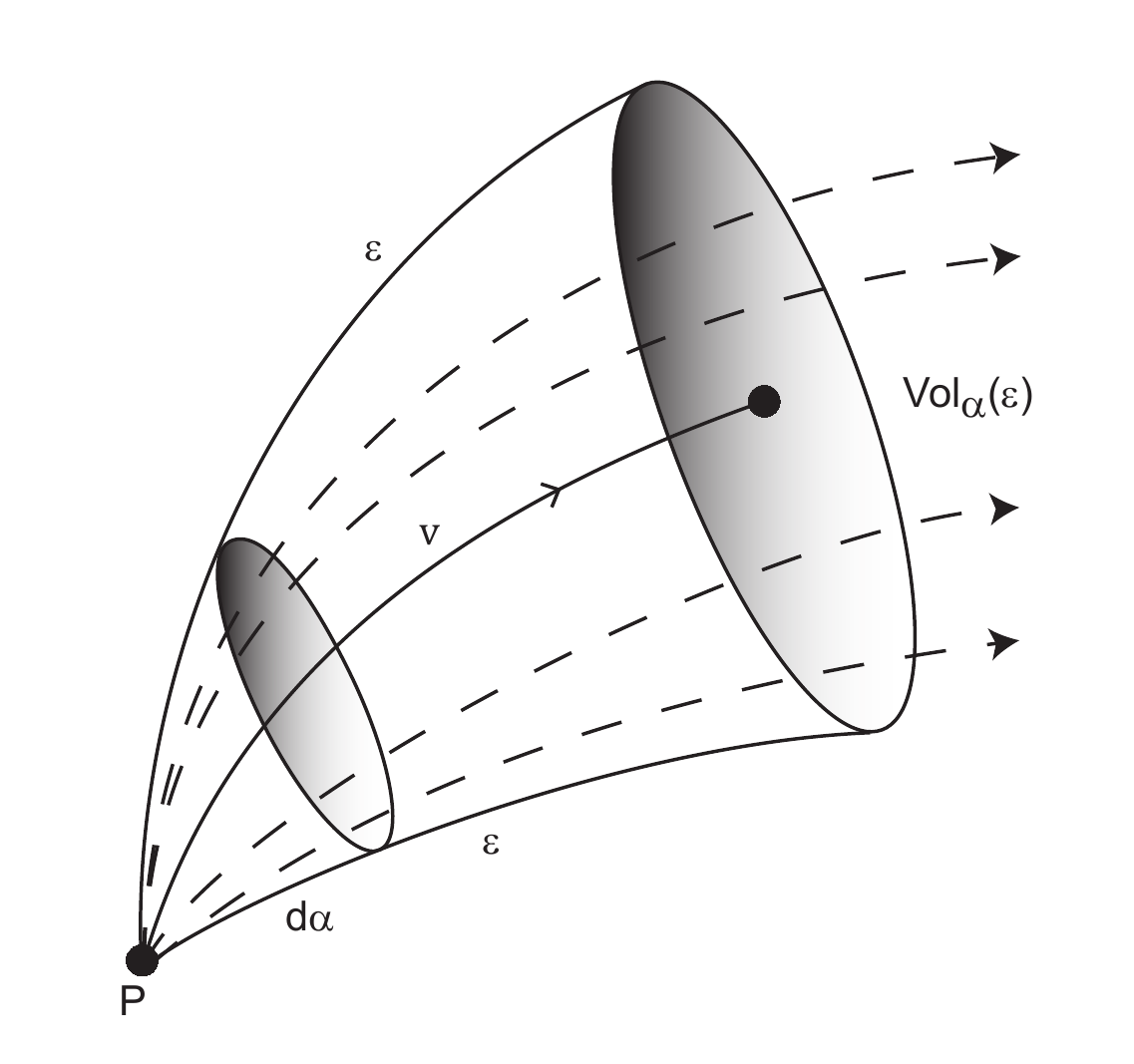}
\end{center}
\caption{Schematic figure on the geometric interpretation of Ricci 
curvature. Ricci curvature quantifies the deviation of the manifold from 
being locally Euclidean in tangent directions. In a $n$-dimensional Riemannian 
manifold, Ricci curvature controls the growth of $(n-1)$-volume 
$\text{Vol}_{\alpha}(\epsilon)$ generated within an $n$-solid angle 
$\mathrm{d}\alpha$ by geodesics of length $\epsilon$ in the direction of 
the vector $v$. Ricci curvature also quantifies the dispersion rate of 
geodesics with the same initial point and contained within the given solid 
angle.}
\end{figure}

\begin{figure}[!htbp]
\begin{center}
\includegraphics[width=0.34\linewidth]{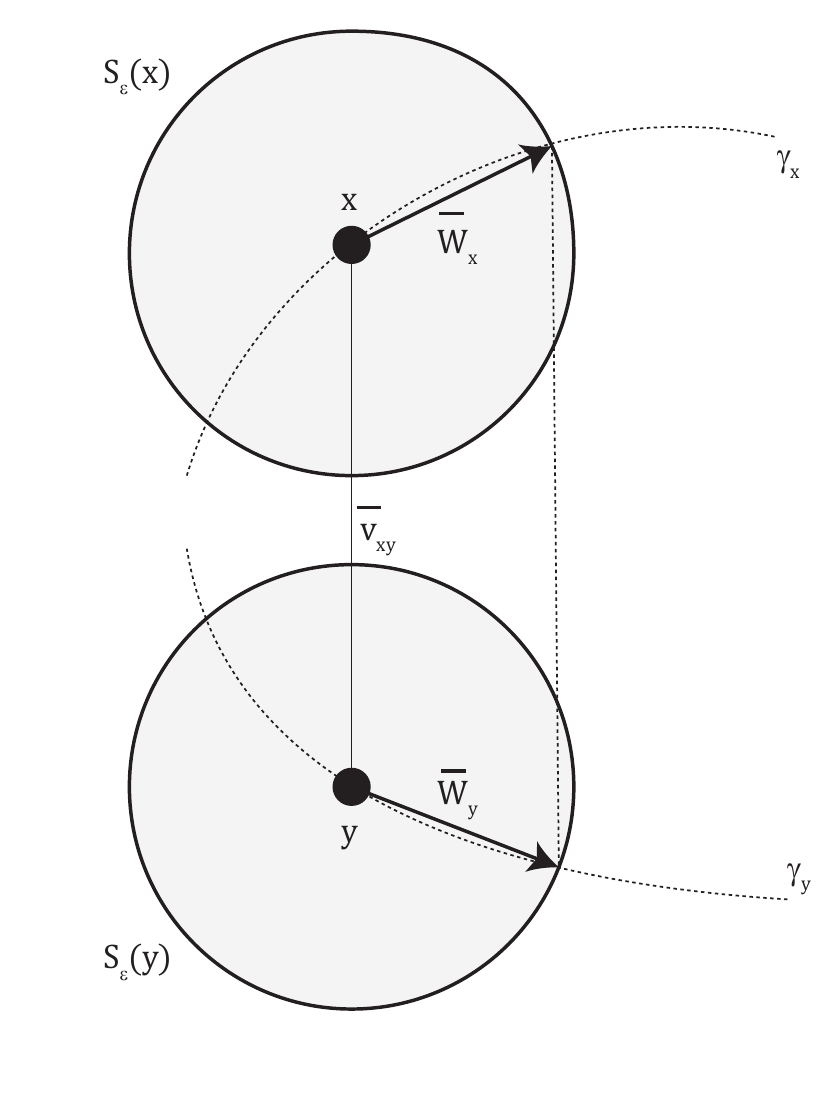}
\end{center}
\caption{Schematic figure on the geometric interpretation of Ollivier-Ricci 
(OR) curvature. Consider two close points $x$ and $y$ in a $n$-dimensional 
Riemannian manifold, defining a tangent vector $\bar{v}_{xy}$. In 
Ollivier-Ricci curvature, one considers the parallel transport in the 
direction $\bar{v}_{xy}$ wherein points on a infinitesimal sphere 
$S_\varepsilon(x)$ centered at $x$, are transported to points on the 
corresponding sphere $S_\varepsilon(y)$. In spaces of positive (respectively, 
negative) Ricci curvature, balls are closer (respectively, farther) than 
their centers.}
\end{figure}

\begin{figure}[!htbp]
\begin{center}
\includegraphics[width=0.34\linewidth]{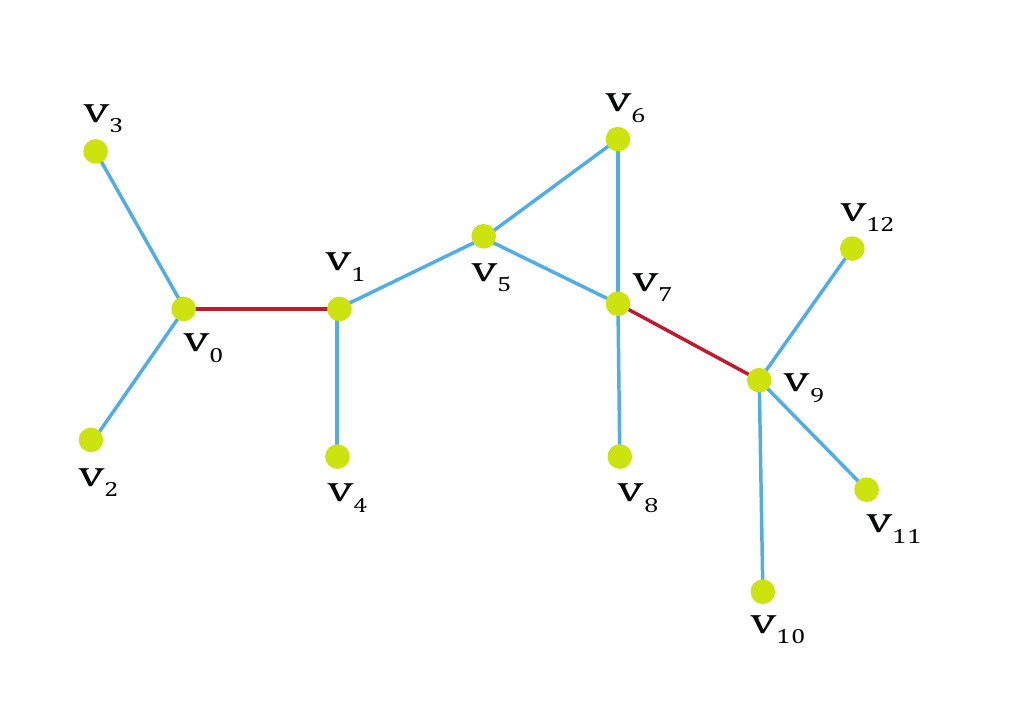}
\end{center}
\caption{An illustrative example of the computation of the Forman-Ricci (FR)
curvature of edges in a unweighted and undirected graph with 13 nodes and 13 
edges. The FR curvature of an edge $e$ depends only on the weights of its 
neighbouring edges and the two nodes constituting the edge (see Eq. 2.4 in main 
text). In the simplest case of unweighted graphs, the weights of nodes and 
edges are equal to $1$, and thus, the FR curvature of an edge $e$ in such an 
undirected graph is given by $\mathbf{F}(e=(v_i, v_j)) = 4 - \text{degree}(v_i) - 
\text{degree}(v_j)$. In this example, the FR curvature of the edge $(v_0, v_1)$ 
depends only on the edges connected to nodes $v_0$ and $v_1$ with degree 3 and 3, 
respectively. Therefore, $\mathbf{F}(v_0, v_1) = 4 - 3 - 3 = -2$. Similarly, the 
FR curvature of the edge $(v_7, v_9)$ depends only on the edges connected to nodes 
$v_7$ and $v_9$ with degree 4 and 4, respectively. Therefore, $\mathbf{F}(v_7, v_9) 
= 4 - 4 - 4 = -4$. As the edge $(v_7, v_9)$ has higher information spread at 
its ends (or anchoring nodes) in comparison to the edge $(v_0, v_1)$, it has more 
negative value of FR curvature.}
\end{figure}

\begin{figure}[!htbp]
\begin{center}
\includegraphics[width=0.52\linewidth]{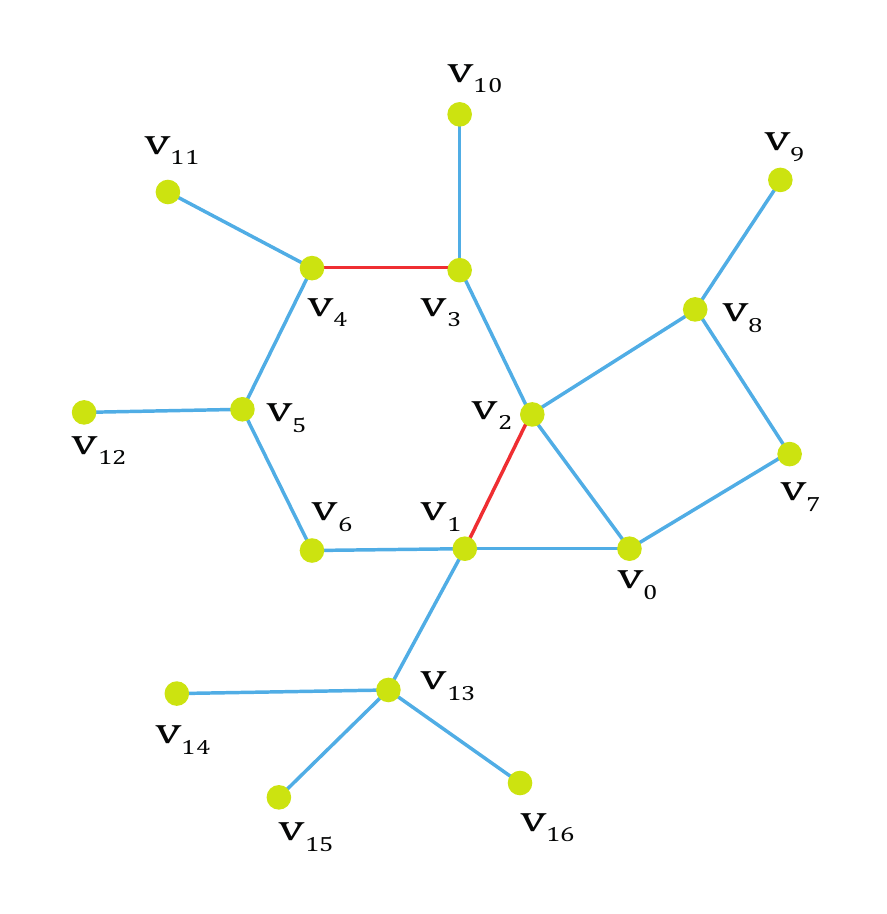}
\end{center}
\caption{An illustrative example of the computation of the Menger-Ricci (MR)
and Haantjes-Ricci (HR) curvature of edges in a unweighted and undirected graph 
with 17 nodes and 19 edges. The MR curvature of an edge $e$ depends only on the 
triangles adjacent to it (see Eq. 2.6 in main text), while the HR curvature 
depends on simple paths of any length that connect the two nodes at ends of
the edge (see Eq. 2.9 in main text). In case of unweighted graphs, each triangle 
adjacent to an edge $e$ contributes $\sqrt{3}/2$ to its MR curvature. In case of 
unweighted graphs, each path of length $n$ between the two nodes at the 
ends of an edge $e$ contributes $\sqrt{n-1}$ to its HR curvature. In this example, 
for the edge $(v_3, v_4)$, the MR curvature is 0 as there are no triangles 
adajacent to the edge, while the HR curvature is $\sqrt{4}$ as there is only a 
path $\pi = v_3, v_2, v_1, v_6, v_5, v_4$ of length 5 other than the edge 
connecting the nodes $v_3$ and $v_4$. For the edge $(v_1, v_2)$, the MR curvature 
is $\sqrt{3}/2$ as there is one triangle $v_0, v_1, v_2$ adjacent to the edge, 
while the HR curvature is $1 + \sqrt{3} + \sqrt{4}$ as there is a path $v_1, v_0, v_2$ 
of length 2, a path $v_1, v_0, v_7, v_8, v_2$ of length 4, and a path $v_1, v_6, v_5, 
v_4, v_3, v_2$ of length 5 other than the edge connecting the nodes $v_1$ and $v_2$.  
It can be seen from this simple example that HR curvature is more general than 
MR curvature as it depends on paths of any length unlike MR curvature which is 
dependent on only triangles.}
\end{figure}

\begin{figure}[!htbp]
\begin{center}
\includegraphics[width=0.9\linewidth]{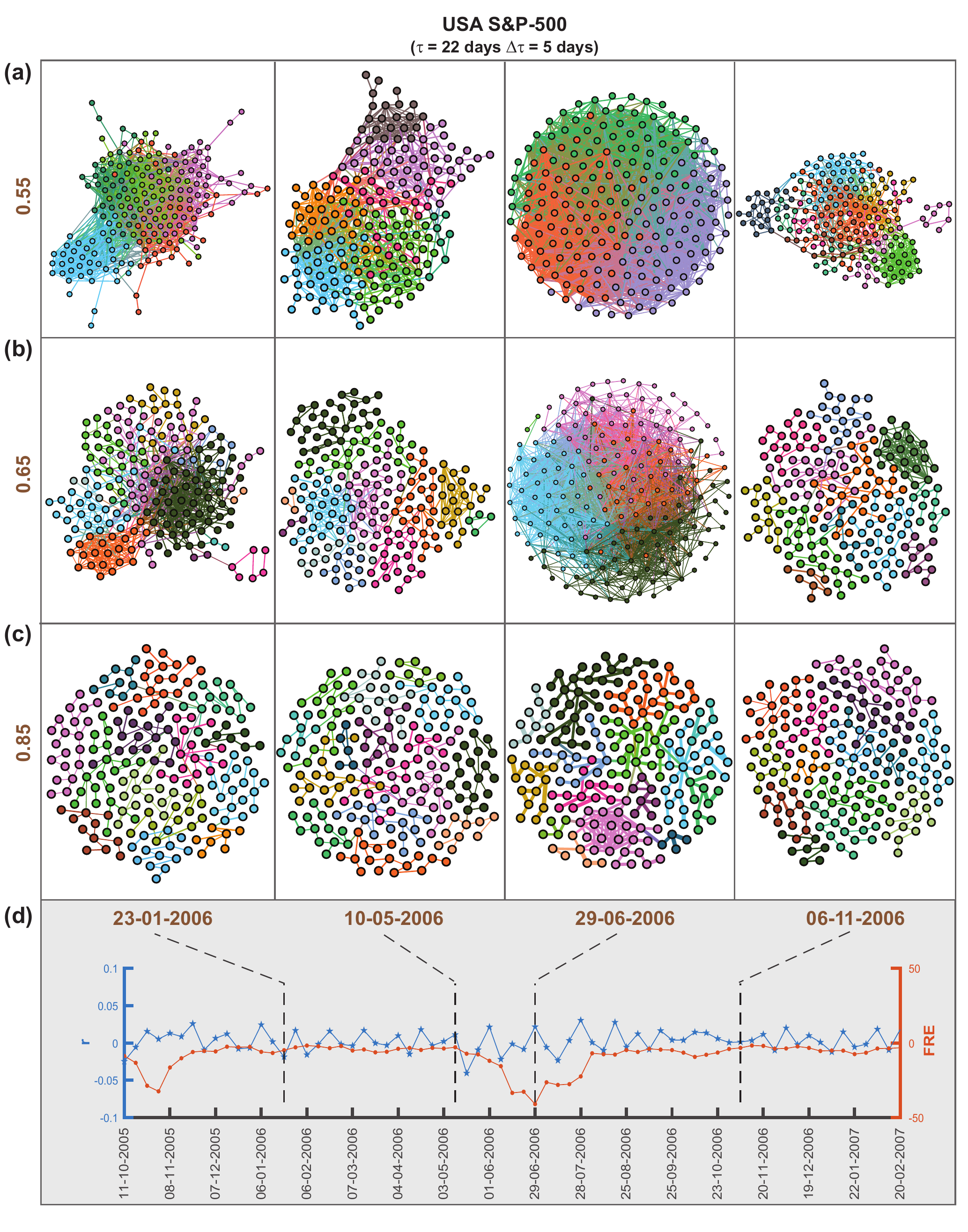}
\end{center}
\caption{Visualization of threshold networks for USA S\&P-500 market 
around the US Housing bubble period (2006-2007) at four distinct epochs 
of $\tau=22$ days ending on trading days 23-01-2006, 10-05-2006, 
29-06-2006, and 06-11-2006, with thresholds (\textbf{a}) $C_{ij(t)}\geq 0.55$, 
(\textbf{b}) $C_{ij}(t)\geq 0.65$, and (\textbf{c}) $C_{ij}(t)\geq 0.85$. Here, 
the colour of the nodes correspond to the different communities determined 
by Louvain method for community detection. The number of edges and communities 
in $S_\tau(t)$ for different thresholds are shown in Table S3. (\textbf{d}) 
Plot shows the evolution of log-returns $r$ of S\&P-500 index (blue color line) 
and average Forman-Ricci curvature of edges (FRE) (sienna color line) 
computed using threshold networks with $C_{ij}(t)\geq 0.75$ for the period 
around the US Housing bubble.}
\label{fig:corr-networks_housing}
\end{figure}

\begin{figure}[!htbp]
\begin{center}
\includegraphics[width=0.9\linewidth]{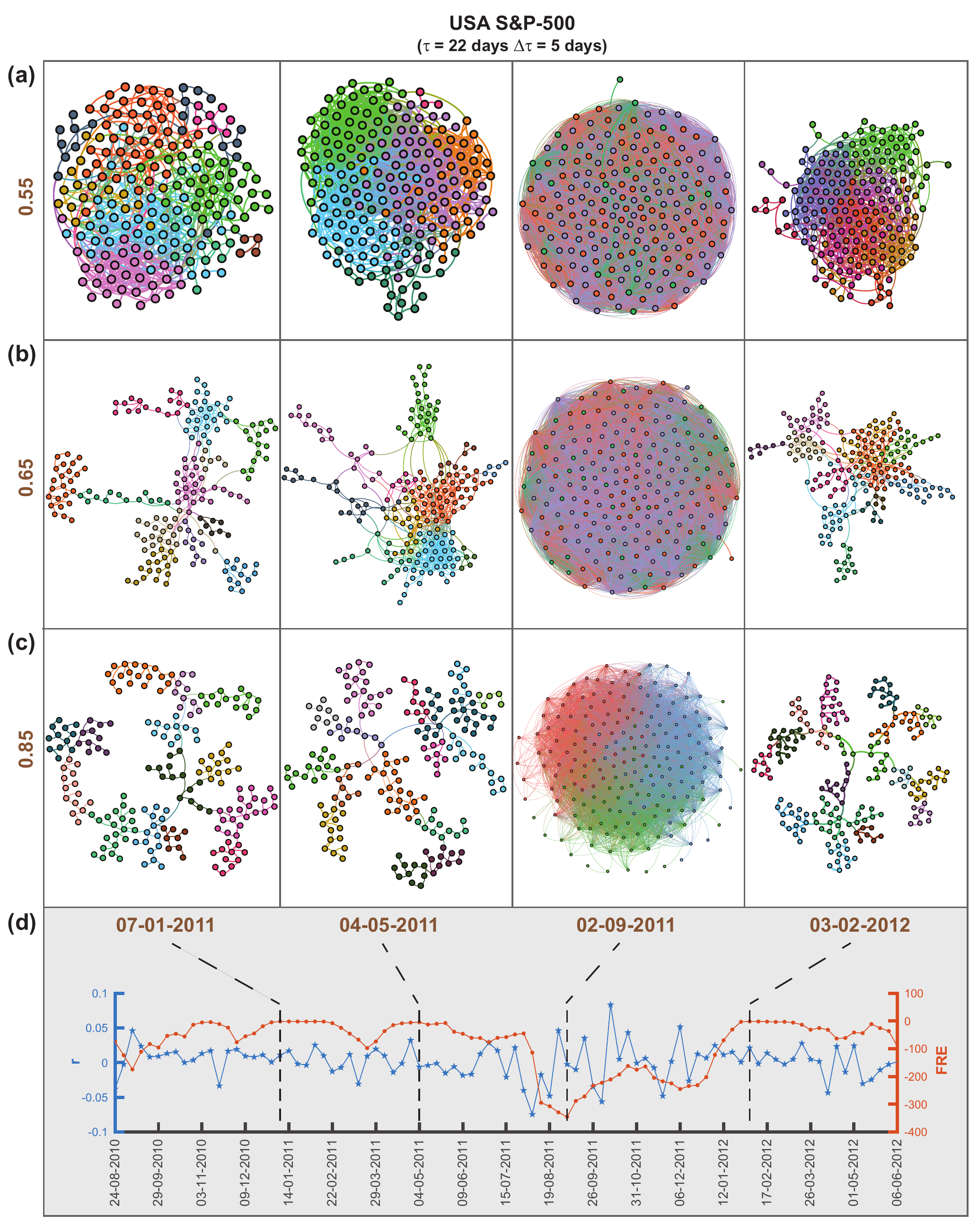}
\end{center}
\caption{Visualization of threshold networks for USA S\&P-500 market around 
the August 2011 stock markets fall at four distinct epochs of $\tau=22$ days 
ending on 07-01-2011, 04-05-2011, 02-09-2011, and 03-02-2012 with thresholds
(\textbf{a}) $C_{ij}(t)\geq 0.55$, (\textbf{b}) $C_{ij}(t)\geq 0.65$, and 
(\textbf{c}) $C_{ij}(t)\geq 0.85$. Here, the colour of the nodes correspond to 
the different communities determined by Louvain method for community detection. 
The number of edges and communities in $S_\tau(t)$ for different thresholds 
are shown in Table S3. (\textbf{d}) Plot shows the evolution of log-returns 
$r$ of S\&P-500 index (blue color line) and average Forman-Ricci curvature of 
edges (FRE) (sienna color line) computed using threshold networks with 
$C_{ij}(t)\geq 0.75$ for the period around the August 2011 stock markets fall 
crisis.}
\label{fig:corr-networks_marketfall}
\end{figure}

\begin{figure}[!htbp]
\begin{center}
\includegraphics[width=0.64\linewidth]{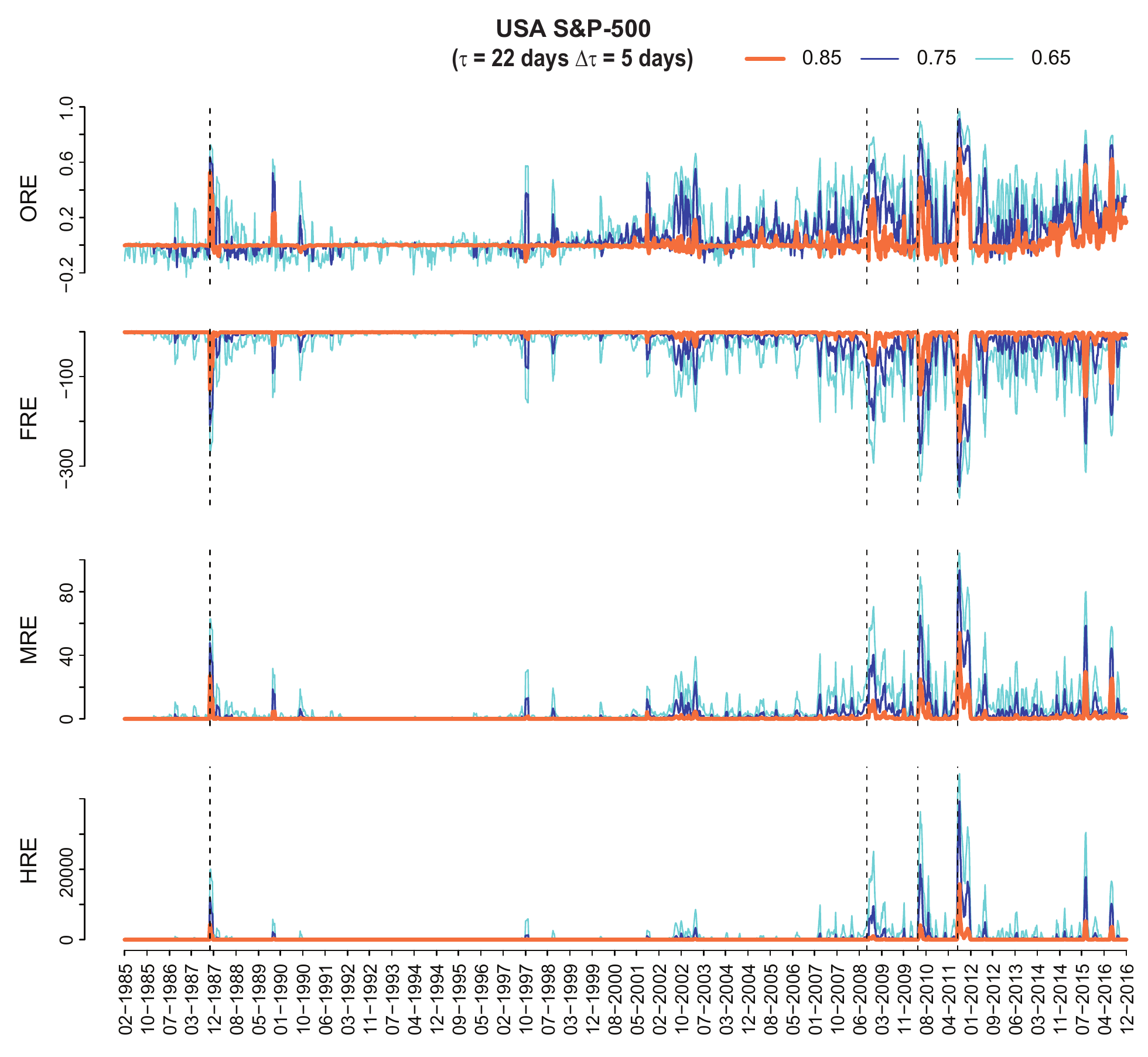}
\llap{\parbox[b]{23cm}{\textbf{(a)}\\\rule{0ex}{3.9in}}}
\includegraphics[width=0.64\linewidth]{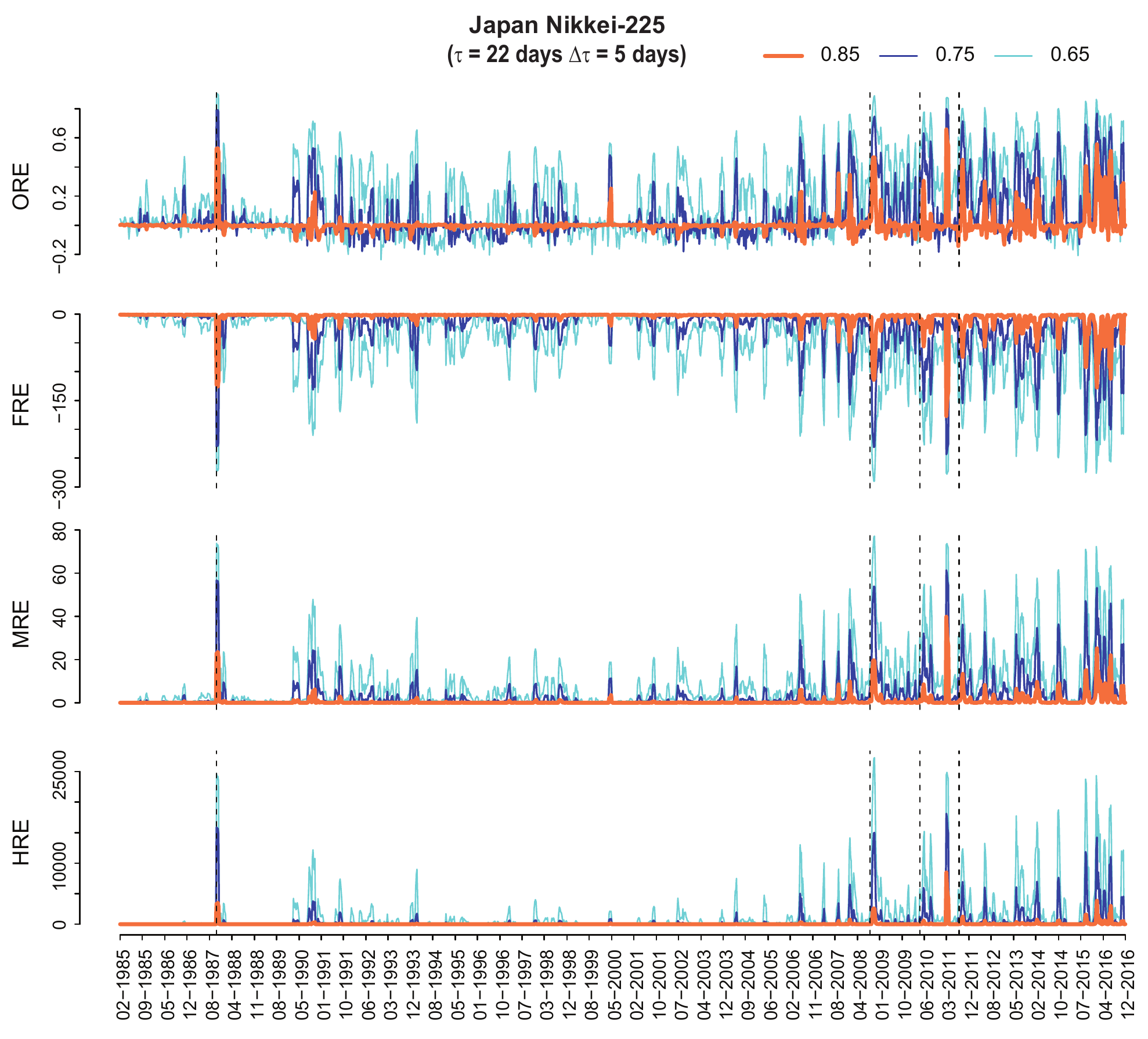}
\llap{\parbox[b]{23cm}{\textbf{(b)}\\\rule{0ex}{3.9in}}}
\end{center}
\caption{Comparison plots for the four edge-centric geometric curvatures, 
namely, Ollivier-Ricci (ORE), Forman-Ricci (FRE), Menger-Ricci (MRE) and 
Haantjes-Ricci (HRE) in threshold networks $S_\tau(t)$ obtained using 
three different thresholds $C_{ij}(t)\geq 0.65$ (cyan color), $C_{ij}(t)\geq 0.75$ 
(dark blue color), and  $C_{ij}(t)\geq 0.85$ (sienna color) for (\textbf{a}) 
USA S\&P-500 and (\textbf{b}) Japan Nikkei-225 markets. The curvature 
measures are calculated for time epochs of $\tau= 22$ days and overlapping 
shift of $\Delta\tau = 5$ days over the period (1985-2016). The absolute 
value of ORE, FRE, MRE and HRE decreases with the increase in the threshold 
$C_{ij}(t)$ used to construct $S_\tau(t)$. Four vertical dashed lines correspond 
to the epochs of four important crashes (Black Monday 1987, Lehman Brothers 
crash 2008, DJ Flash crash 2010, and August 2011 stock markets fall) listed 
in the Table 1 of the main text.}
\label{fig:TS_sensitivity}
\end{figure}

\begin{figure}[!htbp]
\begin{center}
\includegraphics[width=0.65\linewidth]{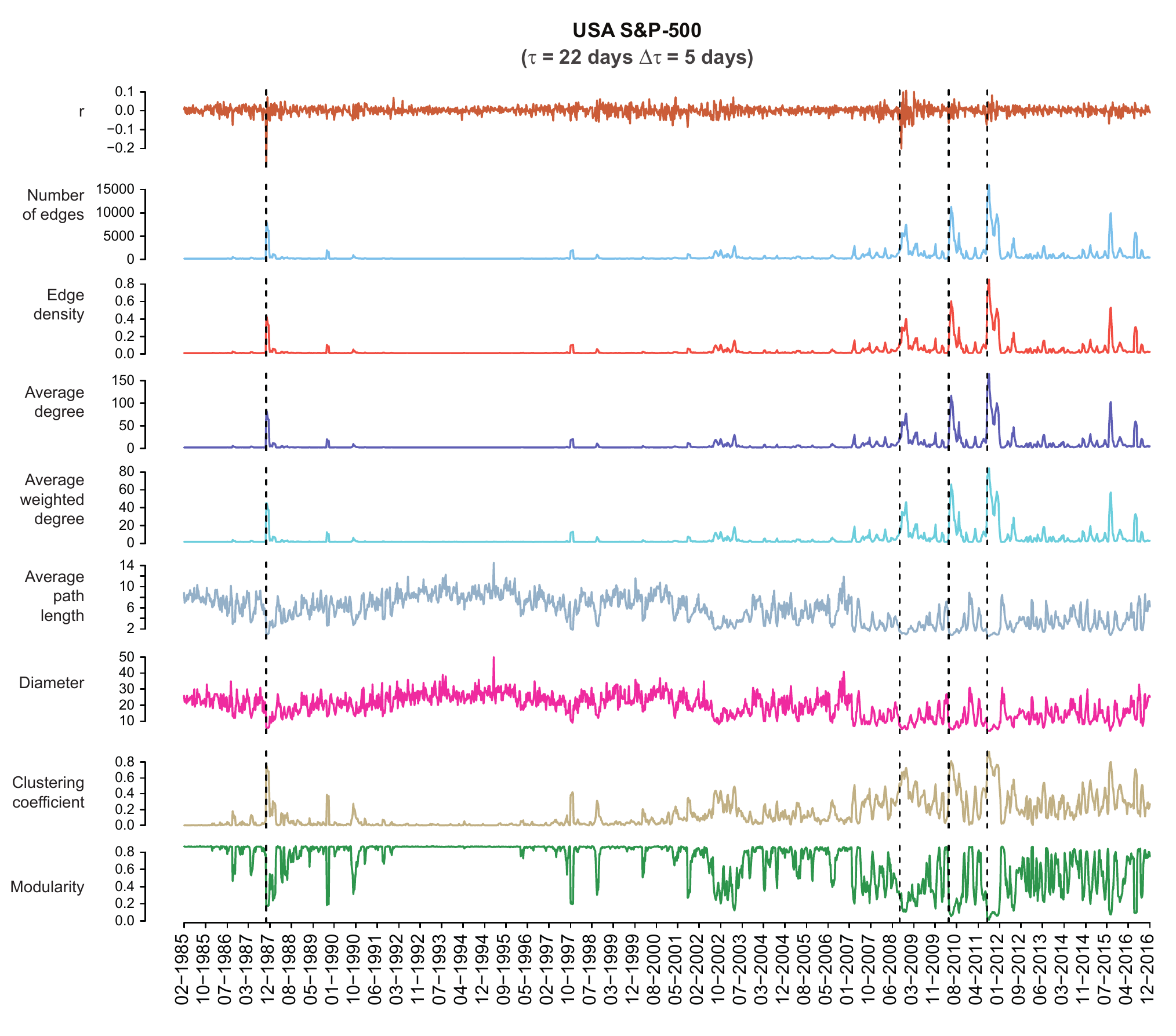}
\llap{\parbox[b]{23cm}{\textbf{(a)}\\\rule{0ex}{3.8in}}}
\includegraphics[width=0.65\linewidth]{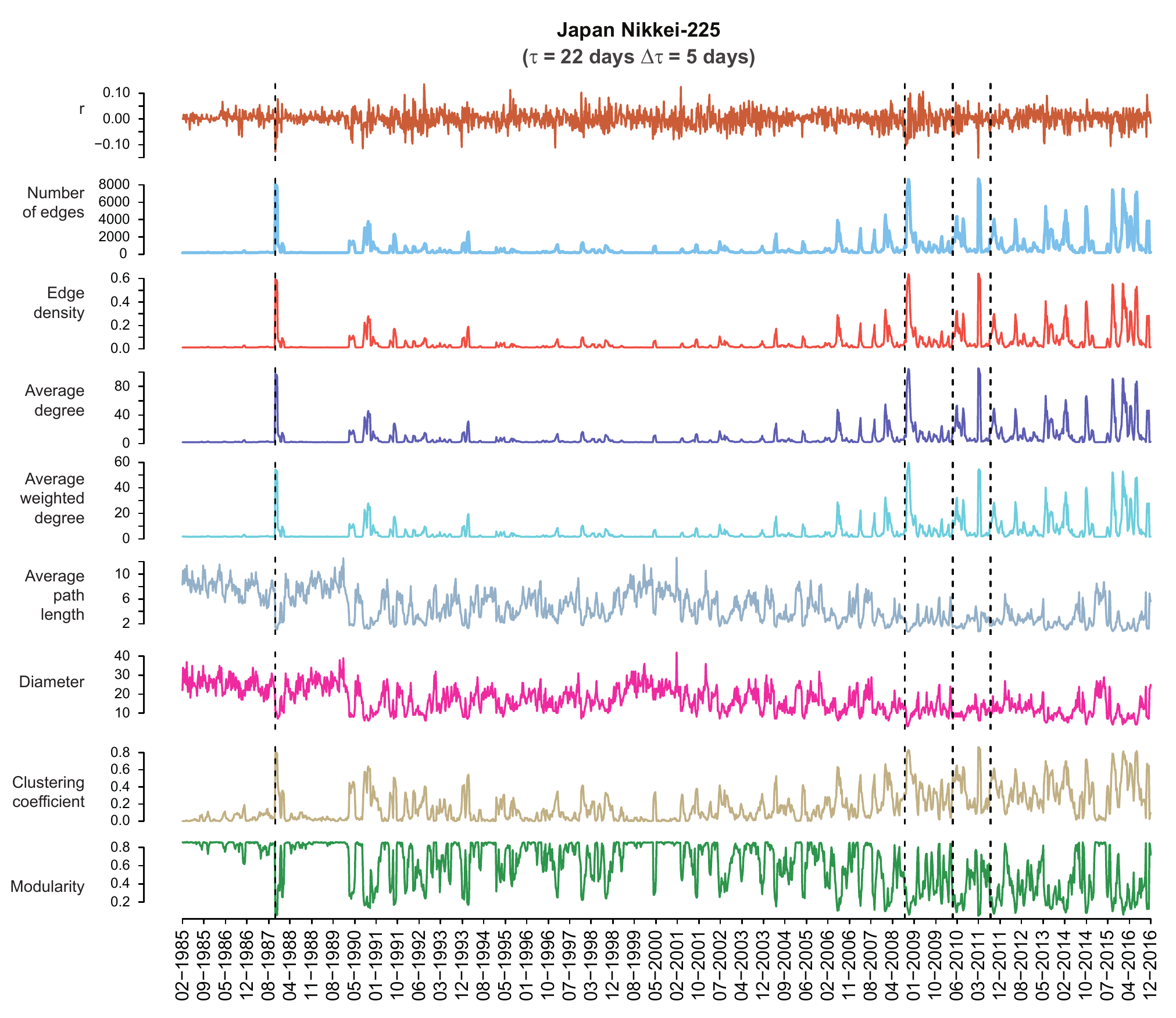}
\llap{\parbox[b]{23cm}{\textbf{(b)}\\\rule{0ex}{3.8in}}}
\end{center}
\caption{Evolution of network properties for (\textbf{a}) USA S\&P-500 
and (\textbf{b}) Japanese Nikkei-225 markets evaluated from the 
correlation matrices $\boldsymbol C_\tau(t)$ of window size $\tau=22$ 
days and an overlapping shift of $\Delta\tau = 5$ days over the period 
(1985-2016). From top to bottom, we compare the plot of index log-returns 
$r$ with common network measures, namely, number of edges, edge density, 
average degree, average weighted degree, average path length, diameter, 
clustering coefficient and modularity.}
\label{fig:TS_networks}
\end{figure}

\begin{figure}[!htbp]
\begin{center}
\includegraphics[width=0.6\linewidth]{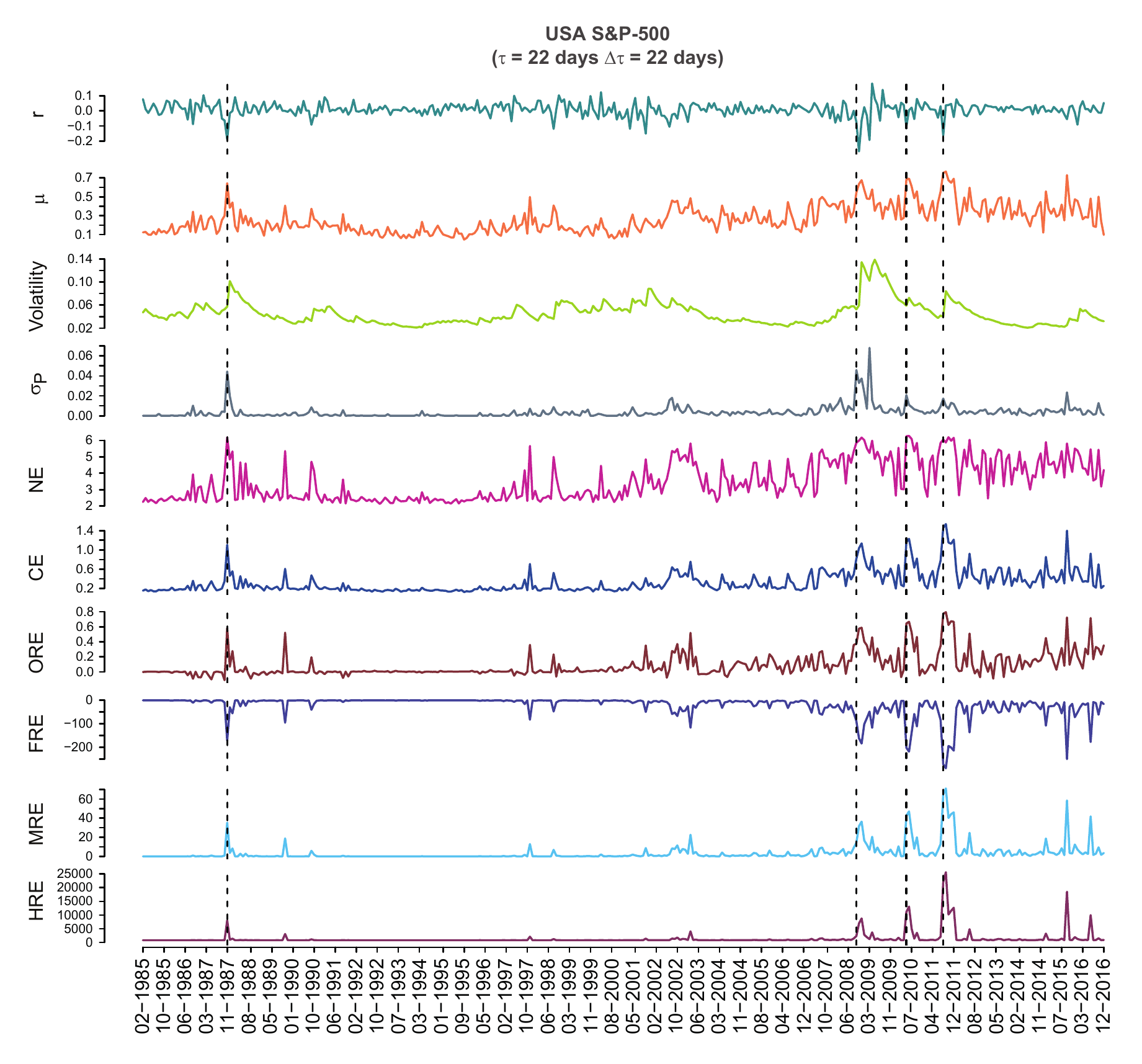}
\llap{\parbox[b]{23cm}{\textbf{(a)}\\\rule{0ex}{3.9in}}}
\includegraphics[width=0.6\linewidth]{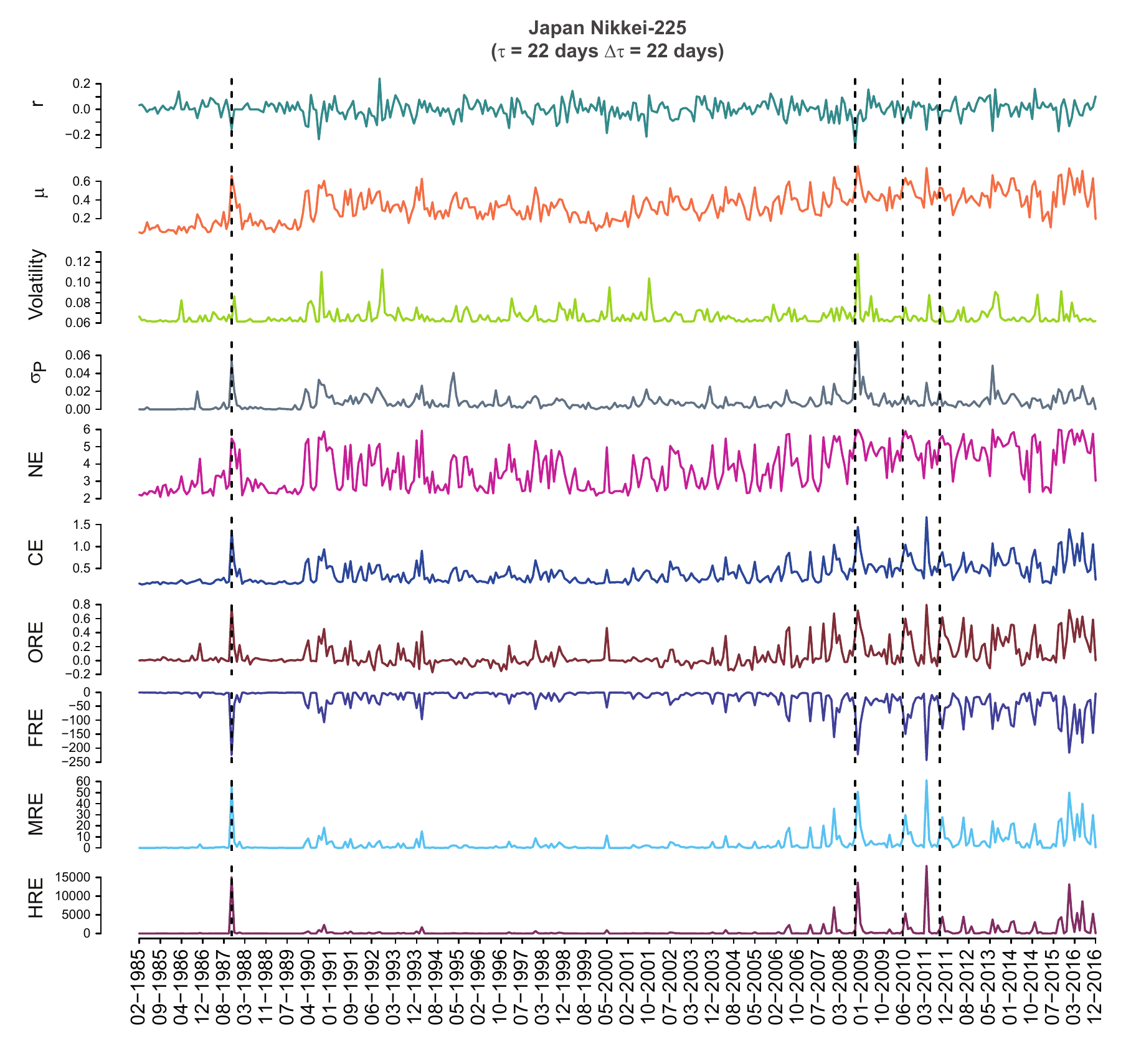}
\llap{\parbox[b]{23cm}{\textbf{(b)}\\\rule{0ex}{3.9in}}}
\end{center}
\caption{Evolution of the market indicators and edge-centric geometric 
curvatures for (\textbf{a}) USA S\&P-500 and (\textbf{b}) Japanese 
Nikkei-225 markets. From top to bottom, we plot the index log-returns $r$,  
mean market correlation $\mu$, volatility of the market index $r$ estimated 
using GARCH(1,1) process, risk $\sigma_P$ corresponding to the minimum risk 
Markowitz portfolio of all the stocks in the market, network entropy (NE), 
communication efficiency (CE), average of Ollivier-Ricci (ORE), Forman-Ricci 
(FRE), Menger-Ricci (MRE), and Haantjes-Ricci (HRE) curvature of edges 
evaluated from the correlation matrices $\boldsymbol C_\tau(t)$ of window 
size $\tau=22$ days and a non-overlapping shift of $\Delta\tau=22$ days. 
Four vertical dashed lines indicate the epochs of four important crashes: 
Black Monday 1987, Lehman Brothers crash 2008, DJ Flash crash 2010, and 
August 2011 stock markets fall.}
\label{fig:TS_Ricci_non}
\end{figure}

\begin{figure}[!htbp]
\includegraphics[width=0.68\linewidth]{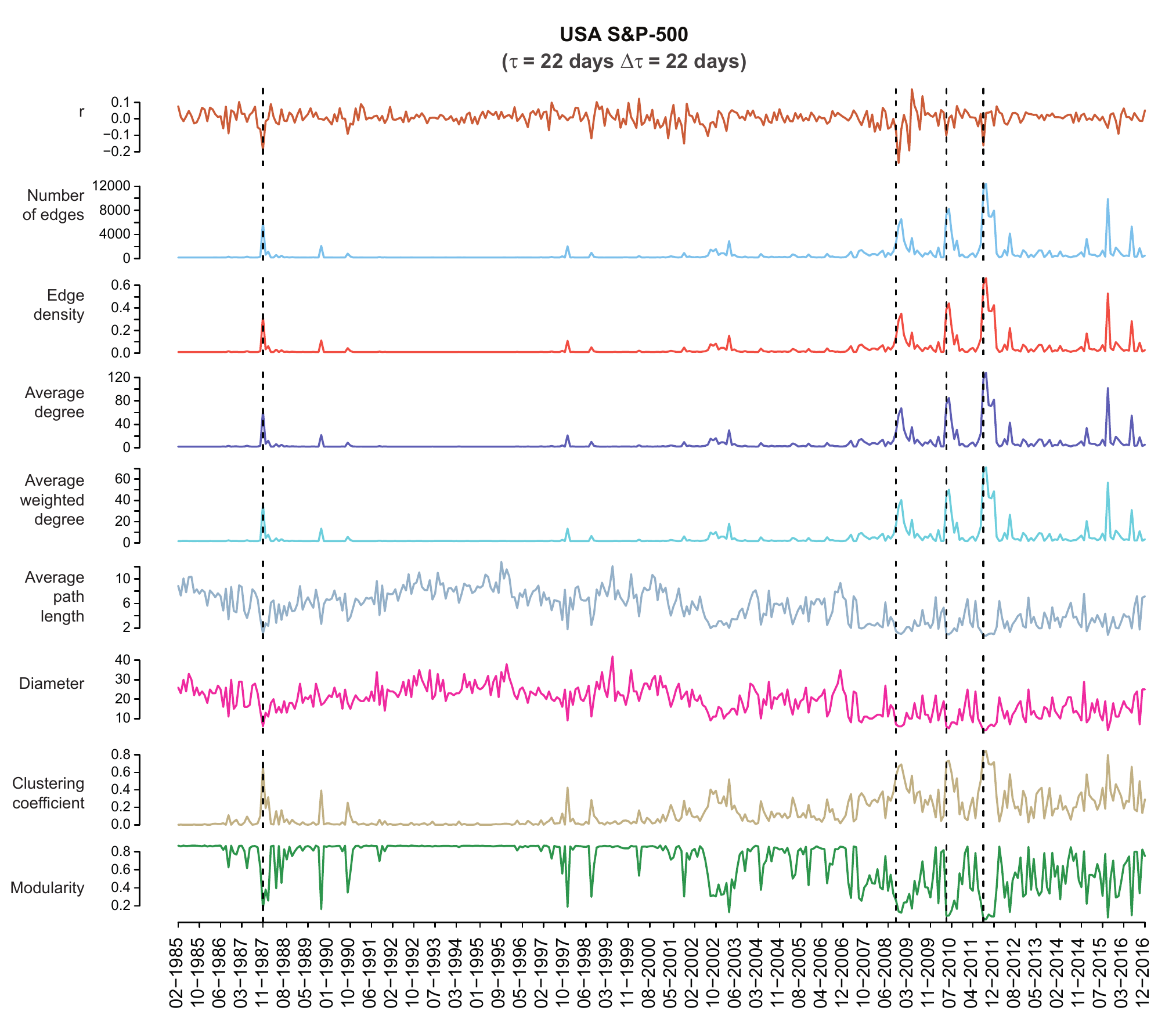}
\llap{\parbox[b]{25cm}{\textbf{(a)}\\\rule{0ex}{3.9in}}}
\includegraphics[width=0.68\linewidth]{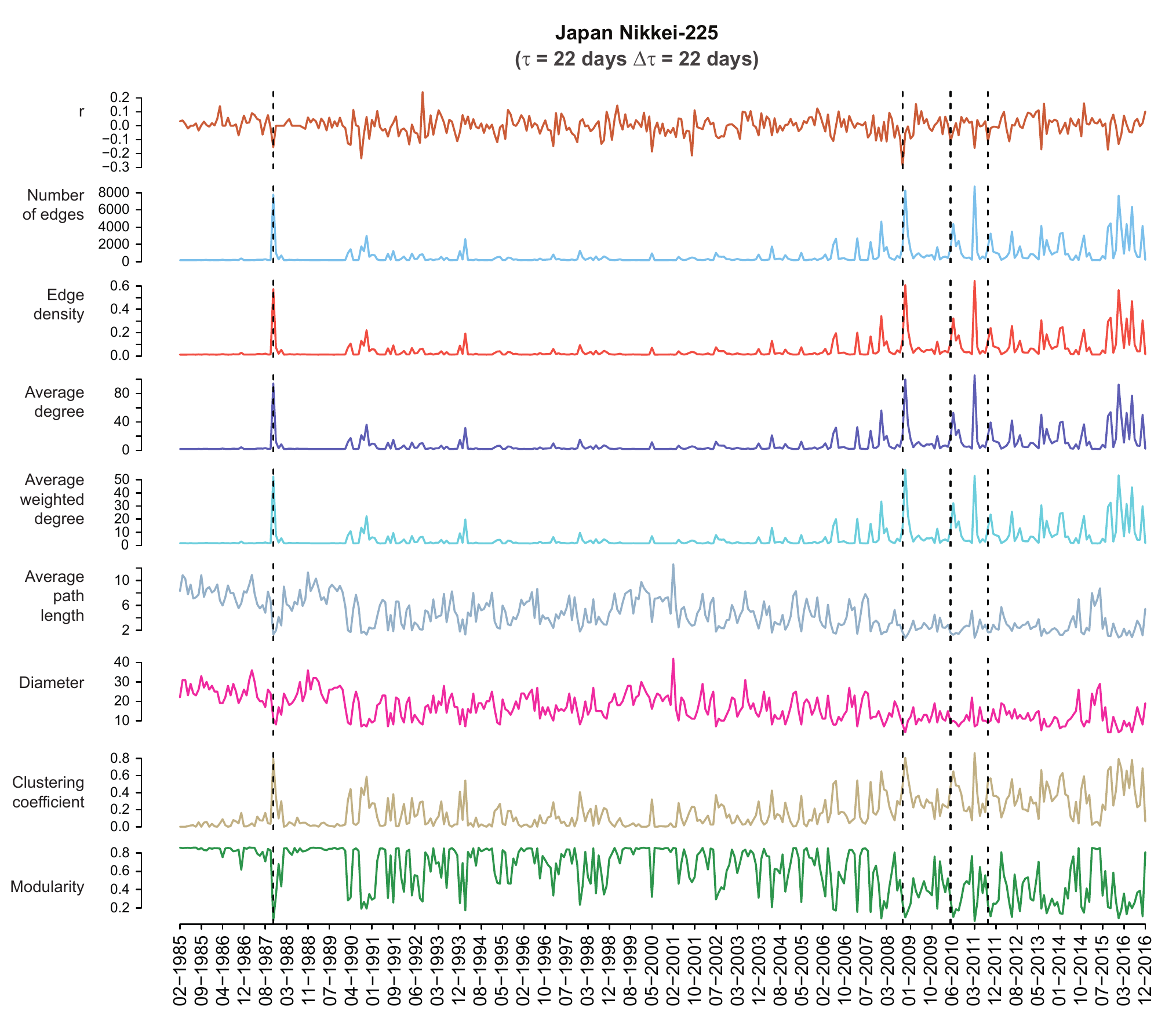}
\llap{\parbox[b]{25cm}{\textbf{(b)}\\\rule{0ex}{3.9in}}}
\caption{Evolution of network properties for (\textbf{a}) USA S\&P-500 
and (\textbf{b}) Japanese Nikkei-225 markets evaluated from the 
correlation matrices $\boldsymbol C_\tau(t)$ of window size $\tau=22$ 
days and a non-overlapping shift of $\Delta\tau = 22$ days over the 
period (1985-2016). From top to bottom, we compare the plot of index 
log-returns $r$ with common network measures, namely, number of edges, 
edge density, average degree, average weighted degree, average path 
length, diameter, clustering coefficient and modularity.}
\label{fig:TS_networks_non}
\end{figure}

\begin{figure}[!htbp]
\begin{center}
\includegraphics[width=0.61\linewidth]{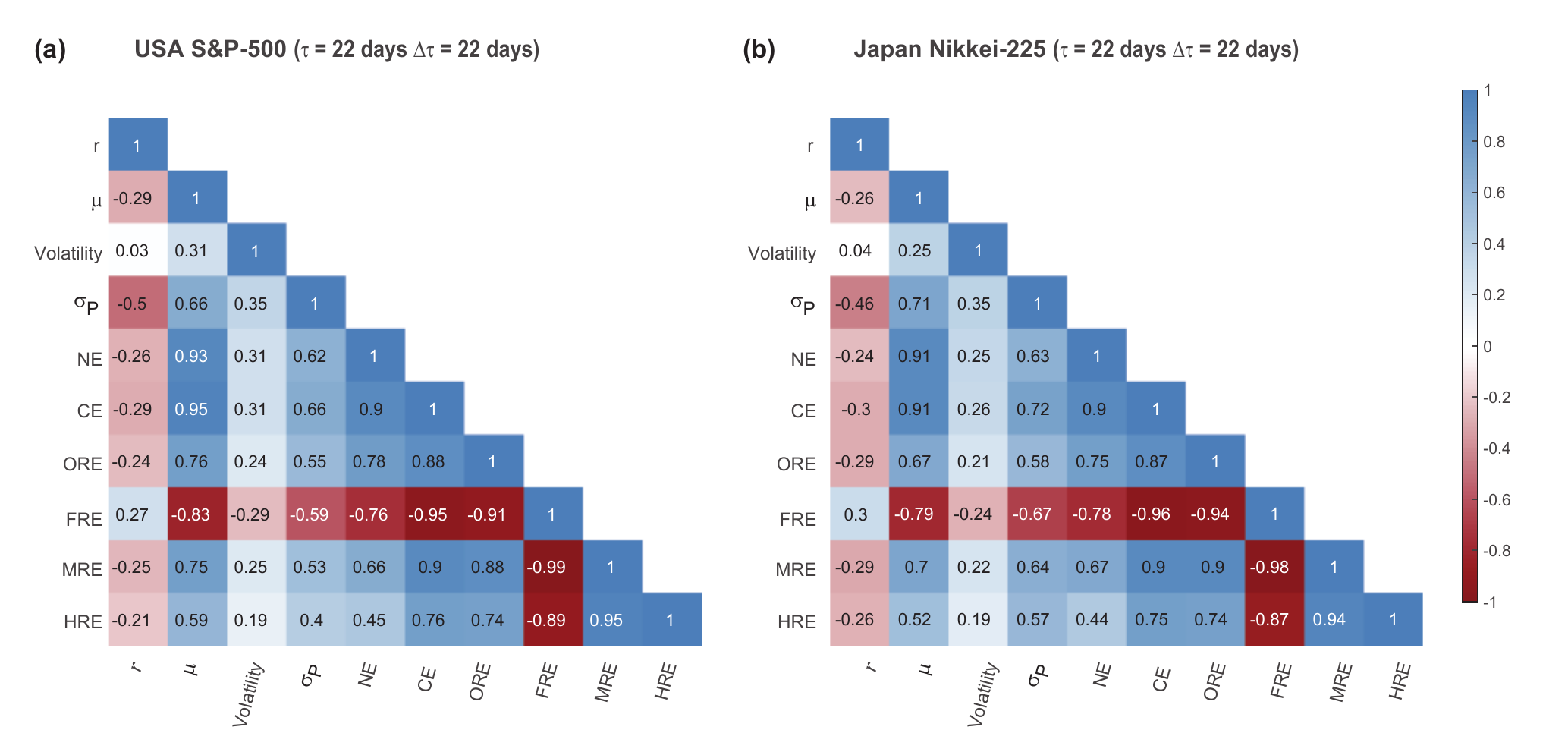}
\end{center}
\caption{Correlogram plots of \textbf{(a)} USA S\&P-500 and \textbf{(b)} 
Japan Nikkei-225 markets, for the traditional market indicators (index 
returns $r$, mean market correlation $\mu$, volatility, and minimum 
portfolio risk $\sigma_P$), network properties (network entropy (NE) and 
communication efficiency (CE)) and discrete Ricci curvatures for edges 
(Ollivier-Ricci ORE, Forman-Ricci FRE, Menger-Ricci MRE, and Haantjes-Ricci 
HRE), computed for epochs of size $\tau=22$ days and non-overlapping shift 
of $\Delta\tau=22$ days. Among the four curvature measures, FRE has the 
highest correlation with the market indicators, and this measure can be 
used as an indicator of market risk as it captures local to global 
system-level fragility of the markets.}
\label{fig:corrologram}
\end{figure}

\begin{figure}[!htbp]
\begin{center}
\includegraphics[width=0.61\linewidth]{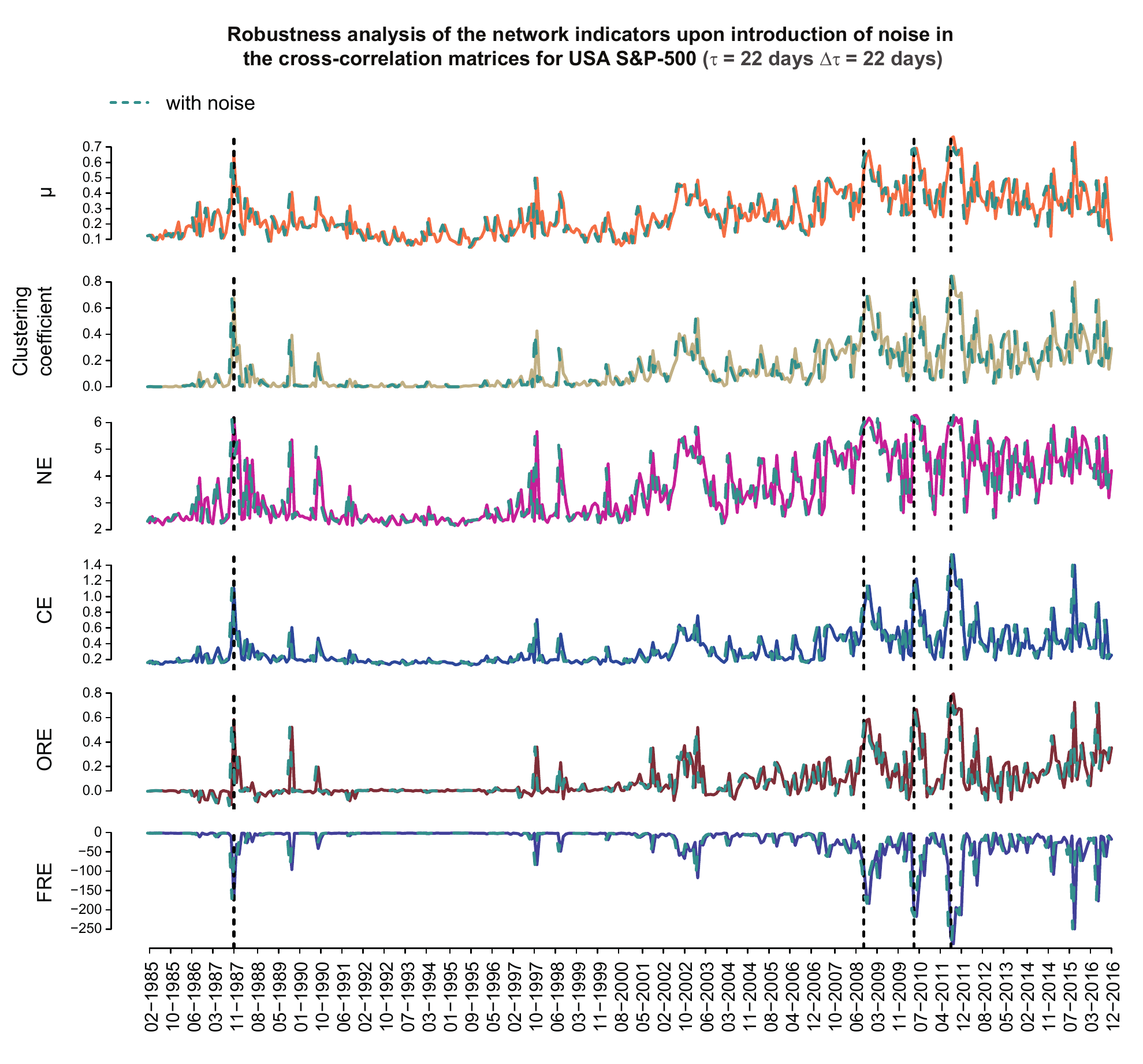}
\end{center}
\caption{Evolution of network properties for USA S\&P-500 market, 
evaluated from the empirical correlation matrices $C_\tau (t)$ and the 
correlation matrices perturbed by small amounts of random Gaussian noise 
(off diagonal elements of Wishart matrix with dimension same as 
$C_\tau (t)$), keeping correlation coefficients bounded in the range 
$[-1, 1]$, for epochs of size $\tau=22$ days and non-overlapping shift 
of $\Delta\tau=22$ days over the 32-year period (1985-2016). From top 
to bottom, we compare the plots of mean correlation $\mu$, clustering 
coefficient, network entropy (NE), communication efficiency (CE), 
average of Ollivier-Ricci (ORE) and Forman-Ricci (FRE) curvature of edges.
In this figure, the solid lines represent the actual empirical values 
and the dashed lines represent the perturbed values. The near overlap of 
these lines, for each of the measures, signifies that our methodology 
is not sensitive to small noise.}
\label{fig:noise}
\end{figure}

\newpage
\begin{table}[!htbp]
\centering
\caption{List of all stocks in USA S\&P-500 market considered in this 
analysis. The first column gives the serial number, the second column 
gives the symbols, the third column gives the full name of the stock, 
the fourth column specifies the sector, and the fifth column has the 
abbreviation of the sector for the S\&P-500 market.}
\label{USA_Table}
\begin{tabular}{|p{0.8cm}|p{1.5cm}|p{7cm}|p{4cm}|p{2cm}|}
\hline
\textbf{S.No.} & \textbf{Code} & \textbf{Company Name} & \textbf{Sector} & \textbf{Abbreviation} \\ \hline
1 & CMCSA & Comcast Corp. & Consumer Discretionary & CD \\ \hline
2 & DIS & The Walt Disney Company & Consumer Discretionary & CD \\ \hline
3 & F & Ford Motor & Consumer Discretionary & CD \\ \hline
4 & GPC & Genuine Parts & Consumer Discretionary & CD \\ \hline
5 & GPS & Gap Inc. & Consumer Discretionary & CD \\ \hline
6 & GT & Goodyear Tire \& Rubber & Consumer Discretionary & CD \\ \hline
7 & HAS & Hasbro Inc. & Consumer Discretionary & CD \\ \hline
8 & HD & Home Depot & Consumer Discretionary & CD \\ \hline
9 & HRB & Block H\&R & Consumer Discretionary & CD \\ \hline
10 & IPG & Interpublic Group & Consumer Discretionary & CD \\ \hline
11 & JCP & J. C. Penney Company, Inc. & Consumer Discretionary & CD \\ \hline
12 & JWN & Nordstrom & Consumer Discretionary & CD \\ \hline
13 & LEG & Leggett \& Platt & Consumer Discretionary & CD \\ \hline
14 & LEN & Lennar Corp. & Consumer Discretionary & CD \\ \hline
15 & LOW & Lowe's Cos. & Consumer Discretionary & CD \\ \hline
16 & MAT & Mattel Inc. & Consumer Discretionary & CD \\ \hline
17 & MCD & McDonald's Corp. & Consumer Discretionary & CD \\ \hline
18 & NKE & Nike & Consumer Discretionary & CD \\ \hline
19 & SHW & Sherwin-Williams & Consumer Discretionary & CD \\ \hline
20 & TGT & Target Corp. & Consumer Discretionary & CD \\ \hline
21 & VFC & V.F. Corp. & Consumer Discretionary & CD \\ \hline
22 & WHR & Whirlpool Corp. & Consumer Discretionary & CD \\ \hline
23 & ADM & Archer-Daniels-Midland Co & Consumer Staples & CS \\ \hline
24 & AVP & Avon Products, Inc. & Consumer Staples & CS \\ \hline
25 & CAG & Conagra Brands & Consumer Staples & CS \\ \hline
26 & CL & Colgate-Palmolive & Consumer Staples & CS \\ \hline
27 & CPB & Campbell Soup & Consumer Staples & CS \\ \hline
28 & CVS & CVS Health & Consumer Staples & CS \\ \hline
29 & GIS & General Mills & Consumer Staples & CS \\ \hline
30 & HRL & Hormel Foods Corp. & Consumer Staples & CS \\ \hline
31 & HSY & The Hershey Company & Consumer Staples & CS \\ \hline
32 & K & Kellogg Co. & Consumer Staples & CS \\ \hline
33 & KMB & Kimberly-Clark & Consumer Staples & CS \\ \hline
34 & KO & Coca-Cola Company (The) & Consumer Staples & CS \\ \hline
35 & KR & Kroger Co. & Consumer Staples & CS \\ \hline
36 & MKC & McCormick \& Co. & Consumer Staples & CS \\ \hline
37 & MO & Altria Group Inc & Consumer Staples & CS \\ \hline
38 & SYY & Sysco Corp. & Consumer Staples & CS \\ \hline
39 & TAP & Molson Coors Brewing Company & Consumer Staples & CS \\ \hline
40 & TSN & Tyson Foods & Consumer Staples & CS \\ \hline
41 & WMT & Wal-Mart Stores & Consumer Staples & CS \\ \hline
42 & APA & Apache Corporation & Energy & EG \\ \hline
43 & COP & ConocoPhillips & Energy & EG \\ \hline
44 & CVX & Chevron Corp. & Energy & EG \\ \hline
45 & ESV & Ensco plc & Energy & EG \\ \hline
46 & HAL & Halliburton Co. & Energy & EG \\ \hline
47 & HES & Hess Corporation & Energy & EG \\ \hline
48 & HP & Helmerich \& Payne & Energy & EG \\ \hline
49 & MRO & Marathon Oil Corp. & Energy & EG \\ \hline
50 & MUR & Murphy Oil Corporation & Energy & EG \\ \hline
\end{tabular}
\end{table}
\begin{table}[!htbp]
\centering
\begin{tabular}{|p{0.8cm}|p{1.5cm}|p{7cm}|p{4cm}|p{2cm}|}
\hline
51 & NBL & Noble Energy Inc & Energy & EG \\ \hline
52 & NBR & Nabors Industries Ltd. & Energy & EG \\ \hline
53 & SLB & Schlumberger Ltd. & Energy & EG \\ \hline
54 & TSO & Tesoro Corp & Energy & EG \\ \hline
55 & VLO & Valero Energy & Energy & EG \\ \hline
56 & WMB & Williams Cos. & Energy & EG \\ \hline
57 & XOM & Exxon Mobil Corp. & Energy & EG \\ \hline
58 & AFL & AFLAC Inc & Financials & FN \\ \hline
59 & AIG & American International Group, Inc. & Financials & FN \\ \hline
60 & AON & Aon plc & Financials & FN \\ \hline
61 & AXP & American Express Co & Financials & FN \\ \hline
62 & BAC & Bank of America Corp & Financials & FN \\ \hline
63 & BBT & BB\&T Corporation & Financials & FN \\ \hline
64 & BEN & Franklin Resources & Financials & FN \\ \hline
65 & BK & The Bank of New York Mellon Corp. & Financials & FN \\ \hline
66 & C & Citigroup Inc. & Financials & FN \\ \hline
67 & CB & Chubb Limited & Financials & FN \\ \hline
68 & CINF & Cincinnati Financial & Financials & FN \\ \hline
69 & CMA & Comerica Inc. & Financials & FN \\ \hline
70 & EFX & Equifax Inc. & Financials & FN \\ \hline
71 & FHN & First Horizon National Corporation & Financials & FN \\ \hline
72 & HBAN & Huntington Bancshares & Financials & FN \\ \hline
73 & HCN & Welltower Inc. & Financials & FN \\ \hline
74 & HST & Host Hotels \& Resorts, Inc. & Financials & FN \\ \hline
75 & JPM & JPMorgan Chase \& Co. & Financials & FN \\ \hline
76 & L & Loews Corp. & Financials & FN \\ \hline
77 & LM & Legg Mason, Inc. & Financials & FN \\ \hline
78 & LNC & Lincoln National & Financials & FN \\ \hline
79 & LUK & Leucadia National Corp. & Financials & FN \\ \hline
80 & MMC & Marsh \& McLennan & Financials & FN \\ \hline
81 & MTB & M\&T Bank Corp. & Financials & FN \\ \hline
82 & PSA & Public Storage & Financials & FN \\ \hline
83 & SLM & SLM Corporation & Financials & FN \\ \hline
84 & TMK & Torchmark Corp. & Financials & FN \\ \hline
85 & TRV & The Travelers Companies Inc. & Financials & FN \\ \hline
86 & USB & U.S. Bancorp & Financials & FN \\ \hline
87 & VNO & Vornado Realty Trust & Financials & FN \\ \hline
88 & WFC & Wells Fargo & Financials & FN \\ \hline
89 & WY & Weyerhaeuser Corp. & Financials & FN \\ \hline
90 & ZION & Zions Bancorp & Financials & FN \\ \hline
91 & ABT & Abbott Laboratories & Health Care & HC \\ \hline
92 & AET & Aetna Inc & Health Care & HC \\ \hline
93 & AMGN & Amgen Inc & Health Care & HC \\ \hline
94 & BAX & Baxter International Inc. & Health Care & HC \\ \hline
95 & BCR & Bard (C.R.) Inc. & Health Care & HC \\ \hline
96 & BDX & Becton Dickinson & Health Care & HC \\ \hline
97 & BMY & Bristol-Myers Squibb & Health Care & HC \\ \hline
98 & CAH & Cardinal Health Inc. & Health Care & HC \\ \hline
99 & CI & CIGNA Corp. & Health Care & HC \\ \hline
100 & HUM & Humana Inc. & Health Care & HC \\ \hline
\end{tabular}
\end{table}
\begin{table}[!htbp]
\centering
\begin{tabular}{|p{0.8cm}|p{1.5cm}|p{7cm}|p{4cm}|p{2cm}|}
\hline
101 & JNJ & Johnson \& Johnson & Health Care & HC \\ \hline
102 & LLY & Lilly (Eli) \& Co. & Health Care & HC \\ \hline
103 & MDT & Medtronic plc & Health Care & HC \\ \hline
104 & MRK & Merck \& Co. & Health Care & HC \\ \hline
105 & MYL & Mylan N.V. & Health Care & HC \\ \hline
106 & SYK & Stryker Corp. & Health Care & HC \\ \hline
107 & THC & Tenet Healthcare Corp & Health Care & HC \\ \hline
108 & TMO & Thermo Fisher Scientific & Health Care & HC \\ \hline
109 & UNH & United Health Group Inc. & Health Care & HC \\ \hline
110 & VAR & Varian Medical Systems & Health Care & HC \\ \hline
111 & AVY & Avery Dennison Corp & Industrials & ID \\ \hline
112 & BA & Boeing Company & Industrials & ID \\ \hline
113 & CAT & Caterpillar Inc. & Industrials & ID \\ \hline
114 & CMI & Cummins Inc. & Industrials & ID \\ \hline
115 & CSX & CSX Corp. & Industrials & ID \\ \hline
116 & CTAS & Cintas Corporation & Industrials & ID \\ \hline
117 & DE & Deere \& Co. & Industrials & ID \\ \hline
118 & DHR & Danaher Corp. & Industrials & ID \\ \hline
119 & DNB & The Dun \& Bradstreet Corporation & Industrials & ID \\ \hline
120 & DOV & Dover Corp. & Industrials & ID \\ \hline
121 & EMR & Emerson Electric Company & Industrials & ID \\ \hline
122 & ETN & Eaton Corporation & Industrials & ID \\ \hline
123 & EXPD & Expeditors International & Industrials & ID \\ \hline
124 & FDX & FedEx Corporation & Industrials & ID \\ \hline
125 & FLS & Flowserve Corporation & Industrials & ID \\ \hline
126 & GD & General Dynamics & Industrials & ID \\ \hline
127 & GE & General Electric & Industrials & ID \\ \hline
128 & GLW & Corning Inc. & Industrials & ID \\ \hline
129 & GWW & Grainger (W.W.) Inc. & Industrials & ID \\ \hline
130 & HON & Honeywell Int'l Inc. & Industrials & ID \\ \hline
131 & IR & Ingersoll-Rand PLC & Industrials & ID \\ \hline
132 & ITW & Illinois Tool Works & Industrials & ID \\ \hline
133 & JEC & Jacobs Engineering Group & Industrials & ID \\ \hline
134 & LMT & Lockheed Martin Corp. & Industrials & ID \\ \hline
135 & LUV & Southwest Airlines & Industrials & ID \\ \hline
136 & MAS & Masco Corp. & Industrials & ID \\ \hline
137 & MMM & 3M Company & Industrials & ID \\ \hline
138 & ROK & Rockwell Automation Inc. & Industrials & ID \\ \hline
139 & RTN & Raytheon Co. & Industrials & ID \\ \hline
140 & TXT & Textron Inc. & Industrials & ID \\ \hline
141 & UNP & Union Pacific & Industrials & ID \\ \hline
142 & UTX & United Technologies & Industrials & ID \\ \hline
143 & AAPL & Apple Inc. & Information Technology & IT \\ \hline
144 & ADI & Analog Devices, Inc. & Information Technology & IT \\ \hline
145 & ADP & Automatic Data Processing & Information Technology & IT \\ \hline
146 & AMAT & Applied Materials Inc & Information Technology & IT \\ \hline
147 & AMD & Advanced Micro Devices Inc & Information Technology & IT \\ \hline
148 & CA & CA, Inc. & Information Technology & IT \\ \hline
149 & HPQ & HP Inc. & Information Technology & IT \\ \hline
150 & HRS & Harris Corporation & Information Technology & IT \\ \hline
\end{tabular}
\end{table}
\begin{table}[!htbp]
\centering
\begin{tabular}{|p{0.8cm}|p{1.5cm}|p{7cm}|p{4cm}|p{2cm}|}
\hline
151 & IBM & International Business Machines & Information Technology & IT \\ \hline
152 & INTC & Intel Corp. & Information Technology & IT \\ \hline
153 & KLAC & KLA-Tencor Corp. & Information Technology & IT \\ \hline
154 & LRCX & Lam Research & Information Technology & IT \\ \hline
155 & MSI & Motorola Solutions Inc. & Information Technology & IT \\ \hline
156 & MU & Micron Technology & Information Technology & IT \\ \hline
157 & TSS & Total System Services, Inc. & Information Technology & IT \\ \hline
158 & TXN & Texas Instruments & Information Technology & IT \\ \hline
159 & WDC & Western Digital & Information Technology & IT \\ \hline
160 & XRX & Xerox Corp. & Information Technology & IT \\ \hline
161 & AA & Alcoa Corporation & Materials & MT \\ \hline
162 & APD & Air Products \& Chemicals Inc & Materials & MT \\ \hline
163 & BLL & Ball Corp & Materials & MT \\ \hline
164 & BMS & Bemis Company, Inc. & Materials & MT \\ \hline
165 & CLF & Cleveland-Cliffs Inc. & Materials & MT \\ \hline
166 & DD & DuPont & Materials & MT \\ \hline
167 & ECL & Ecolab Inc. & Materials & MT \\ \hline
168 & FMC & FMC Corporation & Materials & MT \\ \hline
169 & IFF & Intl Flavors \& Fragrances & Materials & MT \\ \hline
170 & IP & International Paper & Materials & MT \\ \hline
171 & NEM & Newmont Mining Corporation & Materials & MT \\ \hline
172 & PPG & PPG Industries & Materials & MT \\ \hline
173 & VMC & Vulcan Materials & Materials & MT \\ \hline
174 & CTL & CenturyLink Inc & Telecommunication Services & TC \\ \hline
175 & FTR & Frontier Communications Corporation & Telecommunication Services & TC \\ \hline
176 & S & Sprint Nextel Corp. & Telecommunication Services & TC \\ \hline
177 & T & AT\&T Inc & Telecommunication Services & TC \\ \hline
178 & VZ & Verizon Communications & Telecommunication Services & TC \\ \hline
179 & AEP & American Electric Power & Utilities & UT \\ \hline
180 & CMS & CMS Energy & Utilities & UT \\ \hline
181 & CNP & CenterPoint Energy & Utilities & UT \\ \hline
182 & D & Dominion Energy & Utilities & UT \\ \hline
183 & DTE & DTE Energy Co. & Utilities & UT \\ \hline
184 & ED & Consolidated Edison & Utilities & UT \\ \hline
185 & EIX & Edison Int'l & Utilities & UT \\ \hline
186 & EQT & EQT Corporation & Utilities & UT \\ \hline
187 & ETR & Entergy Corp. & Utilities & UT \\ \hline
188 & EXC & Exelon Corp. & Utilities & UT \\ \hline
189 & NEE & NextEra Energy & Utilities & UT \\ \hline
190 & NI & NiSource Inc. & Utilities & UT \\ \hline
191 & PNW & Pinnacle West Capital & Utilities & UT \\ \hline
192 & SO & Southern Co. & Utilities & UT \\ \hline
193 & WEC & Wec Energy Group Inc & Utilities & UT \\ \hline
194 & XEL & Xcel Energy Inc & Utilities & UT \\ \hline
\end{tabular}
\end{table}

\begin{table}[!htbp]
\centering
\caption{List of all stocks in Japan Nikkei-225 market considered in this 
analysis. The first column gives the serial number, the second column 
gives the symbols, the third column gives the full name of the stock, 
the fourth column specifies the sector, and the fifth column has the 
abbreviation of the sector for the Nikkei-225 market.}
\label{JPN_Table}
\begin{tabular}{|p{0.8cm}|p{1.5cm}|p{7cm}|p{4cm}|p{2cm}|}
\hline
\textbf{S.No.} & \textbf{Code} & \textbf{Company Name} & \textbf{Sector} & \textbf{Abbreviation} \\ \hline
1 & S-8801 & MITSUI FUDOSAN CO., LTD. & Capital Goods & CG \\ \hline
2 & S-8802 & MITSUBISHI ESTATE CO., LTD. & Capital Goods & CG \\ \hline
3 & S-8804 & TOKYO TATEMONO CO., LTD. & Capital Goods & CG \\ \hline
4 & S-8830 & SUMITOMO REALTY \& DEVELOPMENT CO., LTD. & Capital Goods & CG \\ \hline
5 & S-7003 & MITSUI ENG. \& SHIPBUILD. CO., LTD. & Capital Goods & CG \\ \hline
6 & S-7012 & KAWASAKI HEAVY IND., LTD. & Capital Goods & CG \\ \hline
7 & S-9202 & ANA HOLDINGS INC. & Capital Goods & CG \\ \hline
8 & S-1801 & TAISEI CORP. & Capital Goods & CG \\ \hline
9 & S-1802 & OBAYASHI CORP. & Capital Goods & CG \\ \hline
10 & S-1803 & SHIMIZU CORP. & Capital Goods & CG \\ \hline
11 & S-1808 & HASEKO CORP. & Capital Goods & CG \\ \hline
12 & S-1812 & KAJIMA CORP. & Capital Goods & CG \\ \hline
13 & S-1925 & DAIWA HOUSE IND. CO., LTD. & Capital Goods & CG \\ \hline
14 & S-1928 & SEKISUI HOUSE, LTD. & Capital Goods & CG \\ \hline
15 & S-1963 & JGC CORP. & Capital Goods & CG \\ \hline
16 & S-5631 & THE JAPAN STEEL WORKS, LTD. & Capital Goods & CG \\ \hline
17 & S-6103 & OKUMA CORP. & Capital Goods & CG \\ \hline
18 & S-6113 & AMADA HOLDINGS CO., LTD. & Capital Goods & CG \\ \hline
19 & S-6301 & KOMATSU LTD. & Capital Goods & CG \\ \hline
20 & S-6302 & SUMITOMO HEAVY IND., LTD. & Capital Goods & CG \\ \hline
21 & S-6305 & HITACHI CONST. MACH. CO., LTD. & Capital Goods & CG \\ \hline
22 & S-6326 & KUBOTA CORP. & Capital Goods & CG \\ \hline
23 & S-6361 & EBARA CORP. & Capital Goods & CG \\ \hline
24 & S-6366 & CHIYODA CORP. & Capital Goods & CG \\ \hline
25 & S-6367 & DAIKIN INDUSTRIES, LTD. & Capital Goods & CG \\ \hline
26 & S-6471 & NSK LTD. & Capital Goods & CG \\ \hline
27 & S-6472 & NTN CORP. & Capital Goods & CG \\ \hline
28 & S-6473 & JTEKT CORP. & Capital Goods & CG \\ \hline
29 & S-7004 & HITACHI ZOSEN CORP. & Capital Goods & CG \\ \hline
30 & S-7011 & MITSUBISHI HEAVY IND., LTD. & Capital Goods & CG \\ \hline
31 & S-7013 & IHI CORP. & Capital Goods & CG \\ \hline
32 & S-7911 & TOPPAN PRINTING CO., LTD. & Capital Goods & CG \\ \hline
33 & S-7912 & DAI NIPPON PRINTING CO., LTD. & Capital Goods & CG \\ \hline
34 & S-7951 & YAMAHA CORP. & Capital Goods & CG \\ \hline
35 & S-1332 & NIPPON SUISAN KAISHA, LTD. & Consumer Goods & CN \\ \hline
36 & S-2002 & NISSHIN SEIFUN GROUP INC. & Consumer Goods & CN \\ \hline
37 & S-2282 & NH FOODS LTD. & Consumer Goods & CN \\ \hline
38 & S-2501 & SAPPORO HOLDINGS LTD. & Consumer Goods & CN \\ \hline
39 & S-2502 & ASAHI GROUP HOLDINGS, LTD. & Consumer Goods & CN \\ \hline
40 & S-2503 & KIRIN HOLDINGS CO., LTD. & Consumer Goods & CN \\ \hline
41 & S-2531 & TAKARA HOLDINGS INC. & Consumer Goods & CN \\ \hline
42 & S-2801 & KIKKOMAN CORP. & Consumer Goods & CN \\ \hline
43 & S-2802 & AJINOMOTO CO., INC. & Consumer Goods & CN \\ \hline
44 & S-2871 & NICHIREI CORP. & Consumer Goods & CN \\ \hline
45 & S-8233 & TAKASHIMAYA CO., LTD. & Consumer Goods & CN \\ \hline
46 & S-8252 & MARUI GROUP CO., LTD. & Consumer Goods & CN \\ \hline
47 & S-8267 & AEON CO., LTD. & Consumer Goods & CN \\ \hline
48 & S-9602 & TOHO CO., LTD & Consumer Goods & CN \\ \hline
49 & S-9681 & TOKYO DOME CORP. & Consumer Goods & CN \\ \hline
50 & S-9735 & SECOM CO., LTD. & Consumer Goods & CN \\ \hline
\end{tabular}
\end{table}
\begin{table}[!htbp]
\centering
\begin{tabular}{|p{0.8cm}|p{1.5cm}|p{7cm}|p{4cm}|p{2cm}|}
\hline
51 & S-8331 & THE CHIBA BANK, LTD. & Financials & FN \\ \hline
52 & S-8355 & THE SHIZUOKA BANK, LTD. & Financials & FN \\ \hline
53 & S-8253 & CREDIT SAISON CO., LTD. & Financials & FN \\ \hline
54 & S-8601 & DAIWA SECURITIES GROUP INC. & Financials & FN \\ \hline
55 & S-8604 & NOMURA HOLDINGS, INC. & Financials & FN \\ \hline
56 & S-3405 & KURARAY CO., LTD. & Materials & MT \\ \hline
57 & S-3407 & ASAHI KASEI CORP. & Materials & MT \\ \hline
58 & S-4004 & SHOWA DENKO K.K. & Materials & MT \\ \hline
59 & S-4005 & SUMITOMO CHEMICAL CO., LTD. & Materials & MT \\ \hline
60 & S-4021 & NISSAN CHEMICAL IND., LTD. & Materials & MT \\ \hline
61 & S-4042 & TOSOH CORP. & Materials & MT \\ \hline
62 & S-4043 & TOKUYAMA CORP. & Materials & MT \\ \hline
63 & S-4061 & DENKA CO., LTD. & Materials & MT \\ \hline
64 & S-4063 & SHIN-ETSU CHEMICAL CO., LTD. & Materials & MT \\ \hline
65 & S-4183 & MITSUI CHEMICALS, INC. & Materials & MT \\ \hline
66 & S-4208 & UBE INDUSTRIES, LTD. & Materials & MT \\ \hline
67 & S-4272 & NIPPON KAYAKU CO., LTD. & Materials & MT \\ \hline
68 & S-4452 & KAO CORP. & Materials & MT \\ \hline
69 & S-4901 & FUJIFILM HOLDINGS CORP. & Materials & MT \\ \hline
70 & S-4911 & SHISEIDO CO., LTD. & Materials & MT \\ \hline
71 & S-6988 & NITTO DENKO CORP. & Materials & MT \\ \hline
72 & S-5002 & SHOWA SHELL SEKIYU K.K. & Materials & MT \\ \hline
73 & S-5201 & ASAHI GLASS CO., LTD. & Materials & MT \\ \hline
74 & S-5202 & NIPPON SHEET GLASS CO., LTD. & Materials & MT \\ \hline
75 & S-5214 & NIPPON ELECTRIC GLASS CO., LTD. & Materials & MT \\ \hline
76 & S-5232 & SUMITOMO OSAKA CEMENT CO., LTD. & Materials & MT \\ \hline
77 & S-5233 & TAIHEIYO CEMENT CORP. & Materials & MT \\ \hline
78 & S-5301 & TOKAI CARBON CO., LTD. & Materials & MT \\ \hline
79 & S-5332 & TOTO LTD. & Materials & MT \\ \hline
80 & S-5333 & NGK INSULATORS, LTD. & Materials & MT \\ \hline
81 & S-5706 & MITSUI MINING \& SMELTING CO. & Materials & MT \\ \hline
82 & S-5707 & TOHO ZINC CO., LTD. & Materials & MT \\ \hline
83 & S-5711 & MITSUBISHI MATERIALS CORP. & Materials & MT \\ \hline
84 & S-5713 & SUMITOMO METAL MINING CO., LTD. & Materials & MT \\ \hline
85 & S-5714 & DOWA HOLDINGS CO., LTD. & Materials & MT \\ \hline
86 & S-5715 & FURUKAWA CO., LTD. & Materials & MT \\ \hline
87 & S-5801 & FURUKAWA ELECTRIC CO., LTD. & Materials & MT \\ \hline
88 & S-5802 & SUMITOMO ELECTRIC IND., LTD. & Materials & MT \\ \hline
89 & S-5803 & FUJIKURA LTD. & Materials & MT \\ \hline
90 & S-5901 & TOYO SEIKAN GROUP HOLDINGS, LTD. & Materials & MT \\ \hline
91 & S-3865 & HOKUETSU KISHU PAPER CO., LTD. & Materials & MT \\ \hline
92 & S-3861 & OJI HOLDINGS CORP. & Materials & MT \\ \hline
93 & S-5101 & THE YOKOHAMA RUBBER CO., LTD. & Materials & MT \\ \hline
94 & S-5108 & BRIDGESTONE CORP. & Materials & MT \\ \hline
95 & S-5401 & NIPPON STEEL \& SUMITOMO METAL CORP. & Materials & MT \\ \hline
96 & S-5406 & KOBE STEEL, LTD. & Materials & MT \\ \hline
97 & S-5541 & PACIFIC METALS CO., LTD. & Materials & MT \\ \hline
98 & S-3101 & TOYOBO CO., LTD. & Materials & MT \\ \hline
99 & S-3103 & UNITIKA, LTD. & Materials & MT \\ \hline
100 & S-3401 & TEIJIN LTD. & Materials & MT \\ \hline
\end{tabular}
\end{table}
\begin{table}[!htbp]
\centering
\begin{tabular}{|p{0.8cm}|p{1.5cm}|p{7cm}|p{4cm}|p{2cm}|}
\hline
101 & S-3402 & TORAY INDUSTRIES, INC. & Materials & MT \\ \hline
102 & S-8001 & ITOCHU CORP. & Materials & MT \\ \hline
103 & S-8002 & MARUBENI CORP. & Materials & MT \\ \hline
104 & S-8015 & TOYOTA TSUSHO CORP. & Materials & MT \\ \hline
105 & S-8031 & MITSUI \& CO., LTD. & Materials & MT \\ \hline
106 & S-8053 & SUMITOMO CORP. & Materials & MT \\ \hline
107 & S-8058 & MITSUBISHI CORP. & Materials & MT \\ \hline
108 & S-4151 & KYOWA HAKKO KIRIN CO., LTD. & Pharmaceuticals & PH \\ \hline
109 & S-4503 & ASTELLAS PHARMA INC. & Pharmaceuticals & PH \\ \hline
110 & S-4506 & SUMITOMO DAINIPPON PHARMA CO., LTD. & Pharmaceuticals & PH \\ \hline
111 & S-4507 & SHIONOGI \& CO., LTD. & Pharmaceuticals & PH \\ \hline
112 & S-4519 & CHUGAI PHARMACEUTICAL CO., LTD. & Pharmaceuticals & PH \\ \hline
113 & S-4523 & EISAI CO., LTD. & Pharmaceuticals & PH \\ \hline
114 & S-7201 & NISSAN MOTOR CO., LTD. & Information Technology & IT \\ \hline
115 & S-7202 & ISUZU MOTORS LTD. & Information Technology & IT \\ \hline
116 & S-7205 & HINO MOTORS, LTD. & Information Technology & IT \\ \hline
117 & S-7261 & MAZDA MOTOR CORP. & Information Technology & IT \\ \hline
118 & S-7267 & HONDA MOTOR CO., LTD. & Information Technology & IT \\ \hline
119 & S-7270 & SUBARU CORP. & Information Technology & IT \\ \hline
120 & S-7272 & YAMAHA MOTOR CO., LTD. & Information Technology & IT \\ \hline
121 & S-3105 & NISSHINBO HOLDINGS INC. & Information Technology & IT \\ \hline
122 & S-6479 & MINEBEA MITSUMI INC. & Information Technology & IT \\ \hline
123 & S-6501 & HITACHI, LTD. & Information Technology & IT \\ \hline
124 & S-6502 & TOSHIBA CORP. & Information Technology & IT \\ \hline
125 & S-6503 & MITSUBISHI ELECTRIC CORP. & Information Technology & IT \\ \hline
126 & S-6504 & FUJI ELECTRIC CO., LTD. & Information Technology & IT \\ \hline
127 & S-6506 & YASKAWA ELECTRIC CORP. & Information Technology & IT \\ \hline
128 & S-6508 & MEIDENSHA CORP. & Information Technology & IT \\ \hline
129 & S-6701 & NEC CORP. & Information Technology & IT \\ \hline
130 & S-6702 & FUJITSU LTD. & Information Technology & IT \\ \hline
131 & S-6703 & OKI ELECTRIC IND. CO., LTD. & Information Technology & IT \\ \hline
132 & S-6752 & PANASONIC CORP. & Information Technology & IT \\ \hline
133 & S-6758 & SONY CORP. & Information Technology & IT \\ \hline
134 & S-6762 & TDK CORP. & Information Technology & IT \\ \hline
135 & S-6770 & ALPS ELECTRIC CO., LTD. & Information Technology & IT \\ \hline
136 & S-6773 & PIONEER CORP. & Information Technology & IT \\ \hline
137 & S-6841 & YOKOGAWA ELECTRIC CORP. & Information Technology & IT \\ \hline
138 & S-6902 & DENSO CORP. & Information Technology & IT \\ \hline
139 & S-6952 & CASIO COMPUTER CO., LTD. & Information Technology & IT \\ \hline
140 & S-6954 & FANUC CORP. & Information Technology & IT \\ \hline
141 & S-6971 & KYOCERA CORP. & Information Technology & IT \\ \hline
142 & S-6976 & TAIYO YUDEN CO., LTD. & Information Technology & IT \\ \hline
143 & S-7752 & RICOH CO., LTD. & Information Technology & IT \\ \hline
144 & S-8035 & TOKYO ELECTRON LTD. & Information Technology & IT \\ \hline
145 & S-4543 & TERUMO CORP. & Information Technology & IT \\ \hline
146 & S-4902 & KONICA MINOLTA, INC. & Information Technology & IT \\ \hline
147 & S-7731 & NIKON CORP. & Information Technology & IT \\ \hline
148 & S-7733 & OLYMPUS CORP. & Information Technology & IT \\ \hline
149 & S-7762 & CITIZEN WATCH CO., LTD. & Information Technology & IT \\ \hline
150 & S-9501 & TOKYO ELECTRIC POWER COMPANY HOLDINGS, I & Transportation \& Utilities & TU \\ \hline
\end{tabular}
\end{table}
\begin{table}[!htbp]
\centering
\begin{tabular}{|p{0.8cm}|p{1.5cm}|p{7cm}|p{4cm}|p{2cm}|}
\hline
151 & S-9502 & CHUBU ELECTRIC POWER CO., INC. & Transportation \& Utilities & TU \\ \hline
152 & S-9503 & THE KANSAI ELECTRIC POWER CO., INC. & Transportation \& Utilities & TU \\ \hline
153 & S-9531 & TOKYO GAS CO., LTD. & Transportation \& Utilities & TU \\ \hline
154 & S-9532 & OSAKA GAS CO., LTD. & Transportation \& Utilities & TU \\ \hline
155 & S-9062 & NIPPON EXPRESS CO., LTD. & Transportation \& Utilities & TU \\ \hline
156 & S-9064 & YAMATO HOLDINGS CO., LTD. & Transportation \& Utilities & TU \\ \hline
157 & S-9101 & NIPPON YUSEN K.K. & Transportation \& Utilities & TU \\ \hline
158 & S-9104 & MITSUI O.S.K.LINES, LTD. & Transportation \& Utilities & TU \\ \hline
159 & S-9107 & KAWASAKI KISEN KAISHA, LTD. & Transportation \& Utilities & TU \\ \hline
160 & S-9001 & TOBU RAILWAY CO., LTD. & Transportation \& Utilities & TU \\ \hline
161 & S-9005 & TOKYU CORP. & Transportation \& Utilities & TU \\ \hline
162 & S-9007 & ODAKYU ELECTRIC RAILWAY CO., LTD. & Transportation \& Utilities & TU \\ \hline
163 & S-9008 & KEIO CORP. & Transportation \& Utilities & TU \\ \hline
164 & S-9009 & KEISEI ELECTRIC RAILWAY CO., LTD. & Transportation \& Utilities & TU \\ \hline
165 & S-9301 & MITSUBISHI LOGISTICS CORP. & Transportation \& Utilities & TU \\ \hline
\end{tabular}
\end{table}

\begin{table}[!htbp]
\caption{Number of edges ($\#$ Edges) and communities ($\#$ Communities) in 
threshold networks $S_\tau(t)$ for USA S\&P-500 market at eight distinct 
epochs of $\tau=22$ days ending on trading days 23-01-2006, 10-05-2006, 
29-06-2006, 06-11-2006, 07-01-2011, 04-05-2011, 02-09-2011, and 03-02-2012
(around the US Housing bubble period (2006-2007) and August 2011 stock markets 
fall crisis), constructed using four different thresholds, $C_{ij}(t)\geq 0.55$, 
$C_{ij}(t)\geq 0.65$, $C_{ij}(t)\geq 0.75$, and $C_{ij}(t)\geq 0.85$.}
\label{table:threshold}
\centering
\begin{tabular}{|l|llll|llll|}
\hline
& \multicolumn{4}{c|}{\textbf{US Housing Bubble Networks}} & \multicolumn{4}{c|}{\textbf{August 2011 Fall Crash Networks}} \\\hline\hline
\textbf{End Date}  & \multicolumn{4}{c|}{\textbf{23-01-2006}} & \multicolumn{4}{c|}{\textbf{07-01-2011}}    \\
Threshold      & 0.55  & 0.65     & 0.75  & 0.85 & 0.55  & 0.65     & 0.75  & 0.85 \\
\# Edges       & 1949   & 733      & 251   & 193  & 645   & 284      & 197   & 193 \\
\# Communities & 5     & 13       & 13    & 14 & 9     & 11       & 14    & 14  \\\hline\hline
\textbf{End date}  & \multicolumn{4}{c|}{\textbf{10-05-2006}} & \multicolumn{4}{c|}{\textbf{04-05-2011}}      \\
Threshold      & 0.55  & 0.65     & 0.75  & 0.85 & 0.55  & 0.65     & 0.75  & 0.85\\
\# Edges       & 838  & 351      & 220   & 193 & 1378  & 542    & 245   & 194 \\
\# Communities & 7     & 12       & 16    & 15  & 6     & 11   & 16    & 16\\\hline\hline
\textbf{End date} & \multicolumn{4}{c|}{\textbf{29-06-2006}}  & \multicolumn{4}{c|}{\textbf{02-09-2011}} \\
Threshold      & 0.55  & 0.65     & 0.75  & 0.85 & 0.55  & 0.65     & 0.75  & 0.85\\
\# Edges       & 6291 & 3138    & 996 & 250 & 18258 & 17697    & 16004 & 9906\\
\# Communities & 3     & 5        & 11     & 15  & 3  & 3  & 4  & 3\\\hline\hline
\textbf{End date} & \multicolumn{4}{c|}{\textbf{06-11-2006}} & \multicolumn{4}{c|}{\textbf{03-02-2012}} \\
Threshold      & 0.55  & 0.65     & 0.75  & 0.85 & 0.55  & 0.65  & 0.75  & 0.85 \\
\# Edges       & 677   & 287      & 220   & 193  & 931   & 328   & 198   & 193 \\
\# Communities & 9     & 12       & 14    & 14 & 8 & 12  & 15    & 16  \\\hline
\end{tabular}
\end{table}


\begin{thebibliography}{87}%
\makeatletter
\providecommand \@ifxundefined [1]{%
 \@ifx{#1\undefined}
}%
\providecommand \@ifnum [1]{%
 \ifnum #1\expandafter \@firstoftwo
 \else \expandafter \@secondoftwo
 \fi
}%
\providecommand \@ifx [1]{%
 \ifx #1\expandafter \@firstoftwo
 \else \expandafter \@secondoftwo
 \fi
}%
\providecommand \natexlab [1]{#1}%
\providecommand \enquote  [1]{``#1''}%
\providecommand \bibnamefont  [1]{#1}%
\providecommand \bibfnamefont [1]{#1}%
\providecommand \citenamefont [1]{#1}%
\providecommand \href@noop [0]{\@secondoftwo}%
\providecommand \href [0]{\begingroup \@sanitize@url \@href}%
\providecommand \@href[1]{\@@startlink{#1}\@@href}%
\providecommand \@@href[1]{\endgroup#1\@@endlink}%
\providecommand \@sanitize@url [0]{\catcode `\\12\catcode `\$12\catcode
  `\&12\catcode `\#12\catcode `\^12\catcode `\_12\catcode `\%12\relax}%
\providecommand \@@startlink[1]{}%
\providecommand \@@endlink[0]{}%
\providecommand \url  [0]{\begingroup\@sanitize@url \@url }%
\providecommand \@url [1]{\endgroup\@href {#1}{\urlprefix }}%
\providecommand \urlprefix  [0]{URL }%
\providecommand \Eprint [0]{\href }%
\providecommand \doibase [0]{http://dx.doi.org/}%
\providecommand \selectlanguage [0]{\@gobble}%
\providecommand \bibinfo  [0]{\@secondoftwo}%
\providecommand \bibfield  [0]{\@secondoftwo}%
\providecommand \translation [1]{[#1]}%
\providecommand \BibitemOpen [0]{}%
\providecommand \bibitemStop [0]{}%
\providecommand \bibitemNoStop [0]{.\EOS\space}%
\providecommand \EOS [0]{\spacefactor3000\relax}%
\providecommand \BibitemShut  [1]{\csname bibitem#1\endcsname}%
\let\auto@bib@innerbib\@empty
\bibitem [{\citenamefont {Anderson}(1972)}]{Anderson1972}%
  \BibitemOpen
  \bibfield  {author} {\bibinfo {author} {\bibfnamefont {P.~W.}\ \bibnamefont
  {Anderson}},\ }\href@noop {} {\bibfield  {journal} {\bibinfo  {journal}
  {Science}\ }\textbf {\bibinfo {volume} {177}},\ \bibinfo {pages} {393}
  (\bibinfo {year} {1972})}\BibitemShut {NoStop}%
\bibitem [{\citenamefont {Vemuri}(1978)}]{Vemuri1978}%
  \BibitemOpen
  \bibfield  {author} {\bibinfo {author} {\bibfnamefont {V.}~\bibnamefont
  {Vemuri}},\ }\href@noop {} {\emph {\bibinfo {title} {Modeling of Complex
  Systems: An Introduction}}}\ (\bibinfo  {publisher} {Academic Press, New
  York},\ \bibinfo {year} {1978})\BibitemShut {NoStop}%
\bibitem [{\citenamefont {Gell-Mann}(1995)}]{Gellmann1995}%
  \BibitemOpen
  \bibfield  {author} {\bibinfo {author} {\bibfnamefont {M.}~\bibnamefont
  {Gell-Mann}},\ }\href@noop {} {\bibfield  {journal} {\bibinfo  {journal}
  {Complexity}\ }\textbf {\bibinfo {volume} {1}},\ \bibinfo {pages} {16}
  (\bibinfo {year} {1995})}\BibitemShut {NoStop}%
\bibitem [{\citenamefont {Mantegna}\ and\ \citenamefont
  {Stanley}(2007)}]{Mantegna2007}%
  \BibitemOpen
  \bibfield  {author} {\bibinfo {author} {\bibfnamefont {R.~N.}\ \bibnamefont
  {Mantegna}}\ and\ \bibinfo {author} {\bibfnamefont {H.~E.}\ \bibnamefont
  {Stanley}},\ }\href@noop {} {\emph {\bibinfo {title} {An introduction to
  econophysics: correlations and complexity in finance}}}\ (\bibinfo
  {publisher} {Cambridge University Press, Cambridge},\ \bibinfo {year}
  {2007})\BibitemShut {NoStop}%
\bibitem [{\citenamefont {Bouchaud}\ and\ \citenamefont
  {Potters}(2003)}]{Bouchaud2003}%
  \BibitemOpen
  \bibfield  {author} {\bibinfo {author} {\bibfnamefont {J.-P.}\ \bibnamefont
  {Bouchaud}}\ and\ \bibinfo {author} {\bibfnamefont {M.}~\bibnamefont
  {Potters}},\ }\href@noop {} {\emph {\bibinfo {title} {{Theory of Financial
  Risk and Derivative Pricing: from Statistical Physics to Risk Management}}}}\
  (\bibinfo  {publisher} {Cambridge University Press},\ \bibinfo {year}
  {2003})\BibitemShut {NoStop}%
\bibitem [{\citenamefont {Sinha}\ \emph {et~al.}(2010)\citenamefont {Sinha},
  \citenamefont {Chatterjee}, \citenamefont {Chakraborti},\ and\ \citenamefont
  {Chakrabarti}}]{Sinha2010}%
  \BibitemOpen
  \bibfield  {author} {\bibinfo {author} {\bibfnamefont {S.}~\bibnamefont
  {Sinha}}, \bibinfo {author} {\bibfnamefont {A.}~\bibnamefont {Chatterjee}},
  \bibinfo {author} {\bibfnamefont {A.}~\bibnamefont {Chakraborti}}, \ and\
  \bibinfo {author} {\bibfnamefont {B.~K.}\ \bibnamefont {Chakrabarti}},\
  }\href@noop {} {\emph {\bibinfo {title} {Econophysics: an introduction}}}\
  (\bibinfo  {publisher} {John Wiley \& Sons},\ \bibinfo {year}
  {2010})\BibitemShut {NoStop}%
\bibitem [{\citenamefont {Chakraborti}\ \emph {et~al.}(2011)\citenamefont
  {Chakraborti}, \citenamefont {Muni~Toke}, \citenamefont {Patriarca},\ and\
  \citenamefont {Abergel}}]{Chakraborti2011a}%
  \BibitemOpen
  \bibfield  {author} {\bibinfo {author} {\bibfnamefont {A.}~\bibnamefont
  {Chakraborti}}, \bibinfo {author} {\bibfnamefont {I.}~\bibnamefont
  {Muni~Toke}}, \bibinfo {author} {\bibfnamefont {M.}~\bibnamefont
  {Patriarca}}, \ and\ \bibinfo {author} {\bibfnamefont {F.}~\bibnamefont
  {Abergel}},\ }\href@noop {} {\bibfield  {journal} {\bibinfo  {journal}
  {Quantitative {F}inance}\ }\textbf {\bibinfo {volume} {11}},\ \bibinfo
  {pages} {991} (\bibinfo {year} {2011})}\BibitemShut {NoStop}%
\bibitem [{\citenamefont {Chakraborti}\ \emph {et~al.}(2015)\citenamefont
  {Chakraborti}, \citenamefont {Challet}, \citenamefont {Chatterjee},
  \citenamefont {Marsili}, \citenamefont {Zhang},\ and\ \citenamefont
  {Chakrabarti}}]{Chakraborti2015}%
  \BibitemOpen
  \bibfield  {author} {\bibinfo {author} {\bibfnamefont {A.}~\bibnamefont
  {Chakraborti}}, \bibinfo {author} {\bibfnamefont {D.}~\bibnamefont
  {Challet}}, \bibinfo {author} {\bibfnamefont {A.}~\bibnamefont {Chatterjee}},
  \bibinfo {author} {\bibfnamefont {M.}~\bibnamefont {Marsili}}, \bibinfo
  {author} {\bibfnamefont {Y.-C.}\ \bibnamefont {Zhang}}, \ and\ \bibinfo
  {author} {\bibfnamefont {B.~K.}\ \bibnamefont {Chakrabarti}},\ }\href@noop {}
  {\bibfield  {journal} {\bibinfo  {journal} {Physics {R}eports}\ }\textbf
  {\bibinfo {volume} {552}},\ \bibinfo {pages} {1} (\bibinfo {year}
  {2015})}\BibitemShut {NoStop}%
\bibitem [{\citenamefont {Battiston}\ \emph {et~al.}(2016)\citenamefont
  {Battiston}, \citenamefont {Farmer}, \citenamefont {Flache}, \citenamefont
  {Garlaschelli}, \citenamefont {Haldane}, \citenamefont {Heesterbeek},
  \citenamefont {Hommes}, \citenamefont {Jaeger}, \citenamefont {May},\ and\
  \citenamefont {Scheffer}}]{battiston2016complexity}%
  \BibitemOpen
  \bibfield  {author} {\bibinfo {author} {\bibfnamefont {S.}~\bibnamefont
  {Battiston}}, \bibinfo {author} {\bibfnamefont {J.~D.}\ \bibnamefont
  {Farmer}}, \bibinfo {author} {\bibfnamefont {A.}~\bibnamefont {Flache}},
  \bibinfo {author} {\bibfnamefont {D.}~\bibnamefont {Garlaschelli}}, \bibinfo
  {author} {\bibfnamefont {A.~G.}\ \bibnamefont {Haldane}}, \bibinfo {author}
  {\bibfnamefont {H.}~\bibnamefont {Heesterbeek}}, \bibinfo {author}
  {\bibfnamefont {C.}~\bibnamefont {Hommes}}, \bibinfo {author} {\bibfnamefont
  {C.}~\bibnamefont {Jaeger}}, \bibinfo {author} {\bibfnamefont
  {R.}~\bibnamefont {May}}, \ and\ \bibinfo {author} {\bibfnamefont
  {M.}~\bibnamefont {Scheffer}},\ }\href@noop {} {\bibfield  {journal}
  {\bibinfo  {journal} {Science}\ }\textbf {\bibinfo {volume} {351}},\ \bibinfo
  {pages} {818} (\bibinfo {year} {2016})}\BibitemShut {NoStop}%
\bibitem [{\citenamefont {Lux}\ and\ \citenamefont
  {Westerhoff}(2009)}]{lux2009economics}%
  \BibitemOpen
  \bibfield  {author} {\bibinfo {author} {\bibfnamefont {T.}~\bibnamefont
  {Lux}}\ and\ \bibinfo {author} {\bibfnamefont {F.}~\bibnamefont
  {Westerhoff}},\ }\href@noop {} {\bibfield  {journal} {\bibinfo  {journal}
  {Nature Physics}\ }\textbf {\bibinfo {volume} {5}},\ \bibinfo {pages} {2}
  (\bibinfo {year} {2009})}\BibitemShut {NoStop}%
\bibitem [{\citenamefont
  {Mantegna}(1999{\natexlab{a}})}]{mantegna1999information}%
  \BibitemOpen
  \bibfield  {author} {\bibinfo {author} {\bibfnamefont {R.~N.}\ \bibnamefont
  {Mantegna}},\ }\href {\doibase 10.1016/S0010-4655(99)00302-1} {\bibfield
  {journal} {\bibinfo  {journal} {Computer Physics Communications}\ }\textbf
  {\bibinfo {volume} {121-122}},\ \bibinfo {pages} {153} (\bibinfo {year}
  {1999}{\natexlab{a}})}\BibitemShut {NoStop}%
\bibitem [{\citenamefont
  {Mantegna}(1999{\natexlab{b}})}]{Mantegna1999hierarchical}%
  \BibitemOpen
  \bibfield  {author} {\bibinfo {author} {\bibfnamefont {R.~N.}\ \bibnamefont
  {Mantegna}},\ }\href@noop {} {\bibfield  {journal} {\bibinfo  {journal} {The
  {E}uropean {P}hysical {J}ournal B}\ }\textbf {\bibinfo {volume} {11}},\
  \bibinfo {pages} {193} (\bibinfo {year} {1999}{\natexlab{b}})}\BibitemShut
  {NoStop}%
\bibitem [{\citenamefont {Laloux}\ \emph {et~al.}(1999)\citenamefont {Laloux},
  \citenamefont {Cizeau}, \citenamefont {Bouchaud},\ and\ \citenamefont
  {Potters}}]{laloux_1999}%
  \BibitemOpen
  \bibfield  {author} {\bibinfo {author} {\bibfnamefont {L.}~\bibnamefont
  {Laloux}}, \bibinfo {author} {\bibfnamefont {P.}~\bibnamefont {Cizeau}},
  \bibinfo {author} {\bibfnamefont {J.-P.}\ \bibnamefont {Bouchaud}}, \ and\
  \bibinfo {author} {\bibfnamefont {M.}~\bibnamefont {Potters}},\ }\href@noop
  {} {\bibfield  {journal} {\bibinfo  {journal} {Physical Review Letters}\
  }\textbf {\bibinfo {volume} {83}},\ \bibinfo {pages} {1467} (\bibinfo {year}
  {1999})}\BibitemShut {NoStop}%
\bibitem [{\citenamefont {Plerou}\ \emph {et~al.}(1999)\citenamefont {Plerou},
  \citenamefont {Gopikrishnan}, \citenamefont {Rosenow}, \citenamefont
  {Nunes~Amaral},\ and\ \citenamefont {Stanley}}]{plerou_1999}%
  \BibitemOpen
  \bibfield  {author} {\bibinfo {author} {\bibfnamefont {V.}~\bibnamefont
  {Plerou}}, \bibinfo {author} {\bibfnamefont {P.}~\bibnamefont
  {Gopikrishnan}}, \bibinfo {author} {\bibfnamefont {B.}~\bibnamefont
  {Rosenow}}, \bibinfo {author} {\bibfnamefont {L.~A.}\ \bibnamefont
  {Nunes~Amaral}}, \ and\ \bibinfo {author} {\bibfnamefont {H.~E.}\
  \bibnamefont {Stanley}},\ }\href@noop {} {\bibfield  {journal} {\bibinfo
  {journal} {Physical Review Letters}\ }\textbf {\bibinfo {volume} {83}},\
  \bibinfo {pages} {1471} (\bibinfo {year} {1999})}\BibitemShut {NoStop}%
\bibitem [{\citenamefont {Gopikrishnan}\ \emph {et~al.}(2001)\citenamefont
  {Gopikrishnan}, \citenamefont {Rosenow}, \citenamefont {Plerou},\ and\
  \citenamefont {Stanley}}]{Gopikrishnan_pre2001}%
  \BibitemOpen
  \bibfield  {author} {\bibinfo {author} {\bibfnamefont {P.}~\bibnamefont
  {Gopikrishnan}}, \bibinfo {author} {\bibfnamefont {B.}~\bibnamefont
  {Rosenow}}, \bibinfo {author} {\bibfnamefont {V.}~\bibnamefont {Plerou}}, \
  and\ \bibinfo {author} {\bibfnamefont {H.~E.}\ \bibnamefont {Stanley}},\
  }\href {\doibase 10.1103/PhysRevE.64.035106} {\bibfield  {journal} {\bibinfo
  {journal} {Physical Review E}\ }\textbf {\bibinfo {volume} {64}},\ \bibinfo
  {pages} {035106} (\bibinfo {year} {2001})}\BibitemShut {NoStop}%
\bibitem [{\citenamefont {Kullmann}\ \emph {et~al.}(2002)\citenamefont
  {Kullmann}, \citenamefont {Kert{\'e}sz},\ and\ \citenamefont
  {Kaski}}]{kullmann2002time}%
  \BibitemOpen
  \bibfield  {author} {\bibinfo {author} {\bibfnamefont {L.}~\bibnamefont
  {Kullmann}}, \bibinfo {author} {\bibfnamefont {J.}~\bibnamefont
  {Kert{\'e}sz}}, \ and\ \bibinfo {author} {\bibfnamefont {K.}~\bibnamefont
  {Kaski}},\ }\href@noop {} {\bibfield  {journal} {\bibinfo  {journal}
  {Physical Review E}\ }\textbf {\bibinfo {volume} {66}},\ \bibinfo {pages}
  {026125} (\bibinfo {year} {2002})}\BibitemShut {NoStop}%
\bibitem [{\citenamefont {Plerou}\ \emph {et~al.}(2002)\citenamefont {Plerou},
  \citenamefont {Gopikrishnan}, \citenamefont {Rosenow}, \citenamefont
  {Amaral}, \citenamefont {Guhr},\ and\ \citenamefont {Stanley}}]{Plerou2002}%
  \BibitemOpen
  \bibfield  {author} {\bibinfo {author} {\bibfnamefont {V.}~\bibnamefont
  {Plerou}}, \bibinfo {author} {\bibfnamefont {P.}~\bibnamefont
  {Gopikrishnan}}, \bibinfo {author} {\bibfnamefont {B.}~\bibnamefont
  {Rosenow}}, \bibinfo {author} {\bibfnamefont {L.~A.~N.}\ \bibnamefont
  {Amaral}}, \bibinfo {author} {\bibfnamefont {T.}~\bibnamefont {Guhr}}, \ and\
  \bibinfo {author} {\bibfnamefont {H.~E.}\ \bibnamefont {Stanley}},\
  }\href@noop {} {\bibfield  {journal} {\bibinfo  {journal} {Physical Review
  E}\ }\textbf {\bibinfo {volume} {65}},\ \bibinfo {pages} {066126} (\bibinfo
  {year} {2002})}\BibitemShut {NoStop}%
\bibitem [{\citenamefont {Onnela}\ \emph {et~al.}(2003)\citenamefont {Onnela},
  \citenamefont {Chakraborti}, \citenamefont {Kaski}, \citenamefont {Kertesz},\
  and\ \citenamefont {Kanto}}]{Onnela2003}%
  \BibitemOpen
  \bibfield  {author} {\bibinfo {author} {\bibfnamefont {J.-P.}\ \bibnamefont
  {Onnela}}, \bibinfo {author} {\bibfnamefont {A.}~\bibnamefont {Chakraborti}},
  \bibinfo {author} {\bibfnamefont {K.}~\bibnamefont {Kaski}}, \bibinfo
  {author} {\bibfnamefont {J.}~\bibnamefont {Kertesz}}, \ and\ \bibinfo
  {author} {\bibfnamefont {A.}~\bibnamefont {Kanto}},\ }\href@noop {}
  {\bibfield  {journal} {\bibinfo  {journal} {Physical Review E}\ }\textbf
  {\bibinfo {volume} {68}},\ \bibinfo {pages} {056110} (\bibinfo {year}
  {2003})}\BibitemShut {NoStop}%
\bibitem [{\citenamefont {Tumminello}\ \emph {et~al.}(2005)\citenamefont
  {Tumminello}, \citenamefont {Aste}, \citenamefont {Di~Matteo},\ and\
  \citenamefont {Mantegna}}]{Tumminello2005}%
  \BibitemOpen
  \bibfield  {author} {\bibinfo {author} {\bibfnamefont {M.}~\bibnamefont
  {Tumminello}}, \bibinfo {author} {\bibfnamefont {T.}~\bibnamefont {Aste}},
  \bibinfo {author} {\bibfnamefont {T.}~\bibnamefont {Di~Matteo}}, \ and\
  \bibinfo {author} {\bibfnamefont {R.~N.}\ \bibnamefont {Mantegna}},\
  }\href@noop {} {\bibfield  {journal} {\bibinfo  {journal} {Proceedings of the
  {N}ational {A}cademy of {S}ciences}\ }\textbf {\bibinfo {volume} {102}},\
  \bibinfo {pages} {10421} (\bibinfo {year} {2005})}\BibitemShut {NoStop}%
\bibitem [{\citenamefont {Pharasi}\ \emph {et~al.}(2018)\citenamefont
  {Pharasi}, \citenamefont {Sharma}, \citenamefont {Chatterjee}, \citenamefont
  {Chakraborti}, \citenamefont {Leyvraz},\ and\ \citenamefont
  {Seligman}}]{Pharasi2018}%
  \BibitemOpen
  \bibfield  {author} {\bibinfo {author} {\bibfnamefont {H.~K.}\ \bibnamefont
  {Pharasi}}, \bibinfo {author} {\bibfnamefont {K.}~\bibnamefont {Sharma}},
  \bibinfo {author} {\bibfnamefont {R.}~\bibnamefont {Chatterjee}}, \bibinfo
  {author} {\bibfnamefont {A.}~\bibnamefont {Chakraborti}}, \bibinfo {author}
  {\bibfnamefont {F.}~\bibnamefont {Leyvraz}}, \ and\ \bibinfo {author}
  {\bibfnamefont {T.~H.}\ \bibnamefont {Seligman}},\ }\href@noop {} {\bibfield
  {journal} {\bibinfo  {journal} {New Journal of Physics}\ }\textbf {\bibinfo
  {volume} {20}},\ \bibinfo {pages} {103041} (\bibinfo {year}
  {2018})}\BibitemShut {NoStop}%
\bibitem [{\citenamefont {Pharasi}\ \emph {et~al.}(2019)\citenamefont
  {Pharasi}, \citenamefont {Sharma}, \citenamefont {Chakraborti},\ and\
  \citenamefont {Seligman}}]{Pharasi2019}%
  \BibitemOpen
  \bibfield  {author} {\bibinfo {author} {\bibfnamefont {H.~K.}\ \bibnamefont
  {Pharasi}}, \bibinfo {author} {\bibfnamefont {K.}~\bibnamefont {Sharma}},
  \bibinfo {author} {\bibfnamefont {A.}~\bibnamefont {Chakraborti}}, \ and\
  \bibinfo {author} {\bibfnamefont {T.~H.}\ \bibnamefont {Seligman}},\
  }\enquote {\bibinfo {title} {Complex market dynamics in the light of random
  matrix theory},}\ in\ \href@noop {} {\emph {\bibinfo {booktitle} {New
  Perspectives and Challenges in Econophysics and Sociophysics}}},\ \bibinfo
  {editor} {edited by\ \bibinfo {editor} {\bibfnamefont {F.}~\bibnamefont
  {Abergel}}, \bibinfo {editor} {\bibfnamefont {B.~K.}\ \bibnamefont
  {Chakrabarti}}, \bibinfo {editor} {\bibfnamefont {A.}~\bibnamefont
  {Chakraborti}}, \bibinfo {editor} {\bibfnamefont {N.}~\bibnamefont {Deo}}, \
  and\ \bibinfo {editor} {\bibfnamefont {K.}~\bibnamefont {Sharma}}}\ (\bibinfo
   {publisher} {Springer International Publishing},\ \bibinfo {address}
  {Cham},\ \bibinfo {year} {2019})\ pp.\ \bibinfo {pages} {13--34}\BibitemShut
  {NoStop}%
\bibitem [{\citenamefont {Chakraborti}\ \emph
  {et~al.}(2020{\natexlab{a}})\citenamefont {Chakraborti}, \citenamefont
  {Sharma}, \citenamefont {Pharasi}, \citenamefont {Shuvo~Bakar}, \citenamefont
  {Das},\ and\ \citenamefont {Seligman}}]{Chakraborti2020}%
  \BibitemOpen
  \bibfield  {author} {\bibinfo {author} {\bibfnamefont {A.}~\bibnamefont
  {Chakraborti}}, \bibinfo {author} {\bibfnamefont {K.}~\bibnamefont {Sharma}},
  \bibinfo {author} {\bibfnamefont {H.~K.}\ \bibnamefont {Pharasi}}, \bibinfo
  {author} {\bibfnamefont {K.}~\bibnamefont {Shuvo~Bakar}}, \bibinfo {author}
  {\bibfnamefont {S.}~\bibnamefont {Das}}, \ and\ \bibinfo {author}
  {\bibfnamefont {T.~H.}\ \bibnamefont {Seligman}},\ }\href {\doibase
  10.1088/1367-2630/ab90d4} {\bibfield  {journal} {\bibinfo  {journal} {New
  Journal of Physics}\ }\textbf {\bibinfo {volume} {22}},\ \bibinfo {pages}
  {063043} (\bibinfo {year} {2020}{\natexlab{a}})}\BibitemShut {NoStop}%
\bibitem [{\citenamefont {Chi}\ \emph {et~al.}(2010)\citenamefont {Chi},
  \citenamefont {Liu},\ and\ \citenamefont {Lau}}]{chi2010network}%
  \BibitemOpen
  \bibfield  {author} {\bibinfo {author} {\bibfnamefont {K.~T.}\ \bibnamefont
  {Chi}}, \bibinfo {author} {\bibfnamefont {J.}~\bibnamefont {Liu}}, \ and\
  \bibinfo {author} {\bibfnamefont {F.~C.}\ \bibnamefont {Lau}},\ }\href@noop
  {} {\bibfield  {journal} {\bibinfo  {journal} {Journal of Empirical Finance}\
  }\textbf {\bibinfo {volume} {17}},\ \bibinfo {pages} {659} (\bibinfo {year}
  {2010})}\BibitemShut {NoStop}%
\bibitem [{\citenamefont {Tumminello}\ \emph {et~al.}(2007)\citenamefont
  {Tumminello}, \citenamefont {Di~Matteo}, \citenamefont {Aste},\ and\
  \citenamefont {Mantegna}}]{tumminello2007correlation}%
  \BibitemOpen
  \bibfield  {author} {\bibinfo {author} {\bibfnamefont {M.}~\bibnamefont
  {Tumminello}}, \bibinfo {author} {\bibfnamefont {T.}~\bibnamefont
  {Di~Matteo}}, \bibinfo {author} {\bibfnamefont {T.}~\bibnamefont {Aste}}, \
  and\ \bibinfo {author} {\bibfnamefont {R.~N.}\ \bibnamefont {Mantegna}},\
  }\href@noop {} {\bibfield  {journal} {\bibinfo  {journal} {The European
  Physical Journal B}\ }\textbf {\bibinfo {volume} {55}},\ \bibinfo {pages}
  {209} (\bibinfo {year} {2007})}\BibitemShut {NoStop}%
\bibitem [{\citenamefont {Tumminello}\ \emph {et~al.}(2010)\citenamefont
  {Tumminello}, \citenamefont {Lillo},\ and\ \citenamefont
  {Mantegna}}]{tumminello2010correlation}%
  \BibitemOpen
  \bibfield  {author} {\bibinfo {author} {\bibfnamefont {M.}~\bibnamefont
  {Tumminello}}, \bibinfo {author} {\bibfnamefont {F.}~\bibnamefont {Lillo}}, \
  and\ \bibinfo {author} {\bibfnamefont {R.~N.}\ \bibnamefont {Mantegna}},\
  }\href@noop {} {\bibfield  {journal} {\bibinfo  {journal} {Journal of
  economic behavior \& organization}\ }\textbf {\bibinfo {volume} {75}},\
  \bibinfo {pages} {40} (\bibinfo {year} {2010})}\BibitemShut {NoStop}%
\bibitem [{\citenamefont {Miccich{\`e}}\ \emph {et~al.}(2003)\citenamefont
  {Miccich{\`e}}, \citenamefont {Bonanno}, \citenamefont {Lillo},\ and\
  \citenamefont {Mantegna}}]{micciche2003degree}%
  \BibitemOpen
  \bibfield  {author} {\bibinfo {author} {\bibfnamefont {S.}~\bibnamefont
  {Miccich{\`e}}}, \bibinfo {author} {\bibfnamefont {G.}~\bibnamefont
  {Bonanno}}, \bibinfo {author} {\bibfnamefont {F.}~\bibnamefont {Lillo}}, \
  and\ \bibinfo {author} {\bibfnamefont {R.~N.}\ \bibnamefont {Mantegna}},\
  }\href@noop {} {\bibfield  {journal} {\bibinfo  {journal} {Physica A:
  Statistical Mechanics and its Applications}\ }\textbf {\bibinfo {volume}
  {324}},\ \bibinfo {pages} {66} (\bibinfo {year} {2003})}\BibitemShut
  {NoStop}%
\bibitem [{\citenamefont {Kumar}\ and\ \citenamefont {Deo}(2012)}]{Kumar2012}%
  \BibitemOpen
  \bibfield  {author} {\bibinfo {author} {\bibfnamefont {S.}~\bibnamefont
  {Kumar}}\ and\ \bibinfo {author} {\bibfnamefont {N.}~\bibnamefont {Deo}},\
  }\href {\doibase 10.1103/PhysRevE.86.026101} {\bibfield  {journal} {\bibinfo
  {journal} {Physical Review E}\ }\textbf {\bibinfo {volume} {86}},\ \bibinfo
  {pages} {026101} (\bibinfo {year} {2012})}\BibitemShut {NoStop}%
\bibitem [{\citenamefont {Almog}\ and\ \citenamefont
  {Shmueli}(2019)}]{almog2019}%
  \BibitemOpen
  \bibfield  {author} {\bibinfo {author} {\bibfnamefont {A.}~\bibnamefont
  {Almog}}\ and\ \bibinfo {author} {\bibfnamefont {E.}~\bibnamefont
  {Shmueli}},\ }\href@noop {} {\bibfield  {journal} {\bibinfo  {journal}
  {Scientific Reports}\ }\textbf {\bibinfo {volume} {9}},\ \bibinfo {pages}
  {10832} (\bibinfo {year} {2019})}\BibitemShut {NoStop}%
\bibitem [{\citenamefont {Wang}\ \emph {et~al.}(2019)\citenamefont {Wang},
  \citenamefont {Lin},\ and\ \citenamefont {Wang}}]{wang2019thermodynamic}%
  \BibitemOpen
  \bibfield  {author} {\bibinfo {author} {\bibfnamefont {J.}~\bibnamefont
  {Wang}}, \bibinfo {author} {\bibfnamefont {C.}~\bibnamefont {Lin}}, \ and\
  \bibinfo {author} {\bibfnamefont {Y.}~\bibnamefont {Wang}},\ }\href@noop {}
  {\bibfield  {journal} {\bibinfo  {journal} {Complexity}\ }\textbf {\bibinfo
  {volume} {2019}},\ \bibinfo {pages} {1817248} (\bibinfo {year}
  {2019})}\BibitemShut {NoStop}%
\bibitem [{\citenamefont {Chakraborti}\ \emph
  {et~al.}(2020{\natexlab{b}})\citenamefont {Chakraborti}, \citenamefont
  {Hrishidev}, \citenamefont {Sharma},\ and\ \citenamefont
  {Pharasi}}]{chakraborti2020phase}%
  \BibitemOpen
  \bibfield  {author} {\bibinfo {author} {\bibfnamefont {A.}~\bibnamefont
  {Chakraborti}}, \bibinfo {author} {\bibnamefont {Hrishidev}}, \bibinfo
  {author} {\bibfnamefont {K.}~\bibnamefont {Sharma}}, \ and\ \bibinfo {author}
  {\bibfnamefont {H.~K.}\ \bibnamefont {Pharasi}},\ }\href@noop {} {\bibfield
  {journal} {\bibinfo  {journal} {Journal of Physics: Complexity}\ }\textbf
  {\bibinfo {volume} {2}},\ \bibinfo {pages} {015002} (\bibinfo {year}
  {2020}{\natexlab{b}})}\BibitemShut {NoStop}%
\bibitem [{\citenamefont {Kukreti}\ \emph {et~al.}(2020)\citenamefont
  {Kukreti}, \citenamefont {Pharasi}, \citenamefont {Gupta},\ and\
  \citenamefont {Kumar}}]{kukreti2020}%
  \BibitemOpen
  \bibfield  {author} {\bibinfo {author} {\bibfnamefont {V.}~\bibnamefont
  {Kukreti}}, \bibinfo {author} {\bibfnamefont {H.~K.}\ \bibnamefont
  {Pharasi}}, \bibinfo {author} {\bibfnamefont {P.}~\bibnamefont {Gupta}}, \
  and\ \bibinfo {author} {\bibfnamefont {S.}~\bibnamefont {Kumar}},\
  }\href@noop {} {\bibfield  {journal} {\bibinfo  {journal} {Frontiers in
  Physics}\ }\textbf {\bibinfo {volume} {8}},\ \bibinfo {pages} {323} (\bibinfo
  {year} {2020})}\BibitemShut {NoStop}%
\bibitem [{\citenamefont {Nie}\ and\ \citenamefont
  {Song}(2018)}]{nie2018relationship}%
  \BibitemOpen
  \bibfield  {author} {\bibinfo {author} {\bibfnamefont {C.-x.}\ \bibnamefont
  {Nie}}\ and\ \bibinfo {author} {\bibfnamefont {F.-t.}\ \bibnamefont {Song}},\
  }\href@noop {} {\bibfield  {journal} {\bibinfo  {journal} {Entropy}\ }\textbf
  {\bibinfo {volume} {20}},\ \bibinfo {pages} {177} (\bibinfo {year}
  {2018})}\BibitemShut {NoStop}%
\bibitem [{\citenamefont {Stepanov}\ \emph {et~al.}(2015)\citenamefont
  {Stepanov}, \citenamefont {Rinn}, \citenamefont {Guhr}, \citenamefont
  {Peinke},\ and\ \citenamefont {Sch{\"a}fer}}]{stepanov2015stability}%
  \BibitemOpen
  \bibfield  {author} {\bibinfo {author} {\bibfnamefont {Y.}~\bibnamefont
  {Stepanov}}, \bibinfo {author} {\bibfnamefont {P.}~\bibnamefont {Rinn}},
  \bibinfo {author} {\bibfnamefont {T.}~\bibnamefont {Guhr}}, \bibinfo {author}
  {\bibfnamefont {J.}~\bibnamefont {Peinke}}, \ and\ \bibinfo {author}
  {\bibfnamefont {R.}~\bibnamefont {Sch{\"a}fer}},\ }\href@noop {} {\bibfield
  {journal} {\bibinfo  {journal} {Journal of Statistical Mechanics: Theory and
  Experiment}\ }\textbf {\bibinfo {volume} {2015}},\ \bibinfo {pages} {P08011}
  (\bibinfo {year} {2015})}\BibitemShut {NoStop}%
\bibitem [{\citenamefont {Rinn}\ \emph {et~al.}(2015)\citenamefont {Rinn},
  \citenamefont {Stepanov}, \citenamefont {Peinke}, \citenamefont {Guhr},\ and\
  \citenamefont {Sch{\"a}fer}}]{rinn2015dynamics}%
  \BibitemOpen
  \bibfield  {author} {\bibinfo {author} {\bibfnamefont {P.}~\bibnamefont
  {Rinn}}, \bibinfo {author} {\bibfnamefont {Y.}~\bibnamefont {Stepanov}},
  \bibinfo {author} {\bibfnamefont {J.}~\bibnamefont {Peinke}}, \bibinfo
  {author} {\bibfnamefont {T.}~\bibnamefont {Guhr}}, \ and\ \bibinfo {author}
  {\bibfnamefont {R.}~\bibnamefont {Sch{\"a}fer}},\ }\href@noop {} {\bibfield
  {journal} {\bibinfo  {journal} {EPL}\ }\textbf {\bibinfo {volume} {110}},\
  \bibinfo {pages} {68003} (\bibinfo {year} {2015})}\BibitemShut {NoStop}%
\bibitem [{\citenamefont {Chetalova}\ \emph {et~al.}(2015)\citenamefont
  {Chetalova}, \citenamefont {Sch{\"a}fer},\ and\ \citenamefont
  {Guhr}}]{chetalova2015zooming}%
  \BibitemOpen
  \bibfield  {author} {\bibinfo {author} {\bibfnamefont {D.}~\bibnamefont
  {Chetalova}}, \bibinfo {author} {\bibfnamefont {R.}~\bibnamefont
  {Sch{\"a}fer}}, \ and\ \bibinfo {author} {\bibfnamefont {T.}~\bibnamefont
  {Guhr}},\ }\href@noop {} {\bibfield  {journal} {\bibinfo  {journal} {Journal
  of Statistical Mechanics: Theory and Experiment}\ }\textbf {\bibinfo {volume}
  {2015}},\ \bibinfo {pages} {P01029} (\bibinfo {year} {2015})}\BibitemShut
  {NoStop}%
\bibitem [{\citenamefont {Heckens}\ \emph {et~al.}(2020)\citenamefont
  {Heckens}, \citenamefont {Krause},\ and\ \citenamefont
  {Guhr}}]{Heckens_2020}%
  \BibitemOpen
  \bibfield  {author} {\bibinfo {author} {\bibfnamefont {A.~J.}\ \bibnamefont
  {Heckens}}, \bibinfo {author} {\bibfnamefont {S.~M.}\ \bibnamefont {Krause}},
  \ and\ \bibinfo {author} {\bibfnamefont {T.}~\bibnamefont {Guhr}},\
  }\href@noop {} {\bibfield  {journal} {\bibinfo  {journal} {Journal of
  Statistical Mechanics: Theory and Experiment}\ }\textbf {\bibinfo {volume}
  {2020}},\ \bibinfo {pages} {P103402} (\bibinfo {year} {2020})}\BibitemShut
  {NoStop}%
\bibitem [{\citenamefont {Wang}\ \emph {et~al.}(2020)\citenamefont {Wang},
  \citenamefont {Gartzke}, \citenamefont {Schreckenberg},\ and\ \citenamefont
  {Guhr}}]{wang2020quasi}%
  \BibitemOpen
  \bibfield  {author} {\bibinfo {author} {\bibfnamefont {S.}~\bibnamefont
  {Wang}}, \bibinfo {author} {\bibfnamefont {S.}~\bibnamefont {Gartzke}},
  \bibinfo {author} {\bibfnamefont {M.}~\bibnamefont {Schreckenberg}}, \ and\
  \bibinfo {author} {\bibfnamefont {T.}~\bibnamefont {Guhr}},\ }\href@noop {}
  {\bibfield  {journal} {\bibinfo  {journal} {Journal of Statistical Mechanics:
  Theory and Experiment}\ }\textbf {\bibinfo {volume} {2020}},\ \bibinfo
  {pages} {P103404} (\bibinfo {year} {2020})}\BibitemShut {NoStop}%
\bibitem [{\citenamefont {Jost}(2017)}]{Jost2017}%
  \BibitemOpen
  \bibfield  {author} {\bibinfo {author} {\bibfnamefont {J.}~\bibnamefont
  {Jost}},\ }\href@noop {} {\emph {\bibinfo {title} {Riemannian {G}eometry and
  {G}eometric {A}nalysis}}},\ \bibinfo {edition} {7th}\ ed.\ (\bibinfo
  {publisher} {Springer {I}nternational {P}ublishing},\ \bibinfo {year}
  {2017})\BibitemShut {NoStop}%
\bibitem [{\citenamefont {Carlsson}(2009)}]{Carlsson2009}%
  \BibitemOpen
  \bibfield  {author} {\bibinfo {author} {\bibfnamefont {G.}~\bibnamefont
  {Carlsson}},\ }\href@noop {} {\bibfield  {journal} {\bibinfo  {journal}
  {Bulletin of the {A}merican {M}athematical {S}ociety}\ }\textbf {\bibinfo
  {volume} {46}},\ \bibinfo {pages} {255} (\bibinfo {year} {2009})}\BibitemShut
  {NoStop}%
\bibitem [{\citenamefont {Krioukov}\ \emph {et~al.}(2010)\citenamefont
  {Krioukov}, \citenamefont {Papadopoulos}, \citenamefont {Kitsak},
  \citenamefont {Vahdat},\ and\ \citenamefont {Bogun{\'a}}}]{Krioukov2010}%
  \BibitemOpen
  \bibfield  {author} {\bibinfo {author} {\bibfnamefont {D.}~\bibnamefont
  {Krioukov}}, \bibinfo {author} {\bibfnamefont {F.}~\bibnamefont
  {Papadopoulos}}, \bibinfo {author} {\bibfnamefont {M.}~\bibnamefont
  {Kitsak}}, \bibinfo {author} {\bibfnamefont {A.}~\bibnamefont {Vahdat}}, \
  and\ \bibinfo {author} {\bibfnamefont {M.}~\bibnamefont {Bogun{\'a}}},\
  }\href@noop {} {\bibfield  {journal} {\bibinfo  {journal} {Physical {R}eview
  {E}}\ }\textbf {\bibinfo {volume} {82}},\ \bibinfo {pages} {036106} (\bibinfo
  {year} {2010})}\BibitemShut {NoStop}%
\bibitem [{\citenamefont {Bianconi}(2015)}]{Bianconi2015}%
  \BibitemOpen
  \bibfield  {author} {\bibinfo {author} {\bibfnamefont {G.}~\bibnamefont
  {Bianconi}},\ }\href@noop {} {\bibfield  {journal} {\bibinfo  {journal}
  {EPL}\ }\textbf {\bibinfo {volume} {111}},\ \bibinfo {pages} {56001}
  (\bibinfo {year} {2015})}\BibitemShut {NoStop}%
\bibitem [{\citenamefont {Sandhu}\ \emph {et~al.}(2015)\citenamefont {Sandhu},
  \citenamefont {Georgiou}, \citenamefont {Reznik}, \citenamefont {Zhu},
  \citenamefont {Kolesov}, \citenamefont {Senbabaoglu},\ and\ \citenamefont
  {Tannenbaum}}]{Sandhu2015}%
  \BibitemOpen
  \bibfield  {author} {\bibinfo {author} {\bibfnamefont {R.}~\bibnamefont
  {Sandhu}}, \bibinfo {author} {\bibfnamefont {T.}~\bibnamefont {Georgiou}},
  \bibinfo {author} {\bibfnamefont {E.}~\bibnamefont {Reznik}}, \bibinfo
  {author} {\bibfnamefont {L.}~\bibnamefont {Zhu}}, \bibinfo {author}
  {\bibfnamefont {I.}~\bibnamefont {Kolesov}}, \bibinfo {author} {\bibfnamefont
  {Y.}~\bibnamefont {Senbabaoglu}}, \ and\ \bibinfo {author} {\bibfnamefont
  {A.}~\bibnamefont {Tannenbaum}},\ }\href@noop {} {\bibfield  {journal}
  {\bibinfo  {journal} {Scientific {R}eports}\ }\textbf {\bibinfo {volume}
  {5}},\ \bibinfo {pages} {12323} (\bibinfo {year} {2015})}\BibitemShut
  {NoStop}%
\bibitem [{\citenamefont {Sreejith}\ \emph {et~al.}(2016)\citenamefont
  {Sreejith}, \citenamefont {Mohanraj}, \citenamefont {Jost}, \citenamefont
  {Saucan},\ and\ \citenamefont {Samal}}]{Sreejith2016}%
  \BibitemOpen
  \bibfield  {author} {\bibinfo {author} {\bibfnamefont {R.~P.}\ \bibnamefont
  {Sreejith}}, \bibinfo {author} {\bibfnamefont {K.}~\bibnamefont {Mohanraj}},
  \bibinfo {author} {\bibfnamefont {J.}~\bibnamefont {Jost}}, \bibinfo {author}
  {\bibfnamefont {E.}~\bibnamefont {Saucan}}, \ and\ \bibinfo {author}
  {\bibfnamefont {A.}~\bibnamefont {Samal}},\ }\href@noop {} {\bibfield
  {journal} {\bibinfo  {journal} {Journal of Statistical Mechanics: Theory and
  Experiment}\ }\textbf {\bibinfo {volume} {2016}},\ \bibinfo {pages} {P063206}
  (\bibinfo {year} {2016})}\BibitemShut {NoStop}%
\bibitem [{\citenamefont {Bianconi}\ and\ \citenamefont
  {Rahmede}(2017)}]{Bianconi2017}%
  \BibitemOpen
  \bibfield  {author} {\bibinfo {author} {\bibfnamefont {G.}~\bibnamefont
  {Bianconi}}\ and\ \bibinfo {author} {\bibfnamefont {C.}~\bibnamefont
  {Rahmede}},\ }\href@noop {} {\bibfield  {journal} {\bibinfo  {journal}
  {{S}cientific {R}eports}\ }\textbf {\bibinfo {volume} {7}},\ \bibinfo {pages}
  {41974} (\bibinfo {year} {2017})}\BibitemShut {NoStop}%
\bibitem [{\citenamefont {Kartun-Giles}\ and\ \citenamefont
  {Bianconi}(2019)}]{Kartun-Giles2019}%
  \BibitemOpen
  \bibfield  {author} {\bibinfo {author} {\bibfnamefont {A.~P.}\ \bibnamefont
  {Kartun-Giles}}\ and\ \bibinfo {author} {\bibfnamefont {G.}~\bibnamefont
  {Bianconi}},\ }\href@noop {} {\bibfield  {journal} {\bibinfo  {journal}
  {Chaos, Solitons \& Fractals: X}\ }\textbf {\bibinfo {volume} {1}},\ \bibinfo
  {pages} {100004} (\bibinfo {year} {2019})}\BibitemShut {NoStop}%
\bibitem [{\citenamefont {Iacopini}\ \emph {et~al.}(2019)\citenamefont
  {Iacopini}, \citenamefont {Petri}, \citenamefont {Barrat},\ and\
  \citenamefont {Latora}}]{Iacopini2019}%
  \BibitemOpen
  \bibfield  {author} {\bibinfo {author} {\bibfnamefont {I.}~\bibnamefont
  {Iacopini}}, \bibinfo {author} {\bibfnamefont {G.}~\bibnamefont {Petri}},
  \bibinfo {author} {\bibfnamefont {A.}~\bibnamefont {Barrat}}, \ and\ \bibinfo
  {author} {\bibfnamefont {V.}~\bibnamefont {Latora}},\ }\href@noop {}
  {\bibfield  {journal} {\bibinfo  {journal} {Nature communications}\ }\textbf
  {\bibinfo {volume} {10}},\ \bibinfo {pages} {2485} (\bibinfo {year}
  {2019})}\BibitemShut {NoStop}%
\bibitem [{\citenamefont {Kannan}\ \emph {et~al.}(2019)\citenamefont {Kannan},
  \citenamefont {Saucan}, \citenamefont {Roy},\ and\ \citenamefont
  {Samal}}]{Kannan2019}%
  \BibitemOpen
  \bibfield  {author} {\bibinfo {author} {\bibfnamefont {H.}~\bibnamefont
  {Kannan}}, \bibinfo {author} {\bibfnamefont {E.}~\bibnamefont {Saucan}},
  \bibinfo {author} {\bibfnamefont {I.}~\bibnamefont {Roy}}, \ and\ \bibinfo
  {author} {\bibfnamefont {A.}~\bibnamefont {Samal}},\ }\href@noop {}
  {\bibfield  {journal} {\bibinfo  {journal} {Scientific reports}\ }\textbf
  {\bibinfo {volume} {9}},\ \bibinfo {pages} {13817} (\bibinfo {year}
  {2019})}\BibitemShut {NoStop}%
\bibitem [{\citenamefont {Samal}\ \emph {et~al.}(2018)\citenamefont {Samal},
  \citenamefont {Sreejith}, \citenamefont {Gu}, \citenamefont {Liu},
  \citenamefont {Saucan},\ and\ \citenamefont {Jost}}]{Samal2018}%
  \BibitemOpen
  \bibfield  {author} {\bibinfo {author} {\bibfnamefont {A.}~\bibnamefont
  {Samal}}, \bibinfo {author} {\bibfnamefont {R.~P.}\ \bibnamefont {Sreejith}},
  \bibinfo {author} {\bibfnamefont {J.}~\bibnamefont {Gu}}, \bibinfo {author}
  {\bibfnamefont {S.}~\bibnamefont {Liu}}, \bibinfo {author} {\bibfnamefont
  {E.}~\bibnamefont {Saucan}}, \ and\ \bibinfo {author} {\bibfnamefont
  {J.}~\bibnamefont {Jost}},\ }\href@noop {} {\bibfield  {journal} {\bibinfo
  {journal} {Scientific {R}eports}\ }\textbf {\bibinfo {volume} {8}},\ \bibinfo
  {pages} {8650} (\bibinfo {year} {2018})}\BibitemShut {NoStop}%
\bibitem [{\citenamefont {Ni}\ \emph {et~al.}(2015)\citenamefont {Ni},
  \citenamefont {Lin}, \citenamefont {Gao}, \citenamefont {Gu},\ and\
  \citenamefont {Saucan}}]{Ni2015}%
  \BibitemOpen
  \bibfield  {author} {\bibinfo {author} {\bibfnamefont {C.}~\bibnamefont
  {Ni}}, \bibinfo {author} {\bibfnamefont {Y.}~\bibnamefont {Lin}}, \bibinfo
  {author} {\bibfnamefont {J.}~\bibnamefont {Gao}}, \bibinfo {author}
  {\bibfnamefont {X.~D.}\ \bibnamefont {Gu}}, \ and\ \bibinfo {author}
  {\bibfnamefont {E.}~\bibnamefont {Saucan}},\ }in\ \href@noop {} {\emph
  {\bibinfo {booktitle} {2015 {IEEE} Conference on Computer Communications
  ({INFOCOM})}}}\ (\bibinfo {organization} {{IEEE}},\ \bibinfo {year} {2015})\
  pp.\ \bibinfo {pages} {2758--2766}\BibitemShut {NoStop}%
\bibitem [{\citenamefont {Sandhu}\ \emph {et~al.}(2016)\citenamefont {Sandhu},
  \citenamefont {Georgiou},\ and\ \citenamefont {Tannenbaum}}]{Sandhu2016}%
  \BibitemOpen
  \bibfield  {author} {\bibinfo {author} {\bibfnamefont {R.~S.}\ \bibnamefont
  {Sandhu}}, \bibinfo {author} {\bibfnamefont {T.~T.}\ \bibnamefont
  {Georgiou}}, \ and\ \bibinfo {author} {\bibfnamefont {A.~R.}\ \bibnamefont
  {Tannenbaum}},\ }\href@noop {} {\bibfield  {journal} {\bibinfo  {journal}
  {Science Advances}\ }\textbf {\bibinfo {volume} {2}},\ \bibinfo {pages}
  {e1501495} (\bibinfo {year} {2016})}\BibitemShut {NoStop}%
\bibitem [{\citenamefont {Ni}\ \emph {et~al.}(2019)\citenamefont {Ni},
  \citenamefont {Lin}, \citenamefont {Luo},\ and\ \citenamefont
  {Gao}}]{Ni2019}%
  \BibitemOpen
  \bibfield  {author} {\bibinfo {author} {\bibfnamefont {C.}~\bibnamefont
  {Ni}}, \bibinfo {author} {\bibfnamefont {Y.}~\bibnamefont {Lin}}, \bibinfo
  {author} {\bibfnamefont {F.}~\bibnamefont {Luo}}, \ and\ \bibinfo {author}
  {\bibfnamefont {J.}~\bibnamefont {Gao}},\ }\href@noop {} {\bibfield
  {journal} {\bibinfo  {journal} {Scientific {R}eports}\ }\textbf {\bibinfo
  {volume} {9}},\ \bibinfo {pages} {9984} (\bibinfo {year} {2019})}\BibitemShut
  {NoStop}%
\bibitem [{\citenamefont {Ollivier}(2007)}]{Ollivier2007}%
  \BibitemOpen
  \bibfield  {author} {\bibinfo {author} {\bibfnamefont {Y.}~\bibnamefont
  {Ollivier}},\ }\href@noop {} {\bibfield  {journal} {\bibinfo  {journal}
  {Comptes Rendus Mathematique}\ }\textbf {\bibinfo {volume} {345}},\ \bibinfo
  {pages} {643} (\bibinfo {year} {2007})}\BibitemShut {NoStop}%
\bibitem [{\citenamefont {Ollivier}(2009)}]{Ollivier2009}%
  \BibitemOpen
  \bibfield  {author} {\bibinfo {author} {\bibfnamefont {Y.}~\bibnamefont
  {Ollivier}},\ }\href@noop {} {\bibfield  {journal} {\bibinfo  {journal}
  {Journal of {F}unctional {A}nalysis}\ }\textbf {\bibinfo {volume} {256}},\
  \bibinfo {pages} {810} (\bibinfo {year} {2009})}\BibitemShut {NoStop}%
\bibitem [{\citenamefont {Aste}\ \emph {et~al.}(2005)\citenamefont {Aste},
  \citenamefont {Di~Matteo},\ and\ \citenamefont {Hyde}}]{aste2005complex}%
  \BibitemOpen
  \bibfield  {author} {\bibinfo {author} {\bibfnamefont {T.}~\bibnamefont
  {Aste}}, \bibinfo {author} {\bibfnamefont {T.}~\bibnamefont {Di~Matteo}}, \
  and\ \bibinfo {author} {\bibfnamefont {S.}~\bibnamefont {Hyde}},\ }\href@noop
  {} {\bibfield  {journal} {\bibinfo  {journal} {Physica A: Statistical
  Mechanics and its Applications}\ }\textbf {\bibinfo {volume} {346}},\
  \bibinfo {pages} {20} (\bibinfo {year} {2005})}\BibitemShut {NoStop}%
\bibitem [{\citenamefont {Saucan}\ \emph {et~al.}(2020)\citenamefont {Saucan},
  \citenamefont {Samal},\ and\ \citenamefont {Jost}}]{Saucan2020}%
  \BibitemOpen
  \bibfield  {author} {\bibinfo {author} {\bibfnamefont {E.}~\bibnamefont
  {Saucan}}, \bibinfo {author} {\bibfnamefont {A.}~\bibnamefont {Samal}}, \
  and\ \bibinfo {author} {\bibfnamefont {J.}~\bibnamefont {Jost}},\ }\href
  {\doibase doi:10.1017/nws.2020.42} {\bibfield  {journal} {\bibinfo  {journal}
  {Network Science}\ ,\ \bibinfo {pages} {1}} (\bibinfo {year}
  {2020})}\BibitemShut {NoStop}%
\bibitem [{\citenamefont {Saucan}(2012)}]{saucan2012isometric}%
  \BibitemOpen
  \bibfield  {author} {\bibinfo {author} {\bibfnamefont {E.}~\bibnamefont
  {Saucan}},\ }\href@noop {} {\bibfield  {journal} {\bibinfo  {journal}
  {Journal of Mathematical Imaging and Vision}\ }\textbf {\bibinfo {volume}
  {43}},\ \bibinfo {pages} {143} (\bibinfo {year} {2012})}\BibitemShut
  {NoStop}%
\bibitem [{\citenamefont {Scheffer}\ \emph {et~al.}(2012)\citenamefont
  {Scheffer}, \citenamefont {Carpenter}, \citenamefont {Lenton}, \citenamefont
  {Bascompte}, \citenamefont {Brock}, \citenamefont {Dakos}, \citenamefont
  {van~de Koppel}, \citenamefont {van~de Leemput}, \citenamefont {Levin},
  \citenamefont {van Nes}, \citenamefont {Pascual},\ and\ \citenamefont
  {Vandermeer}}]{Scheffer2012}%
  \BibitemOpen
  \bibfield  {author} {\bibinfo {author} {\bibfnamefont {M.}~\bibnamefont
  {Scheffer}}, \bibinfo {author} {\bibfnamefont {S.~R.}\ \bibnamefont
  {Carpenter}}, \bibinfo {author} {\bibfnamefont {T.~M.}\ \bibnamefont
  {Lenton}}, \bibinfo {author} {\bibfnamefont {J.}~\bibnamefont {Bascompte}},
  \bibinfo {author} {\bibfnamefont {W.}~\bibnamefont {Brock}}, \bibinfo
  {author} {\bibfnamefont {V.}~\bibnamefont {Dakos}}, \bibinfo {author}
  {\bibfnamefont {J.}~\bibnamefont {van~de Koppel}}, \bibinfo {author}
  {\bibfnamefont {I.~A.}\ \bibnamefont {van~de Leemput}}, \bibinfo {author}
  {\bibfnamefont {S.~A.}\ \bibnamefont {Levin}}, \bibinfo {author}
  {\bibfnamefont {E.~H.}\ \bibnamefont {van Nes}}, \bibinfo {author}
  {\bibfnamefont {M.}~\bibnamefont {Pascual}}, \ and\ \bibinfo {author}
  {\bibfnamefont {J.}~\bibnamefont {Vandermeer}},\ }\href@noop {} {\bibfield
  {journal} {\bibinfo  {journal} {Science}\ }\textbf {\bibinfo {volume}
  {338}},\ \bibinfo {pages} {344} (\bibinfo {year} {2012})}\BibitemShut
  {NoStop}%
\bibitem [{\citenamefont {Lin}\ and\ \citenamefont {Yau}(2010)}]{Lin2010}%
  \BibitemOpen
  \bibfield  {author} {\bibinfo {author} {\bibfnamefont {Y.}~\bibnamefont
  {Lin}}\ and\ \bibinfo {author} {\bibfnamefont {S.-T.}\ \bibnamefont {Yau}},\
  }\href@noop {} {\bibfield  {journal} {\bibinfo  {journal} {Math. Res. Lett.}\
  }\textbf {\bibinfo {volume} {17}},\ \bibinfo {pages} {343} (\bibinfo {year}
  {2010})}\BibitemShut {NoStop}%
\bibitem [{\citenamefont {Lin}\ \emph {et~al.}(2011)\citenamefont {Lin},
  \citenamefont {Lu},\ and\ \citenamefont {Yau}}]{Lin2011}%
  \BibitemOpen
  \bibfield  {author} {\bibinfo {author} {\bibfnamefont {Y.}~\bibnamefont
  {Lin}}, \bibinfo {author} {\bibfnamefont {L.}~\bibnamefont {Lu}}, \ and\
  \bibinfo {author} {\bibfnamefont {S.-T.}\ \bibnamefont {Yau}},\ }\href@noop
  {} {\bibfield  {journal} {\bibinfo  {journal} {Tohoku Mathematical Journal}\
  }\textbf {\bibinfo {volume} {63}},\ \bibinfo {pages} {605} (\bibinfo {year}
  {2011})}\BibitemShut {NoStop}%
\bibitem [{\citenamefont {Bauer}\ \emph {et~al.}(2012)\citenamefont {Bauer},
  \citenamefont {Jost},\ and\ \citenamefont {Liu}}]{Bauer2012}%
  \BibitemOpen
  \bibfield  {author} {\bibinfo {author} {\bibfnamefont {F.}~\bibnamefont
  {Bauer}}, \bibinfo {author} {\bibfnamefont {J.}~\bibnamefont {Jost}}, \ and\
  \bibinfo {author} {\bibfnamefont {S.}~\bibnamefont {Liu}},\ }\href@noop {}
  {\bibfield  {journal} {\bibinfo  {journal} {Math. Res. Lett.}\ }\textbf
  {\bibinfo {volume} {19}},\ \bibinfo {pages} {1185} (\bibinfo {year}
  {2012})}\BibitemShut {NoStop}%
\bibitem [{\citenamefont {Jost}\ and\ \citenamefont {Liu}(2014)}]{Jost2014}%
  \BibitemOpen
  \bibfield  {author} {\bibinfo {author} {\bibfnamefont {J.}~\bibnamefont
  {Jost}}\ and\ \bibinfo {author} {\bibfnamefont {S.}~\bibnamefont {Liu}},\
  }\href@noop {} {\bibfield  {journal} {\bibinfo  {journal} {Discrete \&
  Computational Geometry}\ }\textbf {\bibinfo {volume} {51}},\ \bibinfo {pages}
  {300} (\bibinfo {year} {2014})}\BibitemShut {NoStop}%
\bibitem [{\citenamefont {Sia}\ \emph {et~al.}(2019)\citenamefont {Sia},
  \citenamefont {Jonckheere},\ and\ \citenamefont {Bogdan}}]{Sia2019}%
  \BibitemOpen
  \bibfield  {author} {\bibinfo {author} {\bibfnamefont {J.}~\bibnamefont
  {Sia}}, \bibinfo {author} {\bibfnamefont {E.}~\bibnamefont {Jonckheere}}, \
  and\ \bibinfo {author} {\bibfnamefont {P.}~\bibnamefont {Bogdan}},\
  }\href@noop {} {\bibfield  {journal} {\bibinfo  {journal} {Scientific
  {R}eports}\ }\textbf {\bibinfo {volume} {9}},\ \bibinfo {pages} {9800}
  (\bibinfo {year} {2019})}\BibitemShut {NoStop}%
\bibitem [{\citenamefont {Vaserstein}(1969)}]{Vaserstein1969}%
  \BibitemOpen
  \bibfield  {author} {\bibinfo {author} {\bibfnamefont {L.~N.}\ \bibnamefont
  {Vaserstein}},\ }\href@noop {} {\bibfield  {journal} {\bibinfo  {journal}
  {Probl. Peredachi Inf.}\ }\textbf {\bibinfo {volume} {5}},\ \bibinfo {pages}
  {64} (\bibinfo {year} {1969})}\BibitemShut {NoStop}%
\bibitem [{\citenamefont {Forman}(2003)}]{Forman2003}%
  \BibitemOpen
  \bibfield  {author} {\bibinfo {author} {\bibfnamefont {R.}~\bibnamefont
  {Forman}},\ }\href@noop {} {\bibfield  {journal} {\bibinfo  {journal}
  {Discrete \& {C}omputational {G}eometry}\ }\textbf {\bibinfo {volume} {29}},\
  \bibinfo {pages} {323} (\bibinfo {year} {2003})}\BibitemShut {NoStop}%
\bibitem [{\citenamefont {Saucan}\ \emph
  {et~al.}(2019{\natexlab{a}})\citenamefont {Saucan}, \citenamefont {Sreejith},
  \citenamefont {Vivek-Ananth}, \citenamefont {Jost},\ and\ \citenamefont
  {Samal}}]{Saucan2019a}%
  \BibitemOpen
  \bibfield  {author} {\bibinfo {author} {\bibfnamefont {E.}~\bibnamefont
  {Saucan}}, \bibinfo {author} {\bibfnamefont {R.~P.}\ \bibnamefont
  {Sreejith}}, \bibinfo {author} {\bibfnamefont {R.~P.}\ \bibnamefont
  {Vivek-Ananth}}, \bibinfo {author} {\bibfnamefont {J.}~\bibnamefont {Jost}},
  \ and\ \bibinfo {author} {\bibfnamefont {A.}~\bibnamefont {Samal}},\
  }\href@noop {} {\bibfield  {journal} {\bibinfo  {journal} {Chaos, {S}olitons
  \& {F}ractals}\ }\textbf {\bibinfo {volume} {118}},\ \bibinfo {pages} {347}
  (\bibinfo {year} {2019}{\natexlab{a}})}\BibitemShut {NoStop}%
\bibitem [{\citenamefont {Sreejith}\ \emph {et~al.}(2017)\citenamefont
  {Sreejith}, \citenamefont {Jost}, \citenamefont {Saucan},\ and\ \citenamefont
  {Samal}}]{Sreejith2017}%
  \BibitemOpen
  \bibfield  {author} {\bibinfo {author} {\bibfnamefont {R.~P.}\ \bibnamefont
  {Sreejith}}, \bibinfo {author} {\bibfnamefont {J.}~\bibnamefont {Jost}},
  \bibinfo {author} {\bibfnamefont {E.}~\bibnamefont {Saucan}}, \ and\ \bibinfo
  {author} {\bibfnamefont {A.}~\bibnamefont {Samal}},\ }\href@noop {}
  {\bibfield  {journal} {\bibinfo  {journal} {Chaos, Solitons \& Fractals}\
  }\textbf {\bibinfo {volume} {101}},\ \bibinfo {pages} {50} (\bibinfo {year}
  {2017})}\BibitemShut {NoStop}%
\bibitem [{\citenamefont {Menger}(1930)}]{Menger1930}%
  \BibitemOpen
  \bibfield  {author} {\bibinfo {author} {\bibfnamefont {K.}~\bibnamefont
  {Menger}},\ }\href@noop {} {\bibfield  {journal} {\bibinfo  {journal}
  {Mathematische {A}nnalen}\ }\textbf {\bibinfo {volume} {103}},\ \bibinfo
  {pages} {466} (\bibinfo {year} {1930})}\BibitemShut {NoStop}%
\bibitem [{\citenamefont {Saucan}\ \emph
  {et~al.}(2019{\natexlab{b}})\citenamefont {Saucan}, \citenamefont {Samal},\
  and\ \citenamefont {Jost}}]{Saucan2019b}%
  \BibitemOpen
  \bibfield  {author} {\bibinfo {author} {\bibfnamefont {E.}~\bibnamefont
  {Saucan}}, \bibinfo {author} {\bibfnamefont {A.}~\bibnamefont {Samal}}, \
  and\ \bibinfo {author} {\bibfnamefont {J.}~\bibnamefont {Jost}},\ }in\
  \href@noop {} {\emph {\bibinfo {booktitle} {International {C}onference on
  {C}omplex Networks and their {A}pplications}}}\ (\bibinfo {organization}
  {Springer},\ \bibinfo {year} {2019})\ pp.\ \bibinfo {pages}
  {943--954}\BibitemShut {NoStop}%
\bibitem [{\citenamefont {Haantjes}(1947)}]{Haantjes1947}%
  \BibitemOpen
  \bibfield  {author} {\bibinfo {author} {\bibfnamefont {J.}~\bibnamefont
  {Haantjes}},\ }\href@noop {} {\bibfield  {journal} {\bibinfo  {journal}
  {Proc. {K}on. {N}ed. {A}kad. v. {W}etenseh., Amsterdam}\ }\textbf {\bibinfo
  {volume} {50}},\ \bibinfo {pages} {302} (\bibinfo {year} {1947})}\BibitemShut
  {NoStop}%
\bibitem [{Yah()}]{Yahoo_finance}%
  \BibitemOpen
  \href@noop {} {\enquote {\bibinfo {title} {Yahoo finance database},}\
  }\bibinfo {howpublished} {\url{https://finance.yahoo.co.jp/}},\ \bibinfo
  {note} {accessed on 7th July, 2017.}\BibitemShut {Stop}%
\bibitem [{\citenamefont {Prim}(1957)}]{Prim1957}%
  \BibitemOpen
  \bibfield  {author} {\bibinfo {author} {\bibfnamefont {R.~C.}\ \bibnamefont
  {Prim}},\ }\href@noop {} {\bibfield  {journal} {\bibinfo  {journal} {The Bell
  System Technical Journal}\ }\textbf {\bibinfo {volume} {36}},\ \bibinfo
  {pages} {1389} (\bibinfo {year} {1957})}\BibitemShut {NoStop}%
\bibitem [{\citenamefont {Latora}\ and\ \citenamefont
  {Marchiori}(2001)}]{Latora2001}%
  \BibitemOpen
  \bibfield  {author} {\bibinfo {author} {\bibfnamefont {V.}~\bibnamefont
  {Latora}}\ and\ \bibinfo {author} {\bibfnamefont {M.}~\bibnamefont
  {Marchiori}},\ }\href@noop {} {\bibfield  {journal} {\bibinfo  {journal}
  {Physical Review Letters}\ }\textbf {\bibinfo {volume} {87}},\ \bibinfo
  {pages} {198701} (\bibinfo {year} {2001})}\BibitemShut {NoStop}%
\bibitem [{\citenamefont {Girvan}\ and\ \citenamefont
  {Newman}(2002)}]{Girvan2002}%
  \BibitemOpen
  \bibfield  {author} {\bibinfo {author} {\bibfnamefont {M.}~\bibnamefont
  {Girvan}}\ and\ \bibinfo {author} {\bibfnamefont {M.~E.~J.}\ \bibnamefont
  {Newman}},\ }\href@noop {} {\bibfield  {journal} {\bibinfo  {journal}
  {Proceedings of the National Academy of Sciences USA}\ }\textbf {\bibinfo
  {volume} {99}},\ \bibinfo {pages} {7821} (\bibinfo {year}
  {2002})}\BibitemShut {NoStop}%
\bibitem [{\citenamefont {Blondel}\ \emph {et~al.}(2008)\citenamefont
  {Blondel}, \citenamefont {Guillaume}, \citenamefont {Lambiotte},\ and\
  \citenamefont {Lefebvre}}]{Blondel2008}%
  \BibitemOpen
  \bibfield  {author} {\bibinfo {author} {\bibfnamefont {V.~D.}\ \bibnamefont
  {Blondel}}, \bibinfo {author} {\bibfnamefont {J.-L.}\ \bibnamefont
  {Guillaume}}, \bibinfo {author} {\bibfnamefont {R.}~\bibnamefont
  {Lambiotte}}, \ and\ \bibinfo {author} {\bibfnamefont {E.}~\bibnamefont
  {Lefebvre}},\ }\href@noop {} {\bibfield  {journal} {\bibinfo  {journal}
  {Journal of Statistical Mechanics: Theory and Experiment}\ }\textbf {\bibinfo
  {volume} {2008}},\ \bibinfo {pages} {P10008} (\bibinfo {year}
  {2008})}\BibitemShut {NoStop}%
\bibitem [{\citenamefont {Sol{\'e}}\ and\ \citenamefont
  {Valverde}(2004)}]{Sole2004}%
  \BibitemOpen
  \bibfield  {author} {\bibinfo {author} {\bibfnamefont {R.~V.}\ \bibnamefont
  {Sol{\'e}}}\ and\ \bibinfo {author} {\bibfnamefont {S.}~\bibnamefont
  {Valverde}},\ }in\ \href@noop {} {\emph {\bibinfo {booktitle} {Complex
  Networks}}}\ (\bibinfo  {publisher} {Springer},\ \bibinfo {year} {2004})\
  pp.\ \bibinfo {pages} {189--207}\BibitemShut {NoStop}%
\bibitem [{\citenamefont {Hagberg}\ \emph {et~al.}(2008)\citenamefont
  {Hagberg}, \citenamefont {Swart},\ and\ \citenamefont {Chult}}]{Hagberg2008}%
  \BibitemOpen
  \bibfield  {author} {\bibinfo {author} {\bibfnamefont {A.}~\bibnamefont
  {Hagberg}}, \bibinfo {author} {\bibfnamefont {P.}~\bibnamefont {Swart}}, \
  and\ \bibinfo {author} {\bibfnamefont {D.~S.}\ \bibnamefont {Chult}},\
  }\href@noop {} {\emph {\bibinfo {title} {Exploring network structure,
  dynamics, and function using NetworkX}}},\ \bibinfo {type} {Tech. Rep.}\
  (\bibinfo  {institution} {Los Alamos National Laboratory (LANL), Los Alamos,
  NM (United States)},\ \bibinfo {year} {2008})\BibitemShut {NoStop}%
\bibitem [{\citenamefont {Bollerslev}(1986)}]{Bollerslev1986}%
  \BibitemOpen
  \bibfield  {author} {\bibinfo {author} {\bibfnamefont {T.}~\bibnamefont
  {Bollerslev}},\ }\href@noop {} {\bibfield  {journal} {\bibinfo  {journal}
  {Journal of Econometrics}\ }\textbf {\bibinfo {volume} {31}},\ \bibinfo
  {pages} {307} (\bibinfo {year} {1986})}\BibitemShut {NoStop}%
\bibitem [{Cra({\natexlab{a}})}]{CrashList}%
  \BibitemOpen
  \href@noop {} {\enquote {\bibinfo {title} {List of stock market crashes and
  bear markets},}\ }\bibinfo {howpublished}
  {\url{https://en.wikipedia.org/wiki/List_of_stock_market_crashes_and_bear_markets}}
  ({\natexlab{a}}),\ \bibinfo {note} {accessed on 7th July, 2019.}\BibitemShut
  {Stop}%
\bibitem [{Bul()}]{Bullmarkets}%
  \BibitemOpen
  \href@noop {} {\enquote {\bibinfo {title} {Bull markets},}\ }\bibinfo
  {howpublished} {\url{https://bullmarkets.co/u-s-stock-market-in-1996/}},\
  \bibinfo {note} {accessed on 7th July, 2019.}\BibitemShut {Stop}%
\bibitem [{USh()}]{UShousingbubble}%
  \BibitemOpen
  \href@noop {} {\enquote {\bibinfo {title} {United states housing bubble},}\
  }\bibinfo {howpublished}
  {\url{https://en.wikipedia.org/wiki/United_States_housing_bubble}},\ \bibinfo
  {note} {accessed on 7th July, 2019.}\BibitemShut {Stop}%
\bibitem [{Cra({\natexlab{b}})}]{CrashHistory}%
  \BibitemOpen
  \href@noop {} {\enquote {\bibinfo {title} {A short history of stock market
  crashes},}\ }\bibinfo {howpublished}
  {\url{https://www.cnbc.com/2016/08/24/a-short-history-of-stock-market-crashes.html}}
  ({\natexlab{b}}),\ \bibinfo {note} {accessed on 7th July, 2019.}\BibitemShut
  {Stop}%
\bibitem [{Sel()}]{Selloff}%
  \BibitemOpen
  \href@noop {} {\enquote {\bibinfo {title} {Stock market selloff},}\ }\bibinfo
  {howpublished}
  {\url{https://en.wikipedia.org/wiki/2015-16_stock_market_selloff}},\ \bibinfo
  {note} {accessed on 7th July, 2019.}\BibitemShut {Stop}%
\bibitem [{mar()}]{marketfall2011}%
  \BibitemOpen
  \href@noop {} {\enquote {\bibinfo {title} {August 2011 stock markets fall},}\
  }\bibinfo {howpublished}
  {\url{https://en.wikipedia.org/wiki/August_2011_stock_markets_fall}},\
  \bibinfo {note} {accessed on 13th August, 2020.}\BibitemShut {Stop}%
\bibitem [{\citenamefont {Kenett}\ \emph {et~al.}(2010)\citenamefont {Kenett},
  \citenamefont {Tumminello}, \citenamefont {Madi}, \citenamefont
  {Gur-Gershgoren}, \citenamefont {Mantegna},\ and\ \citenamefont
  {Ben-Jacob}}]{Kenett2010}%
  \BibitemOpen
  \bibfield  {author} {\bibinfo {author} {\bibfnamefont {D.~Y.}\ \bibnamefont
  {Kenett}}, \bibinfo {author} {\bibfnamefont {M.}~\bibnamefont {Tumminello}},
  \bibinfo {author} {\bibfnamefont {A.}~\bibnamefont {Madi}}, \bibinfo {author}
  {\bibfnamefont {G.}~\bibnamefont {Gur-Gershgoren}}, \bibinfo {author}
  {\bibfnamefont {R.~N.}\ \bibnamefont {Mantegna}}, \ and\ \bibinfo {author}
  {\bibfnamefont {E.}~\bibnamefont {Ben-Jacob}},\ }\href@noop {} {\bibfield
  {journal} {\bibinfo  {journal} {PloS {O}ne}\ }\textbf {\bibinfo {volume}
  {5}},\ \bibinfo {pages} {e15032} (\bibinfo {year} {2010})}\BibitemShut
  {NoStop}%
\bibitem [{\citenamefont {San~Miguel}\ \emph {et~al.}(2012)\citenamefont
  {San~Miguel}, \citenamefont {Johnson}, \citenamefont {Kertesz}, \citenamefont
  {Kaski}, \citenamefont {D{\'\i}az-Guilera}, \citenamefont {MacKay},
  \citenamefont {Loreto}, \citenamefont {{\'E}rdi},\ and\ \citenamefont
  {Helbing}}]{San2012}%
  \BibitemOpen
  \bibfield  {author} {\bibinfo {author} {\bibfnamefont {M.}~\bibnamefont
  {San~Miguel}}, \bibinfo {author} {\bibfnamefont {J.~H.}\ \bibnamefont
  {Johnson}}, \bibinfo {author} {\bibfnamefont {J.}~\bibnamefont {Kertesz}},
  \bibinfo {author} {\bibfnamefont {K.}~\bibnamefont {Kaski}}, \bibinfo
  {author} {\bibfnamefont {A.}~\bibnamefont {D{\'\i}az-Guilera}}, \bibinfo
  {author} {\bibfnamefont {R.~S.}\ \bibnamefont {MacKay}}, \bibinfo {author}
  {\bibfnamefont {V.}~\bibnamefont {Loreto}}, \bibinfo {author} {\bibfnamefont
  {P.}~\bibnamefont {{\'E}rdi}}, \ and\ \bibinfo {author} {\bibfnamefont
  {D.}~\bibnamefont {Helbing}},\ }\href@noop {} {\bibfield  {journal} {\bibinfo
   {journal} {The {E}uropean {P}hysical {J}ournal {S}pecial {T}opics}\ }\textbf
  {\bibinfo {volume} {214}},\ \bibinfo {pages} {245} (\bibinfo {year}
  {2012})}\BibitemShut {NoStop}%
\bibitem [{\citenamefont {Millington}\ and\ \citenamefont
  {Niranjan}(2020)}]{Millington2020}%
  \BibitemOpen
  \bibfield  {author} {\bibinfo {author} {\bibfnamefont {T.}~\bibnamefont
  {Millington}}\ and\ \bibinfo {author} {\bibfnamefont {M.}~\bibnamefont
  {Niranjan}},\ }\href@noop {} {\bibfield  {journal} {\bibinfo  {journal}
  {Applied Network Science}\ }\textbf {\bibinfo {volume} {5}},\ \bibinfo
  {pages} {1} (\bibinfo {year} {2020})}\BibitemShut {NoStop}%
\bibitem [{\citenamefont {Sharma}\ \emph {et~al.}(2017)\citenamefont {Sharma},
  \citenamefont {Shah}, \citenamefont {Chakrabarti},\ and\ \citenamefont
  {Chakraborti}}]{Sharma2017}%
  \BibitemOpen
  \bibfield  {author} {\bibinfo {author} {\bibfnamefont {K.}~\bibnamefont
  {Sharma}}, \bibinfo {author} {\bibfnamefont {S.}~\bibnamefont {Shah}},
  \bibinfo {author} {\bibfnamefont {A.~S.}\ \bibnamefont {Chakrabarti}}, \ and\
  \bibinfo {author} {\bibfnamefont {A.}~\bibnamefont {Chakraborti}},\ }in\
  \href {\doibase 10.1007/978-981-10-5705-2_11} {\emph {\bibinfo {booktitle}
  {Economic Foundations for Social Complexity Science}}}\ (\bibinfo
  {publisher} {Springer},\ \bibinfo {year} {2017})\ pp.\ \bibinfo {pages}
  {211--238}\BibitemShut {NoStop}%
\end{thebibliography}
\end{document}